\documentclass[aps,prd,showpacs,nofootinbib,superscriptaddress,preprint,tightenlines,eqsecnum,longbibliography]{revtex4}
\usepackage{amssymb,amsmath}
\usepackage{graphicx}
\usepackage{subfigure}
\usepackage{leftidx}

\usepackage[toc,page]{appendix}

\usepackage[mathscr]{eucal}

\usepackage{pictexwd,dcpic}

\newcommand{\vd}[2]{\frac{\delta #1}{\delta #2}}   

\newcommand{\ord}[2]{\text{o}({#1}^{#2})}  

\newcommand{\Ord}[3]{\text{O}({#1}^{#2}) \text{ #3}}  

\newcommand{\md}{{\mathrm d}}

\newcommand{\dd}{\md  \hspace{-0.9ex} \md}  

\begin{document}

\title{The gravitational Hamiltonian, \\first order action, Poincar\'e charges and surface terms}
\author{Alejandro Corichi}
\email{corichi@matmor.unam.mx}
\affiliation{Centro de Ciencias Matem\'aticas, Universidad Nacional Aut\'onoma de
M\'exico, UNAM-Campus Morelia, A. Postal 61-3, Morelia, Michoac\'an 58090,
Mexico}
\affiliation{Center for Fundamental Theory, Institute for Gravitation and the Cosmos,
Pennsylvania State University, University Park
PA 16802, USA}
\author{Juan D. Reyes}
\email{jdrp75@gmail.com, jdreyes@matmor.unam.mx}
\affiliation{Centro de Ciencias Matem\'aticas, Universidad Nacional Aut\'onoma de
M\'exico, UNAM-Campus Morelia, A. Postal 61-3, Morelia, Michoac\'an 58090,
Mexico}
\affiliation{Instituto de F\'{\i}sica y
Matem\'aticas,  Universidad Michoacana de San Nicol\'as de
Hidalgo, Morelia, Michoac\'an, Mexico}

\begin{abstract}
We consider the issue of attaining a consistent Hamiltonian formulation, after a 3+1 splitting, of a well defined action principle for asymptotically flat gravity. 
More precisely, our starting 
point is the gravitational first order Holst action with surface terms and fall-off conditions that make 
the variational principle and the covariant phase space formulation well defined for asymptotically flat 
spacetimes. Keeping all surface terms and paying due attention to subtleties that arise from the 
different cut-offs at infinity, we give a derivation of the gravitational Hamiltonian starting from this 
action. The 3+1 decomposition and time gauge fixing results in a well defined Hamiltonian action and a 
well defined Hamiltonian formulation for the standard -and more general- asymptotic ADM conditions. Unlike 
the case of the Einstein-Hilbert action with Gibbons-Hawking-York or Hawking-Horowitz terms, here 
we {\it {do}} 
recover the ADM energy-momentum from the covariant surface term also when more general variations 
respecting asymptotic flatness are allowed. Additionally, our strategy yields a derivation of the parity 
conditions for connection variables independent of the conditions given by Regge and Teitelboim for ADM 
variables. Finally, we exhibit the other Poincar\'e generators in terms of real Ashtekar-Barbero 
variables. We complement previous constructions in self-dual variables by pointing out several subtleties 
and refining the argument showing that  -on shell- they coincide with the ADM charges. Our results represent the first consistent treatment of the Hamiltonian formulation for the connection-tetrad gravitational degrees of freedom, starting from a well posed action, in the case of asymptotically flat boundary conditions.
\end{abstract}

\pacs{04.20.Fy, 04.20.Ha, 04.20.Cv}

\maketitle

\section{Introduction}

Ever since the formulation of the general theory of relativity was being finalized, almost 100 years ago,
having a variational principle for it was at the forefront of the theoretical efforts. The original 
formulation of the theory by Einstein \cite{einstein} and Hilbert \cite{hilbert} was, of course, 
in terms of a metric tensor 
$g_{\mu\nu}$. Later on, it was realized that in order to incorporate Fermions into the theory one needs to 
consider instead tetrads $e_\mu^I$ as fundamental variables. On the other hand, a first order variational 
principle, where a connection variable features amongst the fundamental variables, was considered as early 
as 1919 by Palatini \cite{palatini}. A combination of these two elements, namely tetrads and a first order formulation, was completed almost 60 years ago by several authors \cite{utiyama,sciama,kibble}. 
In all these cases, however, the existence of 
boundaries, or fall-off conditions for spacetimes without boundaries was largely ignored. 
It was Gibbons and Hawking who first put forward a modified action principle to account for the presence of 
boundaries, in the second order formulation \cite{GibbonsHawking}.
A treatment of 
general relativity in a Hamiltonian language was completed in the early 60's, for the action principle cast in geometric (ADM) variables \cite{ADM1962}, 
and in the 80's for the connection-tetrad variables \cite{di,aa,jacobson:smolin,samuel}. The consistent incorporation of boundary terms for asymptotically flat boundary 
conditions, in the purely Hamiltonian formulation, was done by Regge-Teitelboim for the ADM variables 
\cite{ReggeTeitelboim}. The analogue treatment for self-dual connection-triad variables was put forward by Thiemann in \cite{Thiemann}.
It is then somewhat surprising that a complete treatment of general relativity in connection-tetrad variables 
with asymptotically flat boundary conditions is still missing. To be precise, a 3+1 splitting of a well defined action principle, and a consistent treatment of the corresponding Hamiltonian theory is yet to be constructed. The purpose of this manuscript is to fill this gap.

The Holst action is a first order covariant action based on orthonormal tetrads and Lorentz connections. 
The `Holst term' adding to the  Einstein-Palatini action was originally introduced in \cite{HojmanMS},  but it was interpreted there as a functional of metric variables instead of tetrads.
It was Holst  \cite{Holst} who first showed that its 3+1 decomposition plus partial gauge fixing gives, for compact spacetimes without boundaries, a Hamiltonian action for general relativity in terms of Ashtekar-Barbero variables. The Holst action is the simplest first order action producing a canonical theory without the complications of second class constraints --as is the case of the simpler Einstein-Palatini action \cite{Romano}--. 
It is, furthermore, the classical starting point for Loop Quantum Gravity and some Spin-foam models.

In \cite{CorichiWilsonEwing}, surface terms were introduced to supplement the Holst action, resulting
in a proper treatment of asymptotically flat spacetimes. These surface terms give a manifestly 
finite action even off-shell\footnote{As we shall see, this is true for so called cylindrical temporal 
cut-offs or asymptotically time-translated spatial boundaries \cite{AshtekarES,MannMarolf}}, 
 and a well defined variational principle, reproducing Einstein's equations under all asymptotically  flat variations. Furthermore, the amended action leads naturally to a well defined covariant phase space in which the Hamiltonians generating asymptotic symmetries provide the total energy-momentum and angular momentum of the spacetime.

The objective of this paper is to continue and amend the work started in \cite{Loops11Proceeding} and to 
extend the results of \cite{Holst} for asymptotically flat boundary conditions. We will show that the 3+1 
decomposition and time gauge fixing of the extended action in \cite{CorichiWilsonEwing} leads (partially 
on-shell) also to a well defined Hamiltonian action in Ashtekar-Barbero variables and where the surface 
term recovers precisely the ADM energy and ADM momentum. As expected as this may be, this is a non-trivial 
result once one takes into account all the technicalities involved in evaluating the asymptotic limit and 
the requirements for a well defined variational principle. One should note that a similar analysis for asymptotically flat 2+1 first order gravity was recently completed in \cite{Cor:Rub}.

As a comparison, in the conventional treatment of the Einstein-Hilbert action for spacetimes with asymptotically flat boundary, two additional terms are  necessary in order to have a well defined variational principle that yields Einstein's equations from the stationary points of the action:
\begin{equation} \label{EHGHaction}
   S_{\textrm{EHGH}}=\frac{1}{2\kappa}\left(\int_{M} \md^4x \sqrt{-g}\,R \,+2\int_{\partial {M}}\md^3y\sqrt{|h|}(K-K_0) \,\right).
\end{equation}
The first term often referred to as the Gibbons-Hawking surface term \cite{GibbonsHawking} is inserted in 
the action so that its variation exactly cancels variations of the first derivatives of the metric, so 
only the metric is required to be fixed at the boundary.  Here $\kappa=8\pi G$, ${M}$ is an 
appropriate portion of spacetime and $\partial {M}$ its boundary, $R$ the Ricci scalar of the 4-metric $g$, $h$ the induced metric on  $\partial {M}$, and $K$ the trace of its extrinsic 
curvature. Since this term proportional to extrinsic curvature is generally divergent for asymptotically flat solutions the last `non-
dynamical' counter-term is required to make the action finite on-shell.

 Hawking and Horowitz \cite{HawkingHorowitz} have shown the surface terms corresponding to ADM energy and momentum may be recovered from the 3+1 decomposition  of (\ref{EHGHaction}), but this is only true if one fixes  variations also at the boundary at spatial infinity.
In a more careful treatment though, for asymptotically flat spacetimes,  the action should be such that asymptotically flat solutions are stationary points under all variations preserving asymptotic flatness, not just under variations of compact support. The Einstein-Hilbert action with Gibbons-Hawking term does not satisfy this requirement. Under all asymptotically flat variations, its variation gives  a non-vanishing surface term when Einstein's equations are satisfied.  Furthermore, the counter term becomes dynamical and since it requires an embedding of $\partial {M}$ in a reference background, which is not always guaranteed, its  variation is not even well defined \cite{MannMarolf}. 

Several proposals or generalizations of a counter-term for the second order action exist in the literature which aim to correct this problem \cite{KrausLarsenSiebelink, MannMarolf}. Most notably in \cite{MannMarolf} a generalization replacing the last term in (\ref{EHGHaction})  was given which does not require an embedding. This addition gives a well defined variational principle for all variations consistent with asymptotic flatness and allows for the computation of conserved charges in the spirit of Brown and York \cite{BrownYork}. Furthermore, although only given implicitly by a Gauss-Codazzi type equation, it was shown in \cite{MannMarolfVirmani} that this surface counter-term (a local function of boundary metric and Ricci curvature) together with the Gibbons-Hawking term do yield the ADM energy and momentum after a 3+1 decomposition. 
Our purpose is to show similar results in the first order formalism for the Holst action. The advantage of the first order treatment is that, unlike \cite{MannMarolf}, the surface term is given explicitly and there are no infinite counter-terms involved. 

Additionally, we want to fill in a gap in the literature that until very recently had not been properly addressed.
We wish to analyse and contrast (draw a clear distinction between) the different asymptotic conditions  necessary for a well defined Hamiltonian action in Ashtekar-Barbero variables and those for a consistent Hamiltonian formulation admitting well defined Poincar\'e generators. The fall-off conditions for connection and triad variables and the form of the Poincar\'e generators and surface terms corresponding to conserved charges were derived directly from the ADM framework in  the canonical treatment in \cite{Thiemann}. This work extended the spinorial description in \cite{AshtekarLectures}. Nevertheless, no careful comparison to the covariant framework had been given and the generators and charges were exhibited only for the self-dual case. Here we amend the results of \cite{Thiemann} and show explicitly the canonical phase space and the generators and charges for arbitrary Barbero-Immirzi parameter. As we shall see some subtleties do arise, particularly in the asymptotic expansions for the triad and connection necessary to ensure the surface counter terms match the ADM angular momentum charges. 
 
Finally, we should note that at the time of preparation of this article, an independent derivation of the canonical phase space in real Ashtekar-Barbero variables,  along with the construction of the generators came out \cite{Campiglia}. Hence, some of our observations and discussion on this matter necessarily overlap. However, we should point out that the expressions we present here extend those of \cite{Thiemann}, while those in \cite{Campiglia} are constructed directly on the real phase space by counter term methods. Of course, both sets of generators necessarily agree modulo a multiple of the Gauss constraint.

The manuscript is organized as follows. In Sec.~\ref{s::preliminaries} we present some preliminaries needed for the rest of the manuscript. In particular, we recall the notion of asymptotically flat spacetimes needed for a consistent variational principle. In Sec.~\ref{sec:3} we spell out in detail the action principle that we shall consider, together with the asymptotic conditions for the basic variables. The 3+1 splitting of spacetime is the subject of Sec.~\ref{sec:4}. In it, we perform the decomposition of the first order action  and arrive at the corresponding Hamiltonian description. The precise Hamiltonian formulation is presented in Sec.~\ref{s::HamiltonianFormulation}, where we analyse in detail the corresponding phase space and the Poincar\'e generators. We present a summary and a discussion in Sec.~\ref{sec:6}. With the aim of making this work self-contained, we have included many details of the calculations in three  appendices. In the first one, we compare the hyperbolic and cylindrical asymptotic expansions of the basic fields. In the second Appendix we present all the specifics of the 3+1 splitting with the time-gauge fixing. In the third Appendix we construct in detail the generators in the connection-triad variables. Here we regard the phase space in terms of Ashtekar-Barbero variables as an extension of the ADM phase space in order to re-derive the generators from the corresponding ADM expressions and to point out several subtleties not mentioned in \cite{Thiemann}.



\section{Preliminaries: asymptotically flat spacetimes and the variational principle}  \label{s::preliminaries}

Since ultimately we wish to perform the 3+1 decomposition of the first order Holst action (\ref{Holst1}) for asymptotically flat spacetimes  and connect with the Hamiltonian formulation, throughout this work  we will take spacetime $\mathcal{M}$ to be globally hyperbolic and such that it may be foliated as $\mathcal{M}\approx\mathbb{R}\times\Sigma$. As most treatments implicitly or explicitly do,  we will also assume the topology of the  hypersurface $\Sigma\approx K\cup V$ to be such that it is homeomorphic to the union of a compact space $K$ and an \emph{asymptotic region} $V$ which itself is diffeomorphic to the complement of a closed ball in $\mathbb{R}^3\,$\footnote{The validity of such assumptions for spacetimes satisfying the geometric definition in \cite{AshtekarHansen} is consistent with the construction or existence, shown thereon in appendix B for such spacetimes, of so-called \emph{asymptotically flat initial data sets} \cite{Geroch}.}. There may be additional asymptotic regions, but since they can all be treated identically, for simplicity we will restrict ourselves to just a single one here.  The asymptotic conditions for the components of the spacetime metric $g_{\mu\nu}$  stated below, and the fall-off conditions for all required additional dynamical fields, will refer to a coordinate patch on the \emph{asymptotic spacetime region} $U\subset\mathcal{M}$, homeomorphic to $\mathbb{R}\times V$.

Generically, a first order local action is  given by a four-dimensional integral over a region $M\subseteq\mathcal{M}$ and a three-dimensional integral over the boundary $\partial M$. The integrand functionals --a Lagrangian density $\mathcal{L}$ and $F$-- depending on fields $\phi^i(x^\mu)$ and their derivatives:  
  \begin{equation} \label{genericAction}
          S=\int_M\md^4x\, \mathcal{L}(\phi^i,\nabla_\mu\phi^i)\,+\int_{\partial M}\md^3y\, F(\phi^i,\nabla_\mu\phi^i)\,.
 \end{equation}
(We take a covariant or Lagrangian formulation here, but similar considerations follow if we perform a $3+1$ decomposition or start with a $3+1$ Hamiltonian theory with spatial fields $\phi^i(t,\vec{x})$, as we will consider later.)
A complete definition of the action demands a detailed description of the spacetime region $M$ as well as of the boundary and/or fall-off conditions of the dynamical fields $\phi^i$. Furthermore, the variational principle also requires a specification of the type of variations $\delta\phi^i$ to be considered.

  \begin{figure}    
     \begin{center}
      \includegraphics[width=5.2cm]{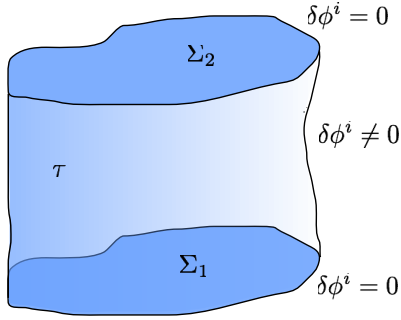}
      \caption{Generic boundary $\partial M=\tau\cup\Sigma_1\cup\Sigma_2$ for spacetime region of integration. Variations may be `fixed' only at the Cuachy surfaces $\Sigma_1$ and $\Sigma_2$ and must remain compatible with asymptotic conditions for the original fields.} \label{generalBoundaries}
      \end{center}
 \end{figure}

The spacetime region $M\subset\mathcal{M}\approx \mathbb{R}\times\Sigma$ may be assumed to be bounded by two Cauchy surfaces $\Sigma_1$ and $\Sigma_2$ and a time-like `world tube' $\tau:=\partial M-(\Sigma_1\cup\Sigma_2)$ (Fig. \ref{generalBoundaries}) (for simplicity here, we will not worry about inner boundaries).  More precisely, for asymptotically flat spacetimes or more general non-compact $\Sigma$, the definition of the action (\ref{genericAction}) and its variation actually involves a limiting process.
As clarified in \cite{MannMarolf}, the region $M$ and its boundary $\partial M$ strictly refer to a one-parameter family of regions of spacetime $\{M_\Omega\}_{\Omega\in\mathcal{I}\subset\mathbb{R}}$ and their associated boundaries $\{\partial M_\Omega\}_{\Omega\in\mathcal{I}\subset\mathbb{R}}$, such that
  $M_\Omega\subset M_{\Omega'}$    for  $\Omega<\Omega'$, and  $M=\bigcup_\Omega M_\Omega$. The action (\ref{genericAction}) is then a shorthand notation for or is rigorously defined as
 \begin{equation} \label{limgenericAction}
 S=\lim_{\Omega\to\infty} S_\Omega:=\lim_{\Omega\to\infty}\, \int_{M_\Omega}\md^4x\, \mathcal{L}(\phi^i,\nabla_\mu\phi^i)\,+\int_{\partial M_\Omega}\md^3y\, F(\phi^i,\nabla_\mu\phi^i)\,,
 \end{equation}
and its variation is defined as $\delta S := \lim_{\Omega\to\infty}\delta S_\Omega$. 

\begin{figure}[h]  
       \begin{center}
       \subfigure{
      \includegraphics[width=4cm]{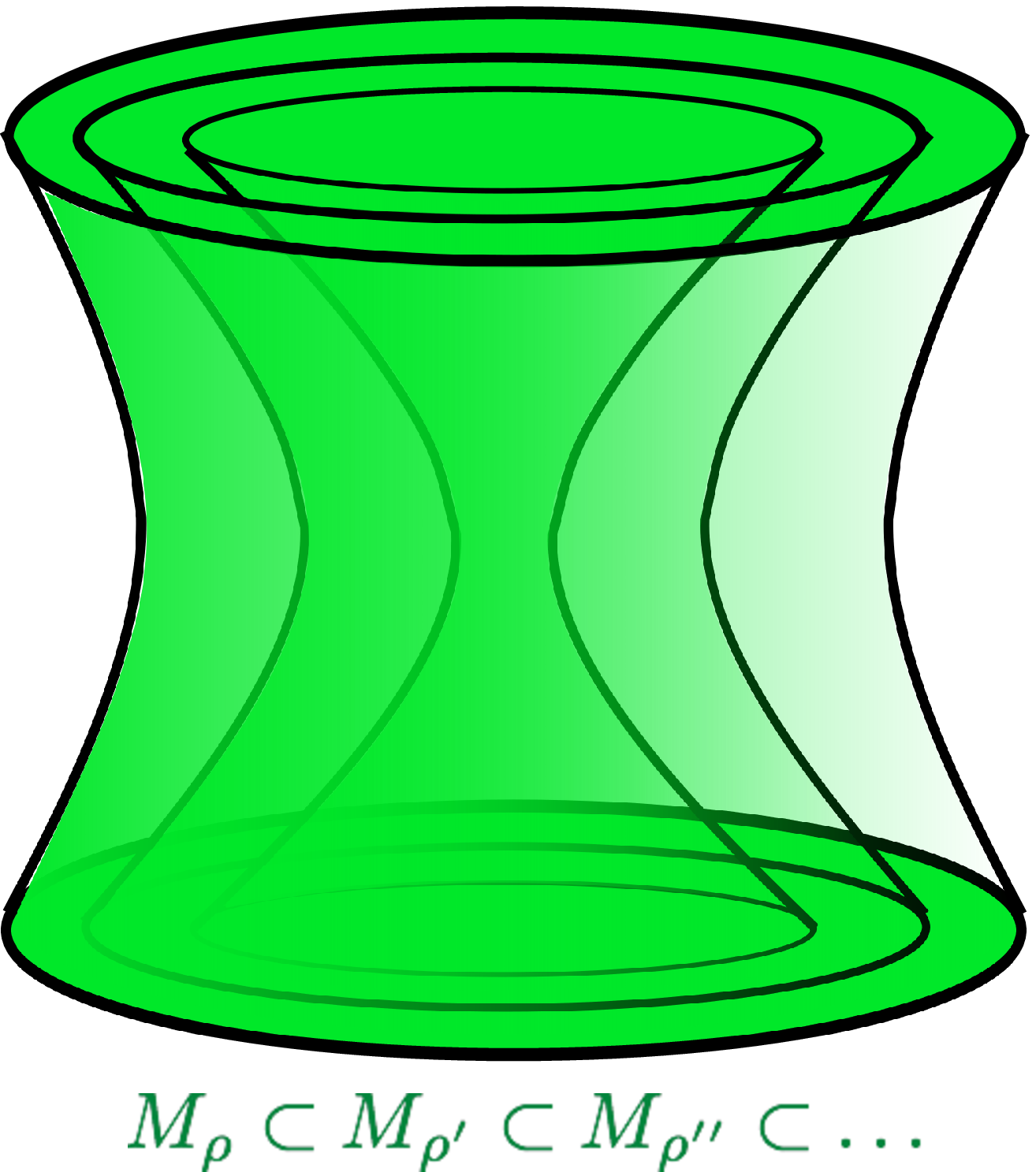}  
     } 
      \hspace{2cm}
    \subfigure{ 
    \includegraphics[width=3.8cm]{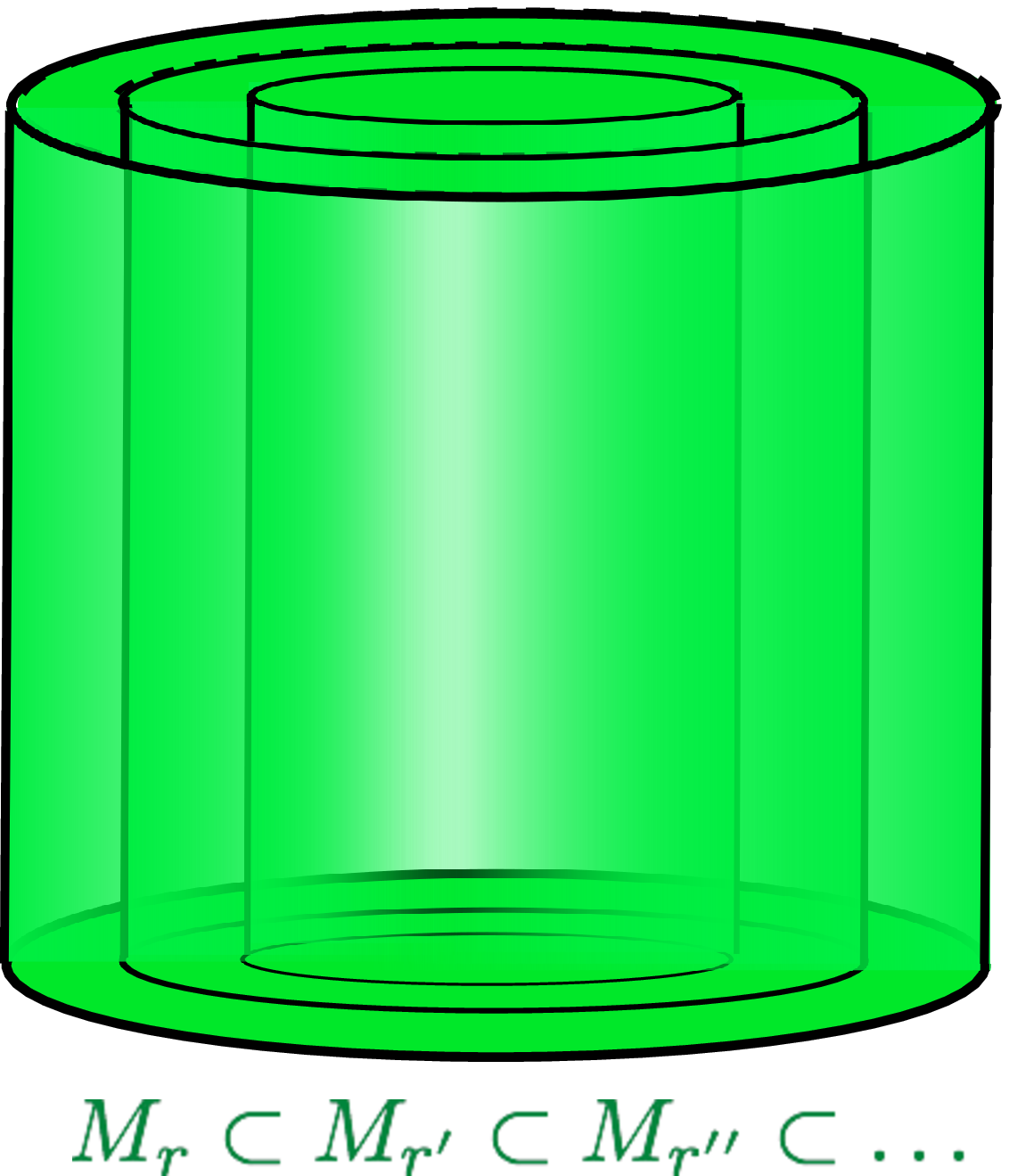}  
    }
    \caption{Hyperbolic and cylindrical cut-offs} \label{foliations}
    \end{center}
    \end{figure}

To exhibit  such families, let us choose a Minkowski or flat metric $\eta$ on the asymptotic region $U\subset\mathbb{R}^4$ which for now we may think of as embedded in Minkowski space.  Let $x^\mu$  or $(t,x^a)$ with $a=1,2,3$, be its corresponding Cartesian coordinates, i.e. $\eta=\text{diag}(-1,1,1,1)$ on this coordinate chart (here and in what follows spacetime indices running from 0 to 3 will be denoted by greek letters, while latin letters from the beginning of the alphabet will correspond to spatial indices).  Associated with these structures, one has \emph{spherical coordinates} $(r,\vartheta,\varphi)$ on the $t=\text{constant}$ slices, with
$r:=(\delta_{ab}x^ax^b)^\frac{1}{2}$ the spatial radial coordinate, and $(\vartheta$, $\varphi)$  the angular coordinates on the unit 2-sphere which may also be parameterized by the three coordinate functions $x^a/r$. Additionally, for the region $\mathcal{R}$ outside the light cone centered at the origin, one has associated \emph{hyperbolic coordinates} $(\rho,\chi,\vartheta,\varphi)$, with spacetime radial coordinate $\rho:=(\eta_{\mu\nu}x^\mu x^\nu)^\frac{1}{2}$, and $(\chi,\vartheta,\varphi)$ the standard angle coordinates on the unit 3-dimensional time-like hyperboloid $\mathcal{H}_1$ which may too be parameterized by  $(t/\rho,x^a/\rho)$.
These sets of coordinates satisfy the standard relations
\begin{equation} \label{hyperCylRelations}
\rho^2=r^2-t^2, \qquad  r=\rho\cosh\chi, \qquad t=\rho\sinh\chi, \qquad \tanh\chi=\frac{t}{r}\,.
\end{equation}
Hyperbolic or spherical coordinates allow one to foliate the region $\mathcal{R}$ of Minkowski space by time-like hyperboloids $\mathcal{H}_\rho$, with $\rho=\text{constant}$, or the the whole of Minkowski space (minus a line) by `cylinders' $\mathbb{R}\times S^2_r$ with $r=\text{constant}$. These hypersurfaces in turn determine the boundary of regions $M_\rho$ or $M_r$ which, when intersected with appropriate Cauchy surfaces $\Sigma_1$ and $\Sigma_2$, supply increasing families of regions parameterized by $\Omega=\rho$ or $\Omega=r$. These two families are the common choices used to define the region of integration $M$ for an action on Minkowski space and consecuently for general asymptotically flat spacetimes (Fig. \ref{foliations}). The family $\{M_\rho\}_{\rho\in\mathbb{R}^+}$ provides a \emph{hyperbolic cut-off} of spacetime and $\tau_\infty=\lim_{\rho\to\infty}\partial M_\rho$ a `hyperbolic description' of spatial infinity $\iota^0\,$\footnote{This is of course related to but not to be confused with the universal structure \emph{Spi}, the 4-dimensional manifold of space-like `geodesics' to infinity, which is a principal $\mathbb{R}$-bundle over the hyperboloid $\mathcal{H}_1$ of unit space-like directions and which is henceforth not a boundary of spacetime \cite{AshtekarHansen}. This `boundary' is more directly related to the treatment in \cite{AshtekarRomano}, which by means of an embedding of spacetime  different from a conformal completion, provides a description of $\iota^0$ as a three-dimensional boundary hypersurface rather than a point.}. This way of foliating spacetime is best suited for covariant treatments such as in \cite{AshtekarES, AshtekarBombelliReula, CorichiWilsonEwing}.  Similarly $\{M_r\}_{r\in\mathbb{R}^+}$ provides a \emph{cylindrical cut-off} of spacetime and $\tau_\infty=\lim_{r\to\infty}\partial M_r$ a `cylindrical description' of $\iota^0$. This is more appropriate, and the choice we will make, for a 3+1 splitting and a canonical formulation.

\begin{figure}[h]  
       \begin{center}
       \subfigure[Boosted $t=$constant hypersurfaces in Minkowski space.]{
       \includegraphics[width=4.8cm]{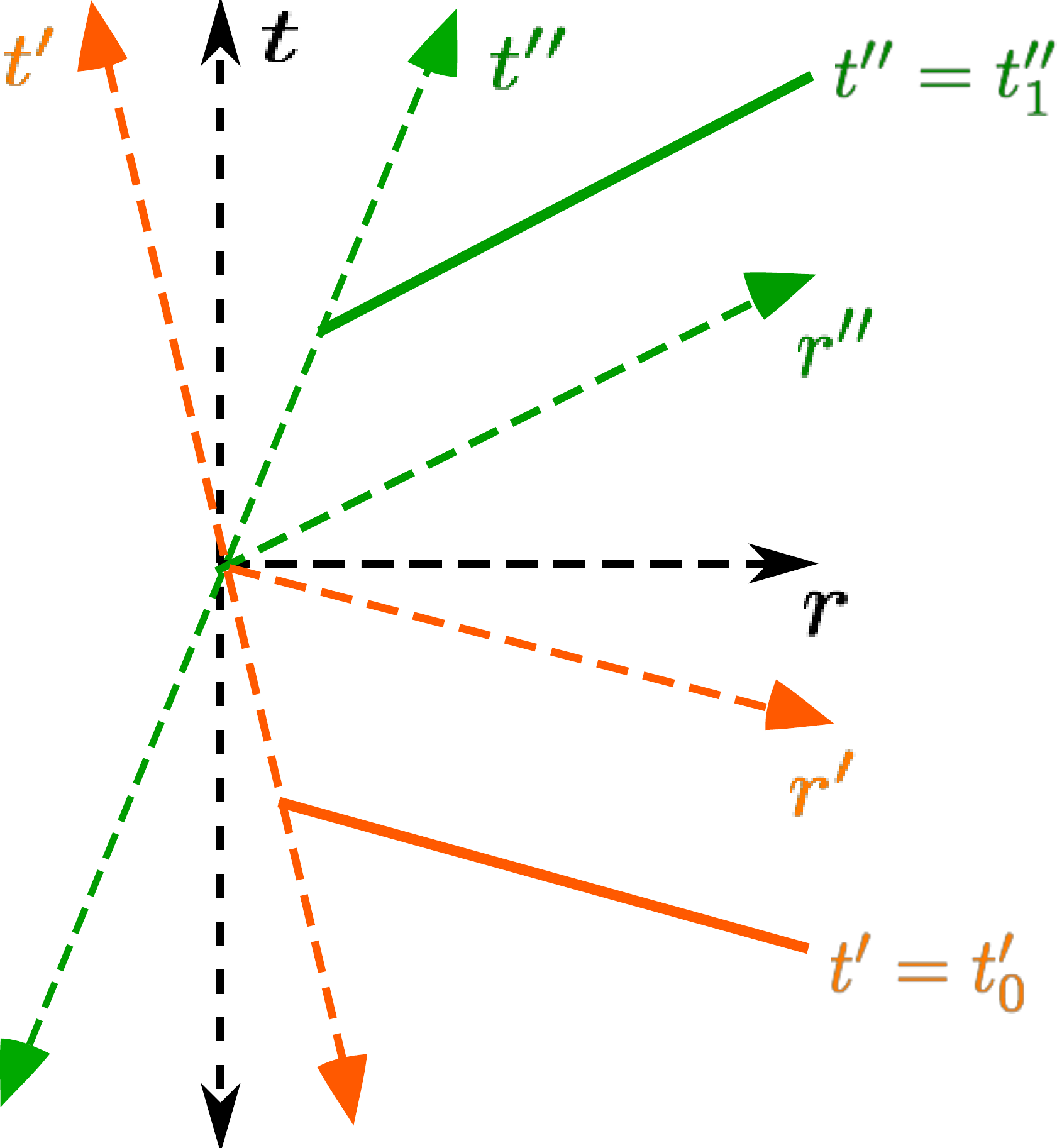}   
       \label{boosts}
     } 
     \hspace{.5cm}
    \subfigure[Corresponding  boosted slabs $\chi=$constant in asymptotic region.]{ 
    \includegraphics[width=4cm]{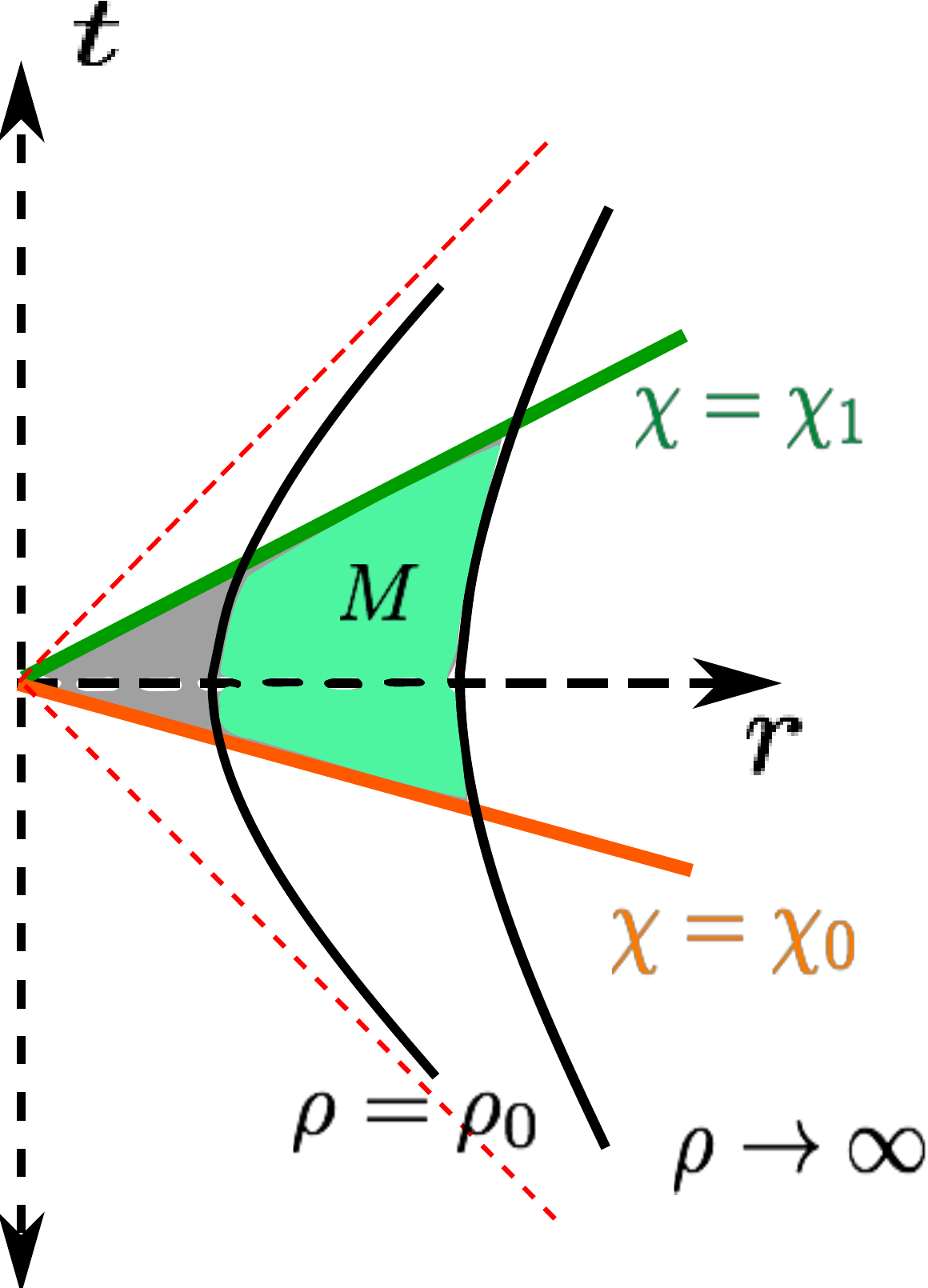}
    \label{boostedSlabs}
    }
     \hspace{.5cm}
    \subfigure[Cylindrical slabs $t=$constant.]{ 
    \includegraphics[width=4cm]{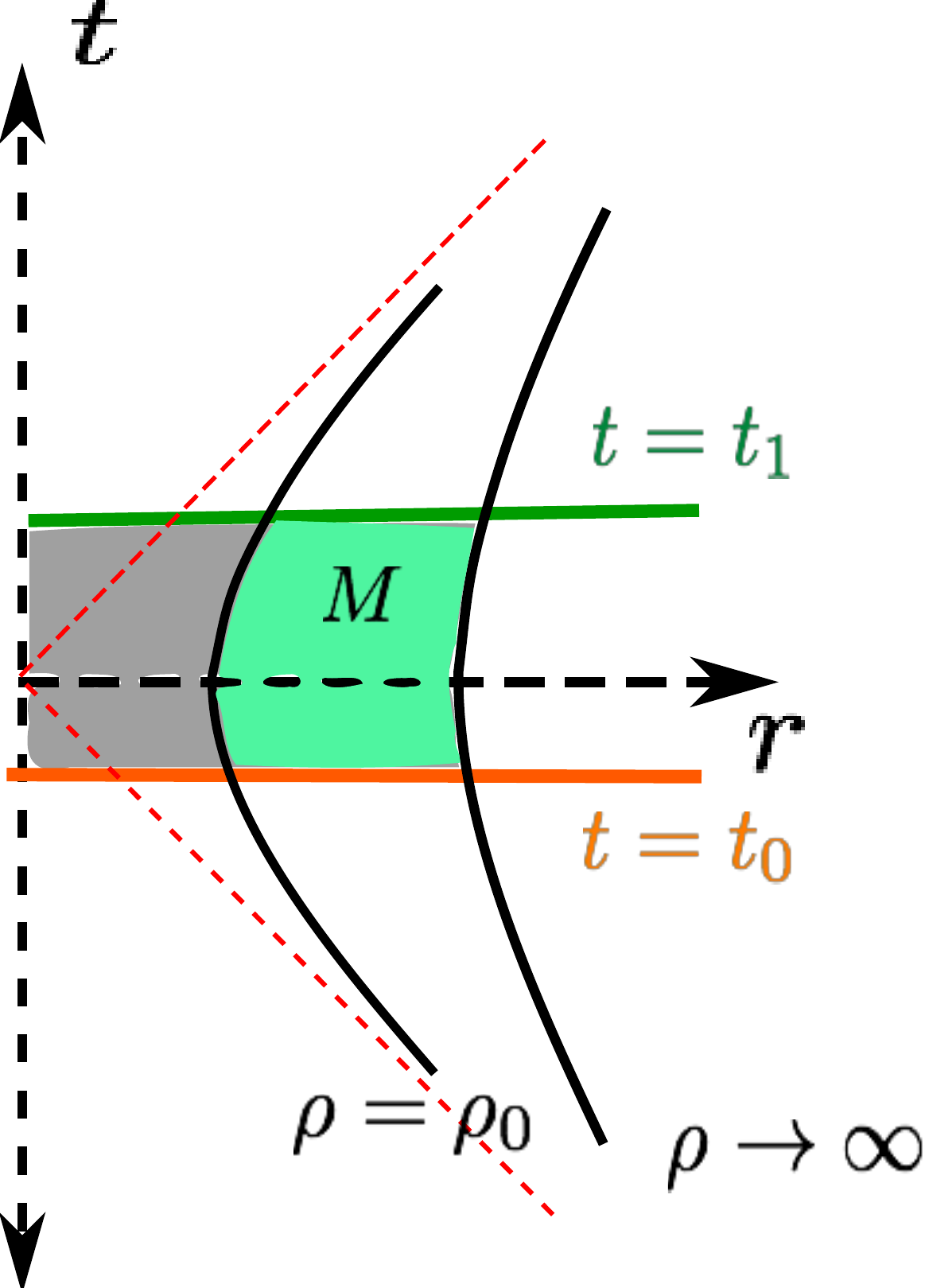}
    \label{cylindricalSlabs}
    }
    \caption{Hyperbolic and cylindrical temporal cut-offs together with hyperbolic cuts defining (asymptotic) spacetime volume integration region $M$ and boundary $\partial M$.} \label{temporalCuts}
    \end{center}
    \end{figure}

Hyperbolic and cylindrical cut-offs provide us with two different ways to cut off spacetime and therefore two seemingly different ways to analyze asymptotic behavior of fields. Intuitively, one would expect the hyperbolic and cylindrical descriptions of spatial infinity to be equivalent, and therefore the behavior of a generic action like (\ref{genericAction})
to be independent of whether one uses hyperbolic or cylindrical cut-offs (or any other cut-off providing an equivalent description of $\iota^0$), however this need not be the case and it ought to be checked since subtleties do arise.  Finally, we remind the reader that $S$ does certainly depend on the way in which spacetime $\mathcal{M}$ is cut-off in time, i.e.  on the choice of Cauchy surfaces $\Sigma_1$ and $\Sigma_2$.
For the asymptotically flat setting there are two natural choices: \emph{cylindrical temporal cut-offs} also known as \emph{cylindrical slabs} where $\Sigma_1$ and $\Sigma_2$ may be related by a translation in time in the asymptotic region, e.g. if  they are defined by $t=t_0$ and $t=t_1$ constant surfaces, and \emph{hyperbolic temporal cut-offs} or \emph{boosted slabs} where $\Sigma_1$ and $\Sigma_2$ may be related by a boost. In the asymptotic region $\rho\to\infty$ the latter may be given by the spatial hypersurfaces $\chi=\chi_0$ and $\chi=\chi_1\,$ (Fig. \ref{temporalCuts}).

Having described the different regions of integration for our covariant or canonical actions we turn to fall-off conditions for dynamical fields. 
Functions of interest to describe (the components of) tensors on asymptotically flat spacetimes will admit an \emph{asymptotic expansion to order $n$}, that is, they will be functions $f:U\subset\mathbb{R}^4\to\mathbb{R}$ admitting a series expansion of the type
\begin{equation} \label{rhoExpansion}
f(\rho,\Phi)=\sum_{m=0}^n\frac{\leftidx{^m}{f}(\Phi)}{\rho^m}+\ord{\rho}{-n}
\end{equation}
for $0<\rho_0<\rho<\infty$, or one of the type 
\begin{equation} \label{rExpansion}
f(t,r,\vartheta,\varphi)=\sum_{m=0}^n\frac{\leftidx{^m}{f}(t,\vartheta,\varphi)}{r^m}+\ord{r}{-n}
\end{equation}
for $0<r_0<r<\infty$, where we have used the shorthand notation $\Phi:=(\chi,\vartheta,\varphi)$ for angles on the hyperboloid $\mathcal{H}_1$ and where the remainders $\ord{\rho}{-n}$ and $\ord{r}{-n}$ have the property that $\lim_{\rho\to\infty}\rho^n\ord{\rho}{-n}=0$ and 
$\lim_{r\to\infty}r^n\ord{r}{-n}=0$ respectively. The remainder $\ord{r}{-n}$ may include terms of the form $h(t,\vartheta,\varphi)/r^{n+\epsilon}$, with $\epsilon>0$,  but it may also contain terms of the form $h(t,\vartheta,\varphi)\ln r/r^{n+1}$ so that $\lim_{r\to\infty}r^{n+1}\ord{r}{-n}$ need not exist. Similar considerations apply to $\ord{\rho}{-n}$.
Later on, we will also use the notation: $\Ord{r}{-n}{\!}$, for terms that fall-off as $r^{-n}$ or faster. (With this notation, the remainders for asymptotic expansions in analyses such as \cite{BeigMurchadha,Thiemann} are  $\Ord{r}{-n-\epsilon}{\!}$ and therefore, slightly less general.)

Notice that since $\leftidx{^m}{f}(\Phi)$ and $\leftidx{^m}{f}(t,\vartheta,\varphi)$ are respectively functions on (a subset of) $\mathcal{H}_1$ and the world tube of the unit sphere or `cylinder' $\mathbb{R}\times S^2_1$, we may also expand (\ref{rhoExpansion}) and (\ref{rExpansion}) as\footnote{with $\leftidx{^m}{f}(\Phi)=\leftidx{^m}{\tilde{f}}(\sinh\chi,\cosh\chi\sin\vartheta\cos\varphi,\cosh\chi\sin\vartheta\sin\varphi,\cosh\chi\cos\vartheta)$, and $\leftidx{^m}{f}(t,\vartheta,\varphi)=\leftidx{^m}{\tilde{f}}(t,\sin\vartheta\cos\varphi,\sin\vartheta\sin\varphi,\cos\vartheta)$. We will however, for simplicity, from now on use the same notation for both $\leftidx{^m}{f}$ and $\leftidx{^m}{\tilde{f}}$.}
\begin{equation} \label{rhoExpansion2}
f(\rho,\Phi)=\sum_{m=0}^n\frac{\leftidx{^m}{\tilde{f}}\left(t/\rho,x^a/\rho\right)}{\rho^m}+\ord{\rho}{-n}
\end{equation}
and
\begin{equation} \label{rExpansion2}
f(t,r,\vartheta,\varphi)=\sum_{m=0}^n\frac{\leftidx{^m}{\tilde{f}}\left(t,x^a/r\right)}{r^m}+\ord{r}{-n}\,.
\end{equation}

By definition, a tensor field $T^{\mu_1\cdots\mu_k}\,_{\nu_1\cdots\nu_l}$ admits an asymptotic expansion to order $n$ if all its components in the Cartesian chart $(t,x^a)$ do so. Derivatives $\partial T^{\mu_1\cdots\mu_k}\,_{\nu_1\cdots\nu_l}/\partial x^\mu$ of components of such tensors with respect to Cartesian coordinates $x^\mu$ will be assumed to admit an asymptotic expansion to order $n+1$. This last supposition is certainly true if the corresponding asymptotic components on the right hand side of (\ref{rhoExpansion}) or (\ref{rhoExpansion2})  are differentiable in their whole domain of definition. For (\ref{rExpansion}) or (\ref{rExpansion2}), it is also an assumption on the specific form of $t$-dependence of expansion functions $\leftidx{^m}{f}$.

We will take a spacetime $\mathcal{M}$ to be \emph{asymptotically flat at spatial infinity} if there exists an open region $U\subset\mathcal{M}$ which is the complement of a spatially compact world tube, and a Minkowski metric $\eta$ on $U$ such that $g-\eta$ admits an asymptotic expansion to order 1 and $\lim_{\rho\to\infty}(g-\eta)=0$ (or $\lim_{r\to\infty}(g-\eta)=0$). This means that in the Cartesian coordinate chart associated with $\eta$, the components of the metric admit an asymptotic expansion of the form
\begin{equation}  \label{cartesianAFCondition1}
g_{\mu\nu}=\eta_{\mu\nu}+\frac{{f}_{\mu\nu}(t/\rho,x^a/\rho)}{\rho}+\ord{\rho}{-1}
\end{equation}
or one of the form
\begin{equation} \label{cartesianAFCondition2}
g_{\mu\nu}=\eta_{\mu\nu}+\frac{{f}_{\mu\nu}(t,x^a/r)}{r}+\ord{r}{-1}\,.
\end{equation}
The choice of flat metric $\eta$ along with (\ref{cartesianAFCondition1}) or (\ref{cartesianAFCondition2}) has associated with it a particular, hyperbolic or cylindrical, cut-off of spacetime $\mathcal{M}$.
In appendix \ref{a::fromHyper2CylExpansion}, it is shown that in the limit where Cartesian time coordinate $t$ is kept bounded and $\rho\to\infty$ or $r\to\infty$, the existence of an expansion of the form (\ref{cartesianAFCondition1}) implies the existence of (\ref{cartesianAFCondition2}) and viceversa.

Conditions (\ref{cartesianAFCondition1}) or (\ref{cartesianAFCondition2}) essentially follow from the coordinate independent definition of asymptotic flatness given by  Ashtekar and Hansen in terms of the existence of a conformal completion of the spacetime $\mathcal{M}$ \cite{AshtekarHansen, AshtekarMagnon}. Conversely, one can show (appendix C of  \cite{AshtekarHansen}) that a spacetime with a metric that can be written  in the forms (\ref{cartesianAFCondition1}) or (\ref{cartesianAFCondition2}) satisfies the `local' conditions in the more geometric definition of Ashtekar-Hansen (and with our topology assumptions the global definition as well). 
However, for the well posedness  of covariant action principles and/or Hamiltonian formulations, and most prominently for the correct definition of conserved Poincar\'e charges, additional conditions need to be imposed\footnote{To distinguish, in \cite{AshtekarES} general spacetimes satisfying (\ref{cartesianAFCondition1}) or (\ref{cartesianAFCondition2}) only are called \emph{weakly} asymptotically flat at spatial infinity.}.
In particular, it is well-known that the asymptotic symmetry group associated with asymptotically flat spaces is much larger than the Poincar\'e group, containing angle or `direction dependent translations' or \emph{supertranslations}  and so called \emph{logarithmic translations} \cite{Bergmann1961, Ashtekar85}. This is because the condition $\mathcal{L}_{\vec{\xi}}\,g_{\mu\nu}=0$ for asymptotic symmetry vector fields $\xi^\mu$ should be fulfilled only to leading order, or equivalently because there is an ambiguity in the choice of background flat metric $\eta$ (for an in depth discussion see \cite{ourReview} or \cite{AshtekarES}). It  then follows that not only transformations of the form $\bar{x}^\mu=\Lambda^\mu\,_\nu x^\nu+T^\mu$, with $\Lambda^\mu\,_\nu$ a Lorentz transformation $\Lambda^T\eta\,\Lambda=\eta$, and $T^\mu$ a `constant'  translation, but the more general coordinate transformation
\begin{equation} \label{coordTransformations}
\bar{x}^\mu=\Lambda^\mu\,_\nu x^\nu+C^\mu\ln\rho+T^\mu+S^\mu(\Phi)+\ord{\rho}{0}
\end{equation}
preserves the forms (\ref{cartesianAFCondition1}) or (\ref{cartesianAFCondition2}). Transformations of the form $\bar{x}^\mu=x^\mu+C^\mu\ln\rho$, with $C^\mu$ constant, are called \emph{logarithmic translations}, and $\bar{x}^\mu=x^\mu+S^\mu(\Phi)$ are the \emph{supertranslations}\footnote{In the more geometric language of \cite{AshtekarHansen}, supertranslations are directly related to the freedom in the choice or re-scaling of the conformal factor in a conformal completion of $\mathcal{M}$. Logarithmic translations on the other hand, refer to the existence of a 4-parameter family of inequivalent, logarithmically related conformal completions \cite{Ashtekar85}. The group containing supertranslations, but no logarithmic translations, is called the \emph{Spi group}, and may be characterized as the set of (bundle) diffeomorphisms preserving the structures on Spi \cite{AshtekarHansen}.}.

There is no known way to canonically select a unique Poncar\'e subgroup of the asymptotic symmetry group to compute conserved charges \cite{Ashtekar85}. An strategy to eliminate such supertranslation and logarithmic ambiguities is to truncate the gravitational configuration or phase space by imposing additional conditions for the fall-off of the components of the metric, hence restricting the allowed diffeomorphisms preserving asymptotic conditions and effectively reducing the symmetry group.

For asymptotically flat spacetimes the behavior of the gravitational field near infinity  is completely determined by the conformal or trace-free part of the curvature tensor \cite{AshtekarHansen,BeigSchmidt,Beig}. Therefore one may truncate the gravitational configuration space by imposing conditions on the fall-off of the `magnetic' an `electric' parts of the Weyl tensor of $g$ as it is done in most covariant treatments. What these conditions imply for the components of the metric can best be seen if one switches to hyperbolic coordinates. 
Using (\ref{hyperCylRelations}), condition (\ref{cartesianAFCondition1}) implies that in the associated hyperbolic coordinates, the line element of $g$ can be expanded as
\begin{align}  
ds^2=g_{\mu\nu}\md x^\mu\md x^\nu=&\left(1+\frac{2\sigma(\Phi)}{\rho}+\ord{\rho}{-1}\right)\md\rho^2+2\rho\left(\frac{A_i(\Phi)}{\rho}+\ord{\rho}{-1}\right)\md\rho\,\md\Phi^i \notag \\
&+\rho^2\left(h_{ij}+\frac{\leftidx{^1}{h}{_{ij}}}{\rho}+\ord{\rho}{-1}\right)\md\Phi^i\,\md\Phi^j   \label{weakAFLineElement}
\end{align}
where $h_{ij}$ is the metric on $\mathcal{H}_1$ and $\Phi^i$ denote its angular coordinates $\Phi=(\chi,\vartheta,\varphi)$:
\[
h_{ij}\md\Phi^i\,\md\Phi^j=-\md\chi^2+\cosh^2\chi\,\md\Omega^2=-\md\chi^2+\cosh^2\chi\,(\md\vartheta^2+\sin^2\vartheta\,\md\varphi^2)\,.
\]
One can assume  \cite{BeigSchmidt} that the chosen Minkowski metric is such that the off-diagonal terms in (\ref{weakAFLineElement}) vanish. To eliminate logarithmic translations one requires that $\sigma(\Phi)$ be symmetric or \emph{even} under reflexions $(t/\rho,x^a/\rho)\to(-t/\rho,-x^a/\rho)$ on the unit hyperboloid, that is
\begin{equation} \label{BeigAshtekarCond} 
\sigma(t/\rho,x^a/\rho)=\sigma(-t/\rho,-x^a/\rho)\,.
\end{equation}
This in turn is equivalent to requiring that the first non-zero order $\rho^{-3}$ electric part of the Weyl tensor be even. More precisely, only the four lowest order harmonics on $\mathcal{H}_1$ of $\sigma$ are required to be parity even in order to eliminate logarithmic ambiguities \cite{CompereDehouckVirmani}. Supertranslations on the other hand are eliminated by requiring
\begin{equation} \label{AshtekarHansenCond}
 \leftidx{^1}{h}{_{ij}}=-2\sigma h_{ij}\,.
\end{equation}
This is essentially equivalent to the vanishing of the order $\rho^{-3}$ magnetic part of the Weyl tensor\footnote{For a slight generalization of (\ref{AshtekarHansenCond}) see \cite{Virmani2011}.}.

Most covariant formulations for an action principle (like those in \cite{AshtekarES, AshtekarBombelliReula, CorichiWilsonEwing})  therefore take or define a spacetime to be \emph{asymptotically flat at spatial infinity} if the metric $g$ can be expanded as
\begin{equation} \label{strongAFLineElement}
ds^2=g_{\mu\nu}\md x^\mu\md x^\nu=\left(1+\frac{2\sigma(\Phi)}{\rho}+\ord{\rho}{-1}\right)\md\rho^2+\rho^2h_{ij}\left(1-\frac{2\sigma(\Phi)}{\rho}+\ord{\rho}{-1}\right)\md\Phi^i\Phi^j
\end{equation}    
with $\sigma$ reflexion symmetric on the unit hyperboloid $\mathcal{H}_1$.
Switching back to Cartesian coordinates, the asymptotic expansion for the Cartesian components of the restricted metrics is
\begin{equation} \label{strongAFmetric}
g_{\mu\nu}=\eta_{\mu\nu}+\frac{2\sigma\left(2\rho_\mu\rho_\nu-\eta_{\mu\nu}\right)}{\rho}+\ord{\rho}{-1}\,.
\end{equation}
with $\rho_\mu:=\partial_\mu\rho=\eta_{\mu\nu}\rho^\nu$ and $\rho^\mu:=x^\mu/\rho$.
 These more restricted conditions not only allow to select an essentially unique Poincar\'e group to compute asymptotic symmetries but they are also necessary to have  well defined covariant phase space formulations \cite{AshtekarES, AshtekarBombelliReula, CorichiWilsonEwing} with finite conserved charges. 

As we will see in section \ref{s::HamiltonianFormulation}, for a 3+1 canonical formulation one may (and we will) take a more general definition because the ambiguity in the choice of asymptotic  Poincar\'e group may be seen partly as  gauge freedom \cite{BeigMurchadha}.

To complete the specification of the variational principle for the general action (\ref{genericAction}) we need to specify the type of variations $\delta\phi^i$ to be considered. We may assume, in analogy with the mechanical case, that $\delta\phi^i=0$ on the Cauchy surfaces $\Sigma_1$ and $\Sigma_2$. However, this condition is no longer natural on the world tube $\tau$. Indeed, if one thinks of $\delta\phi^i$ as `tangent vectors' on configuration space satisfying linearized equations of motion, then these variations must be compatible with the boundary and fall-off conditions specified for the fields $\phi^i$, and therefore $\delta\phi^i$ may not necessarily vanish on the boundary $\tau$ (Fig. \ref{generalBoundaries}). This generalization is also necessary for the semiclassical approximation of the path integral in the quantum theory:  in order to dominate the path integral, the action must be stationary under the full class of variations corresponding to the space of paths  over which the integral is performed  \cite{MannMarolf}.
So, for an action principle for general relativity, whatever the configuration variables chosen to represent the gravitational degrees of freedom are, their fall-off conditions and their variations on $\tau$ must be compatible with asymptotic conditions (\ref{cartesianAFCondition1}) or (\ref{cartesianAFCondition2}) for the metric and their variations:
\begin{equation} \label{AFvariation1}
\delta g_{\mu\nu}=\frac{\delta {f}_{\mu\nu}(t/\rho,x^a/\rho)}{\rho}+\ord{\rho}{-1}
\end{equation}
or
\begin{equation} \label{AFvariation2}
\delta g_{\mu\nu}=\frac{\delta f_{\mu\nu}(t,x^a/r)}{r}+\ord{r}{-1}\,.
\end{equation}

Finally, we shall say we have a well defined action with a well defined variational principle for asymptotically flat spacetimes if the following two conditions hold:
 \begin{enumerate}
 \item  Since the volume of region $M$ is generally infinite, the action (\ref{limgenericAction}) may diverge for arbitrary fields $\phi^i$ in its domain of definition. However, for the well-posedness of the semiclassical approximation from the path integral,  the action $S$ must be finite when evaluated on its stationary points, that is, on solutions to the equations of motion. 
  \item The action must be \emph{differentiable}, in the sense that its variation must be expressible in the form
\begin{equation} \label{differentiableAction}
\delta S=\int_M \frac{\delta S}{\delta \phi^i}\;\delta\phi^i\,.
\end{equation}
This must be true for \emph{all} `directions' in configuration or phase space, that is, 
for \emph{all} variations compatible with boundary or fall-off conditions of the original fields, and not just for those vanishing on the boundary $\partial M$. This implies that the action and asymptotic conditions of the fields must be such that any surface integral in (\ref{differentiableAction}) must be zero in the limit $\Omega\to\infty$. The expression $\frac{\delta S}{\delta \phi^i(x^\mu)}$ multiplying the variation $\delta\phi^i(x^\mu)$ in (\ref{differentiableAction}) may then be formally taken as the functional derivative of $S$ with respect to $\phi^i$. 
The critical or stationary points are the solutions to the equations:
\[
\frac{\delta S}{\delta \phi^i(x^\mu)}=0
\]
and should reproduce the dynamical evolution of the system.
\end{enumerate}

\section{First order action with boundary terms}
\label{sec:3}

Our starting point is the covariant and Lorentz-gauge invariant first order Holst action \cite{Holst} with the additional surface integral term first put forward in \cite{CorichiWilsonEwing} (see also \cite{AshtekarES}). 

The independent fields of this action are (the components of) the co-tetrad one-form $e^I$ and a Lorentz connection potential $\omega^I\,_J$.
 The co-tetrad one-form $e^I=e_{\mu}^I\md x^\mu$  determines a spacetime metric via 
 \begin{equation}
 g_{\mu\nu}=\eta_{IJ}e_\mu^Ie_\nu^J\,,
 \end{equation}
and defines a vector space isomorphism between the tangent space $T_p\mathcal{M}$ at any spacetime point $p$ and a fixed \emph{internal}  Minkowski space with metric $\eta_{IJ}$. Here and in the rest of this work, latin capital letters from the middle of the alphabet will henceforth denote abstract or actual component indices with respect to some fixed basis of this internal Minkowski space.  
The connection potential one-form $\omega^I\,_J=\omega^I_{\mu\,J}\,\md x^\mu$ is $\mathfrak{so}(1,3)$ Lie algebra-valued, so that $\omega_\mu^{IJ}=-\omega_\mu^{JI}$, and it defines a covariant derivative acting on internal indices
\begin{equation} \label{connectionPotentialDef}
\leftidx{^\omega}{\mathcal{D}}{_\mu}V^I:=\bar{\partial}_\mu V^I+\omega^I_{\mu\,J}V^J
\end{equation}
for any other flat and torsion free covariant derivative operator $\bar{\partial}_\mu$.

The full action we consider is\footnote{Notice overall minus sign as compared to \cite{CorichiWilsonEwing}.}
 \begin{align}
 S_{Holst}(e,\omega)=&\frac{1}{2\kappa}\left[\,\int_{M}\Sigma^{IJ}\wedge\left(F_{IJ}+\frac{1}{\gamma}\star F_{IJ}\right) 
 -\int_{\partial{M}}\Sigma^{IJ}\wedge\left(\omega_{IJ}+\frac{1}{\gamma}\star\omega_{IJ}\right) \,\right]\,,  \label{Holst1}
 \end{align}
 with the two-form
\[
\Sigma^{IJ}:=\star (e^I\wedge e^J)=\frac{1}{2}\epsilon^{IJ}\,_{KL}\,e^K\wedge e^L
\]
constructed  from the co-tetrad $e_{\mu}^I$, with $\star$ denoting the Hodge dual in the internal space, so that $\epsilon_{IJKL}$ is the Levi-Civita symbol in four dimensions with the convention $\epsilon_{0123}=1$, 
and where 
\[
F^I\,_J=d\omega^I\,_J+\omega^I\,_K\wedge\omega^K\,_J
\]
is the curvature of the Lorentz connection $\omega^I_\mu\,_J$. The  constant $\gamma$ is an arbitrary real or complex parameter called the Barbero-Immirzi parameter \cite{Barbero,Immirzi}.

From the general asymptotic expansion (\ref{cartesianAFCondition1}) for the metric, one infers co-tetrads whose Cartesian components admit an asymptotic expansion to order  one
\begin{equation} \label{tetradExpansion1}
e_\mu^I=\leftidx{^0}{e}{_\mu^I}(\Phi)+\frac{\leftidx{^1}{e}{_\mu^I}(\Phi)}{\rho}+\ord{\rho}{-1}\,,
\end{equation}
and from this expression (restricting to a region where $t$ is bounded) or directly from (\ref{cartesianAFCondition2}), an expansion of the form
 \begin{equation} \label{tetradExpansion2}
e_\mu^I=\leftidx{^0}{e}{_\mu^I}(t,x^a/r)+\frac{\leftidx{^1}{e}{_\mu^I}(t,x^a/r)}{r}+\ord{r}{-1}\,.
\end{equation}
In either case the angle-dependent asymptotic co-frame $\leftidx{^0}{e}{_\mu^I}$ satisfies $\eta_{IJ}\,\leftidx{^0}{e}{_\mu^I}\,\leftidx{^0}{e}{_\nu^J}=\eta_{\mu\nu}$, and $2\eta_{IJ}\,\leftidx{^0}{e}{_{(\mu}^I}\,\leftidx{^1}{e}{_{\nu)}^J}={f}_{\mu\nu}$ which solving for $\leftidx{^1}{e}{_{\mu}^I}$ implies\footnote{Strictly, according to formulas (\ref{rho2r0}) and (\ref{rho2r1}) in appendix \ref{a::fromHyper2CylExpansion}, if (\ref{tetradExpansion2}) is derived from (\ref{tetradExpansion1}) $\leftidx{^0}{e}{_\mu^I}$ is independent of $t$ and $ \leftidx{^1}{e}{_{\mu}^I}=\frac{1}{2}\eta^{IJ}\,\leftidx{^0}{e}{^\nu_J}{f}_{\mu\nu}|_{\chi=0}+\partial_\chi\left(\leftidx{^0}{e}{_\mu^I}\right)|_{\chi=0}\,t$. The term linear in $t$ is however small in the limit $r\to\infty$ and also later we will take expressions at $t=0$ to infer the general behavior of cylindrical expansions from hyperbolic ones.}:
 \begin{equation} \label{tetradExpansionCoef}
 \leftidx{^1}{e}{_{\mu}^I}=\frac{1}{2}\eta^{IJ}\,\leftidx{^0}{e}{^\nu_J}{f}_{\mu\nu}\,.
 \end{equation}
Here, $\leftidx{^0}{e}{^\nu_J}$ denotes the \emph{asymptotic tetrad} or leading term of  the inverse tetrad $e^\nu_J$, satisfying $e_\mu^Ie^\nu_I=\delta^\nu_\mu$ and $e_\mu^Ie^\mu_J=\delta^I_J$.  Since the choice of asymptotic frame must necessarily be pure gauge, one fixes $\leftidx{^0}{e}{^I_\mu}$ once and for all and only considers fields that admit an asymptotic expansion with respect to this fixed co-tetrad. For (\ref{tetradExpansion2}) we will additionally require that the asymptotic frame is independent of the time variable $t$. More precisely, $\mathcal{L}_{\vec{t}}\,\leftidx{^0}{e}{^\mu_I}=0$, where $\vec{t}:=\frac{\partial}{\partial t}$ is the vector field defined by the Cartesian time coordinate. One could also consider a more general configuration space which for each orthonormal frame $\leftidx{^0}{e}{^I_\mu}$ satisfying $\eta_{IJ}\,\leftidx{^0}{e}{_\mu^I}\,\leftidx{^0}{e}{_\nu^I}=\eta_{\mu\nu}$,  includes all fields admitting an asymptotic expansion with respect to this $\leftidx{^0}{e}{^I_\mu}$. However, as we will see shortly, to guarantee differentiability of action (\ref{Holst1}) and  Lorentz-gauge invariance of its surface term on $\tau_\infty$, the asymptotic co-tetrad must be fixed. This is also necessary for the 3+1 decomposition to be fully consistent with the ADM formulation. The covariant derivative $\bar{\partial}_\mu$ in (\ref{connectionPotentialDef}) is chosen to coincide --in the asymptotic chart-- with the spin connection compatible with the asymptotic co-tetrad: $\bar{\partial}_\mu\leftidx{^0}{e}{^I_\nu}=0$\footnote{Unless $\leftidx{^0}{e}{^I_\mu}$ is the `constant tetrad' $\delta^I_\mu$, this connection is different from the one given by the coordinate derivatives, denoted here as $\partial_\mu$. We use the `bar' in our notation to emphasize this point.}.

Since equations of motion imply (\ref{connectionPotentialDef}) is the covariant derivative or spin connection compatible with the full co-tetrad $e^I$, to derive asymptotic conditions for the connection potential $\omega^I\,_J$ one requires compatibility with the tetrad to appropriate leading orders. Inserting (\ref{tetradExpansion1}) in the formula for the spin connection in terms of the tetrad
\begin{equation} \label{spinConnectionFormula1}
\omega_{\mu IJ}=e_\mu^K\,e^\nu_J\,e^\sigma_I\,\bar{\partial}_{[\nu}\,e_{\sigma]K}-e^\nu_I\,\bar{\partial}_{[\nu}\,e_{\mu]J}-e^\nu_J\,\bar{\partial}_{[\mu}\,e_{\nu]I}\,,
\end{equation}
one gets
\[
\omega_{\mu IJ}=\leftidx{^0}{e}{_\mu^K}\,\leftidx{^0}{e}{^\nu_J}\,\leftidx{^0}{e}{^\sigma_I}\,\bar{\partial}_{[\nu}\left(\frac{\leftidx{^1}{e}{_{\sigma]K}}}{\rho}\right)-\leftidx{^0}{e}{^\nu_I}\,\bar{\partial}_{[\nu}\left(\frac{\leftidx{^1}{e}{_{\mu]J}}}{\rho}\right)-\leftidx{^0}{e}{^\nu_J}\,\bar{\partial}_{[\mu}\left(\frac{\leftidx{^1}{e}{_{\nu]I}}}{\rho}\right)+\ord{\rho}{-2}
\]
and a similar $r$-expansion for (\ref{tetradExpansion2}). So compatibility with the co-tetrad (\ref{tetradExpansion1}) or (\ref{tetradExpansion2}) requires the connection potential to admit an asymptotic $\rho$-expansion to order two, with nonzero leading term of order $\rho^{-2}$:
\begin{equation}  \label{connectionExpansion1}
\omega_\mu^{IJ}=\frac{\leftidx{^2}{\omega}{_\mu^{IJ}}(\Phi)}{\rho^2}+\ord{\rho}{-2}
\end{equation}
and
\begin{equation} \label{connectionExpansionCoef1}
\leftidx{^2}{\omega}{_{\mu IJ}}=\leftidx{^0}{e}{^\sigma_{[I}}\,\leftidx{^0}{e}{^\nu_{J]}}\left(\rho\,{\partial}_{\nu}{f}_{\sigma\mu}-\rho_{\nu}{f}_{\sigma\mu}\right)\,,
\end{equation}
or and $r$-expansion
\begin{equation} \label{connectionExpansion2}
\omega_\mu^{IJ}=\frac{\leftidx{^2}{\omega}{_\mu^{IJ}}(t,x^a/r)}{r^2}+\ord{r}{-2}
\end{equation}
with
\begin{equation} \label{connectionExpansionCoef2}
\leftidx{^2}{\omega}{_{\mu IJ}}=\leftidx{^0}{e}{^\sigma_{[I}}\,\leftidx{^0}{e}{^\nu_{J]}}\left(r\,{\partial}_{\nu}{f}_{\sigma\mu}-(\partial_{\nu}r){f}_{\sigma\mu}\right)\,.
\end{equation}
From appendix \ref{a::fromHyper2CylExpansion}, if one only considers cylindrical temporal cut-offs for the action (\ref{Holst1}), $\rho$-expansions (\ref{tetradExpansion1}) and (\ref{connectionExpansion1}) for the co-tetrad and connection directly imply $r$-expansions (\ref{tetradExpansion2}) and (\ref{connectionExpansion2}). This is the case of interest for a 3+1 decomposition\footnote{Recall that since $\bar{\partial}_{\nu}{f}_{\sigma\mu}$ is order $\rho^{-1}$, (\ref{connectionExpansionCoef1}) is indeed an order $\rho^{0}$ function on the unit hyperboloid. By assumption, or if (\ref{connectionExpansionCoef2}) is derived from the $\rho$-expansion, $\bar{\partial}_tf_{\mu\nu}$ is $\Ord{r}{-1}{\!}$ so also (\ref{connectionExpansionCoef2}) is an order $\Ord{r}{0}{\!}$  function on the cylinder.}.

On the time-like boundary $\tau_\infty$ gauge transformations are frozen because one has fixed the asymptotic frame. After a $3+1$ decomposition, the fall-off conditions above will result in the surface term of (\ref{Holst1}) on this boundary giving the gauge invariant ADM energy and momentum on-shell. This surface term however is not gauge invariant  on the Cauchy surfaces $\Sigma_1$ and $\Sigma_2$, even if, as in \cite{AshtekarES}, one only considers histories such that the pull-back of $\omega$ to $\Sigma_1$ and $\Sigma_2$ is determined by the pull-back of $e$ and furthermore, one requires $\bar{\partial}_\mu n^I=0$, for $n^I:=n^\mu e_\mu^I$ and $n^\mu$ the unit normal to $\Sigma_1$ and $\Sigma_2$.
This nevertheless, just like in covariant analyses, is of no consequence for the 3+1 decomposition.
We point out that in \cite{Wieland, Bodendorfer2013} another manifestly gauge invariant surface integral was proposed which generalizes the term for Palatini introduced by  Obukhov \cite{Obukhov}. However, given that such integral agrees with the Gibbons-Hawking term on-shell, it is not finite and furthermore requires a fixed induced metric at the boundary to ensure differentiability.

For  a cylindrical cut-off of spacetime (to approach spatial infinity $\iota^0$) finiteness and differentiability of action (\ref{Holst1}) follow trivially if one integrates by parts and rewrites
\begin{equation} \label{HolstFiniteForm0}
S_{Holst}(e,\omega)=-\frac{1}{2\kappa}\int_{M}\left(\,\md\Sigma^{IJ}\wedge\leftidx{^\gamma}{\omega}{_{IJ}}
 -\Sigma^{IJ}\wedge\omega_I\,^K\wedge\leftidx{^\gamma}{\omega}{_{KJ}}\,\right)\,,
\end{equation}
with the shorthand notation:
\[
\leftidx{^\gamma}{\omega}{_{IJ}}:=\omega_{IJ}+\frac{1}{\gamma}\star\omega_{IJ}\,.
\]
Explicitly, in component form
\begin{equation} \label{HolstFiniteForm}
S_{Holst}(e,\omega)=-\frac{1}{4\kappa}\int_{M}\md^4x\left(\,\bar{\partial}_\nu\Sigma_{\rho\sigma}^{IJ}\,\leftidx{^\gamma}{\omega}{_{\mu IJ}}-\Sigma_{\rho\sigma}^{IJ}\,\omega_{\nu I}\,^K\,\leftidx{^\gamma}{\omega}{_{\mu KJ}}\right)\epsilon^{\mu\nu\rho\sigma}
\end{equation}
each term is $\Ord{r}{-4}{\!}$, whereas  the volume element $\md^4x:=\md x^0\wedge\md x^1\wedge\md x^2\wedge\md x^3$ goes as $r^2$. Hence, for cylindrical slabs --where $t$ is bounded-- the integral (\ref{HolstFiniteForm}) converges even off-shell. This, as we shall see, is consistent with a well defined Hamiltonian formulation.

For differentiability, variation of terms without derivatives in (\ref{HolstFiniteForm0})  can only yield new expressions with equal or lower order leading terms in an asymptotic expansion. The potentially problematic term is the one containing derivatives.  Variation of (\ref{HolstFiniteForm0}) results in a finite expression of the form (\ref{differentiableAction}) if the surface integral coming from integration by parts of this derivative term vanishes in the limit $r\to\infty$. The surface integral is
\begin{equation} \label{HolstSurfVar}
-\frac{1}{2\kappa}\int_{\partial M}\delta\Sigma^{IJ}\wedge\,\leftidx{^\gamma}{\omega}{_{IJ}}=\frac{1}{4\kappa}\int_{\partial M}\sqrt{|h|}\,\md^3y\left(r_\sigma\frac{\epsilon^{\sigma\mu\nu\rho}}{\sqrt{-g}}\,\delta\Sigma_{\nu\rho}^{IJ}\,\leftidx{^\gamma}{\omega}{_{\mu IJ}}\right)
\end{equation}
with $y^\alpha$ coordinates on $\partial M$, $r_\sigma$ its normal vector and $h$ the (determinant of the) induced metric. On the Cauchy surfaces $\Sigma_1$ and $\Sigma_2$  we are setting variations to zero, so the integral vanishes. On the cylinder $\tau_\infty$, to leading order $\sqrt{|h|}\,\md^3y=r^2\md t\,\md\Omega^2$, with $\md\Omega^2$ the standard volume element of the unit 2-sphere. Since the term in parentheses in (\ref{HolstSurfVar}) is $\Ord{r}{-3}{\!}$, the surface integral vanishes too in the limit $r\to\infty$. Had we not fixed the asymptotic frame and considered variations of the form $\delta e_\mu^I=\delta\,\leftidx{^0}{e}{_\mu^I}+\Ord{r}{-1}{\!}$, the term in parenthesis would be only $\Ord{r}{-2}{\!}$.

Finiteness and differentiability of the action is more subtle for hyperbolic cut-offs of spacetime, even if one restricts to cylindrical temporal cuts. For cylindrical slabs, finiteness off-shell also follows from (\ref{HolstFiniteForm}) \cite{CorichiRV}. For boosted slabs, the action is finite only on-shell \cite{AshtekarES, ourReview}. However, for the general fall-off conditions with coefficients (\ref{tetradExpansionCoef}) and (\ref{connectionExpansionCoef1}), action (\ref{Holst1}) is not differentiable for hyperbolic cut-offs.

Action (\ref{Holst1}) turns out to be differentiable\footnote{Modulo some subtleties explained in \cite{CorichiRV,ourReview}.} for the reduced configuration space consisting of metrics with asymptotic expansion (\ref{strongAFmetric}). This restriction as we have said also eliminates supertranslations and furthermore allows for a well defined pre-symplectic structure on the covariant phase space of \cite{AshtekarES,CorichiWilsonEwing}. In this case (\ref{tetradExpansionCoef})  further specializes to 
\begin{equation} \label{covariantTetradCoef}
\leftidx{^1}{e}{_\mu^I}(\Phi)=\sigma(\Phi)(2\rho_\mu\rho^I-\leftidx{^0}{e}{_\mu^I})
\end{equation}
and accordingly (\ref{connectionExpansionCoef1}) becomes
\begin{equation} \label{covariantConnectionCoef}
\leftidx{^2}{\omega}{_{\mu IJ}}=2\left[\rho(2\rho_\mu\rho_{[I}-\leftidx{^0}{e}{_{\mu[I}})\bar{\partial}_{J]}\sigma-\sigma\,\leftidx{^0}{e}{_{\mu[I}}\rho_{J]}\right]
\end{equation}
with $\rho_I:=\leftidx{^0}{e}{^\mu_I}\rho_\mu$ and $\bar{\partial}_J\sigma:=\leftidx{^0}{e}{^\nu_{J}}\bar{\partial}_\nu\sigma$. 

In summary,  while general fall-off conditions (\ref{tetradExpansion2}) and (\ref{connectionExpansion2}) alone are sufficient to ensure finiteness and differentiability of the Holst action (\ref{Holst1}) for a geometry consistent with a 3+1 decomposition (i.e. for cylindrical cut-offs of spacetime with cylindrical temporal cuts),  further specialization to expansion coefficients (\ref{covariantTetradCoef}) and (\ref{covariantConnectionCoef}) is needed to guarantee its differentiability for hyperbolic cut-offs most adapted to covariant formulations.
Hence, in  covariant treatments  \cite{CorichiWilsonEwing, AshtekarES}  these conditions are imposed from the outset.
 Additionally, in their corresponding covariant Hamiltonian formulation, in order to have well defined  relativistic angular momentum, these treatments require the symmetry of the mass function $\sigma$ that eliminates logarithmic ambiguities, and further, an asymptotic expansion to order at least two for the co-tetrad 
\begin{equation} \label{tetradExpansion3}
e_\mu^I=\leftidx{^0}{e}{_\mu^I}(\Phi)+\frac{\leftidx{^1}{e}{_\mu^I}(\Phi)}{\rho}+\frac{\leftidx{^2}{e}{_\mu^I}(\Phi)}{\rho^2}+\ord{\rho}{-2}
\end{equation}
which from compatibility of the connection with this co-tetrad, requires $\omega^I\,_J$ to admit an asymptotic expansion to order three 
\begin{equation} \label{connectionExpansion3}
\omega_\mu^{IJ}=\frac{\leftidx{^2}{\omega}{_\mu^{IJ}}(\Phi)}{\rho^2}+\frac{\leftidx{^3}{\omega}{_\mu^{IJ}}(\Phi)}{\rho^3}+\ord{\rho}{-3}\,.
\end{equation}

As we shall see, general fall-off conditions (\ref{tetradExpansion2}) and (\ref{connectionExpansion2}) are also sufficient to guarantee finiteness and differentiability of the Hamiltonian action (\ref{HolstHamiltonian}) in terms of Ashtekar-Barbero variables and derived from (\ref{Holst1}). Nevertheless, conditions (\ref{covariantTetradCoef}) and (\ref{covariantConnectionCoef})  along with the symmetry of the mass function (\ref{BeigAshtekarCond}) will actually allow us to infer parity conditions for the (asymptotic expansion coefficients of) Ashtekar-Barbero variables independently of the well-known Regge-Teitelboim parity conditions \cite{ReggeTeitelboim} for ADM variables.
Furthermore, the slightly more restrictive expansions (\ref{tetradExpansion3}) and (\ref{connectionExpansion3})  would be consistent with Hamiltonian results of section \ref{s::HamiltonianFormulation}, where we shall see that, even though they are not needed for most constructions of the canonical Hamiltonian phase space, analogous $r$-expansions in fact seem to be required for a `cotangent bundle' interpretation of the phase space and for a direct proof that surface integrals of generators actually match ADM Poincar\'e charges.


\section{3+1 splitting}
\label{sec:4}

We now generalize the results of \cite{Holst} for asymptotically flat spacetimes by showing that the 3+1 decomposition and time-gauge fixing of action (\ref{Holst1}) leads to a well defined canonical action in terms of Ashtekar-Barbero variables.
Details of the decomposition are relegated to appendix \ref{a::HolstDecomposition}.

As usual for a canonical or Hamiltonian formulation of a covariant theory, we consider a time function $t$ whose level curves $\Sigma_t$ are Cauchy surfaces and furnish a foliation of spacetime $\mathcal{M}\simeq\mathbb{R}\times\Sigma$. Spacetime fields are split into tangential (spatial) and normal components with respect to this foliation. Additionally, one chooses a time-like \emph{evolution} vector field $t^\mu$ such that $t^\mu\nabla_\mu t=1$ and along which spatial fields are defined to `evolve'. Decomposition for the spacetime metric splits the 10 independent components of $g_{\mu\nu}$ into the six independent components of the Euclidean spatial metric $q_{\mu\nu}$, the lapse function $N$ and the  shift vector $N^\mu$. Lapse and shift being respectively the normal and tangential components of the evolution vector field $t^\mu=Nn^\mu+N^\mu$. In coordinates $(t,y^a)$ adapted to the foliation the line element reads
\begin{equation} \label{ADMmetric}
g_{\mu\nu}\md x^\mu \md x^\nu=(-N^2+q_{ab}N^aN^b)\md t^2+2q_{ab}N^b\md t\,\md y^a + q_{ab}\md y^a \,\md y^b\,.
\end{equation}

Without loss of generality, we may assume the Cartesian coordinates associated with expansion (\ref{cartesianAFCondition2}) coincide asymptotically with the coordinates in (\ref{ADMmetric}) adapted to the 3+1 foliation of spacetime. Hence  one infers fall-off conditions for the spatial metric in this chart:
\begin{equation} \label{falloffqmetric1}
q_{ab}=\delta_{ab}+\frac{f_{ab}(t,x^c/r)}{r}+\ord{r}{-1}\,,
\end{equation}
and for lapse and the three nonzero components of the shift vector: 
\begin{equation} \label{falloffCondLapseShift1}
N=1+\frac{f_{tt}}{2r}+\ord{r}{-1}\,, \qquad N^a=\frac{f_{ta}}{r}+\ord{r}{-1}\,.
\end{equation}
This is the expected result since assuming adapted coordinates to be asymptotically Lorentzian implies $t^\mu$ is (asymptotically) orthogonal to the foliation.

\begin{figure}
     \begin{center}
      \includegraphics[width=5cm]{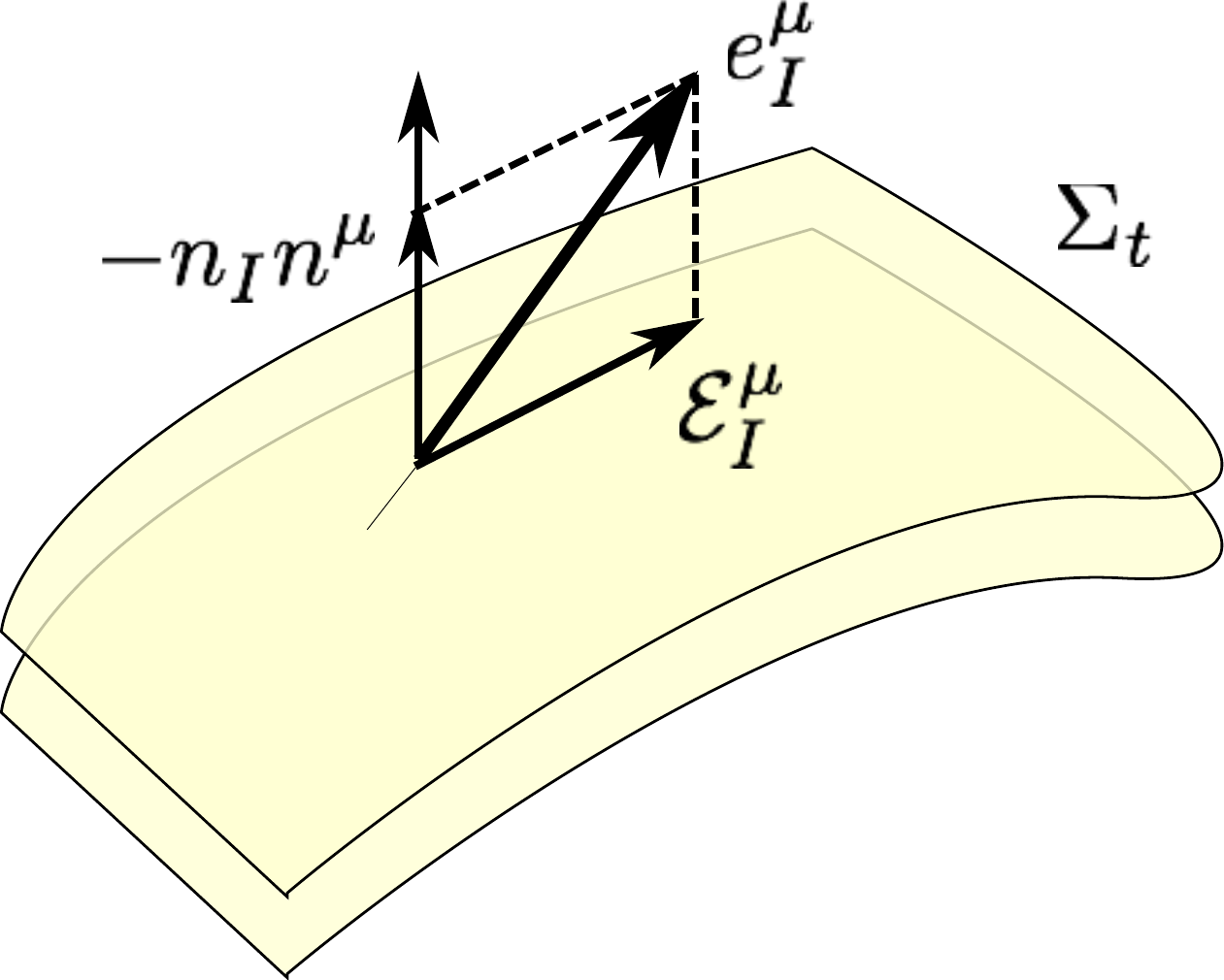}
      \caption{\label{tetrad3plus1} Orthogonal $3+1$ decomposition of tetrad $e^\mu_I$.}
      \end{center}
\end{figure}

 We now decompose the tetrad $e^\mu_I$ into normal and tangential components with respect to the 3+1 foliation.
 Define $\mathcal{E}^\mu_I$ to be the orthogonal projection of the tetrad to each leaf of the foliation, i.e.
 \[
\mathcal{E}^\mu_I:=q^\mu_\nu\,e^\nu_I:=(\delta^\mu_\nu+n^\mu n_\nu)e^\nu_I=e^\mu_I+n^\mu n_\nu e^\nu_I
\]  
so that $\mathcal{E}^\mu_I\in T_p\Sigma\subset T_p\mathcal{M}$ is purely spatial at each point $p$, and the tetrad is decomposed as
\[
e^\mu_I=-(e^\nu_In_\nu)n^\mu+\mathcal{E}^\mu_I=:-n_In^\mu+\mathcal{E}^\mu_I\,,
\]
where one defines $n_I:=e^\nu_In_\nu$ as (minus) the normal component of the tetrad $e^\mu$ and which also happens to be the image on internal Minkowski space of the normal gradient $n_\mu$, under the isomorphism defined by the tetrad (Fig. \ref{tetrad3plus1}). 

Continuing with standard procedures, if $\{y^a\}_{a=1,2,3}$ are (arbitrary) coordinates on $\Sigma_t$, and $\{x^\mu\}_{\mu=0,1,2,3}$ arbitrary coordinates on $\mathcal{M}$, the vectors
\[
\tilde{e}^\mu_a:=\frac{\partial x^\mu}{\partial y^a}
\]
(not to be confused with the tetrads here) form a basis on $T_p\Sigma$, and we may expand purely spatial vectors in terms of this basis:
\[
\mathcal{E}^\mu_I=:\mathcal{E}^a_I\,\tilde{e}^\mu_a
\qquad \text{ and } \qquad N^\mu=:N^a\,\tilde{e}^\mu_a
\]
so $\{n^\mu,\,\tilde{e}^\mu_a\}$ or also $\{t^\mu,\,\tilde{e}^\mu_a\}$ span the tangent space $T_p\mathcal{M}$ and
\begin{align}
e^\mu_I&=-n_In^\mu+\mathcal{E}^a_I\,\tilde{e}^\mu_a \notag\\
&=n_I\frac{\left(N^\mu-t^\mu\right)}{N}+\mathcal{E}^a_I\,\tilde{e}^\mu_a \notag\\
&=-\frac{n_I}{N}t^\mu+\left(\frac{N^an_I}{N}+\mathcal{E}^a_I\right)\tilde{e}^\mu_a\,.  \label{tetradDecomposition}
\end{align}

We may now write the 3+1 decomposition of (\ref{Holst1}) to derive a canonical action. 
Again using the short hand notation
\[
\leftidx{^\gamma}{F}{^{IJ}}:=F^{IJ}+\frac{1}{\gamma}\star F^{IJ}   \qquad\text{ and } \qquad
\leftidx{^\gamma}{\omega}{^{IJ}}:=\omega^{IJ}+\frac{1}{\gamma}\star\omega^{IJ}\,,
\]
the action reads:
\[
 S_\text{Holst}=\frac{1}{2\kappa}\left(\int_{M}\Sigma_{IJ}\wedge\leftidx{^\gamma}{F}{^{IJ}}
 -\int_{\partial{M}}\Sigma_{IJ}\wedge\leftidx{^\gamma}{\omega}{^{IJ}}\,\right) \,.
\]
Equivalently, using index notation this is re-written as
\begin{equation} \label{HolstActionIndex}
 S_\text{Holst}=\frac{1}{2\kappa}\left(\int_{M}\md^4x\,|e|e^\mu_Ie^\nu_J\,\leftidx{^\gamma}{F}{^{IJ}_{\mu\nu}}
 +2\int_{\partial{M}}\md^3y\sqrt{|h|} r_\nu e^\mu_Ie^\nu_J\,\leftidx{^\gamma}{\omega}{^{IJ}_\mu}\,\right),
 \end{equation}
where $e$ denotes the determinant of the co-tetrad $e_\mu^I$, $h$ is the determinant of the induced metric on the boundary $\partial M$, and $r_\mu$ its co-normal.
Substituting $|e|=\sqrt{|\det g|}=N\sqrt{\det q}$ and the tetrad decomposition (\ref{tetradDecomposition}) in the bulk integrand renders:
\begin{align}
|e|e^\mu_Ie^\nu_J\,\leftidx{^\gamma}{F}{^{IJ}_{\mu\nu}}&=\left(-2t^\mu n_I+2N^an_I\tilde{e}^\mu_a+N\mathcal{E}^a_I\tilde{e}^\mu_a\right)\sqrt{\det q}\,\mathcal{E}^b_J\,\tilde{e}^\nu_b\,\leftidx{^\gamma}{F}{^{IJ}_{\mu\nu}}\,. \label{bulkDecomp1}
\end{align}
The first term in parenthesis will give rise to both the kinetic term and the Gauss constraint, the second term corresponds to the diffeomorphism constraint, and the last term will reproduce the Hamiltonian constraint.
The new contribution from the boundary term is
\begin{align}
2\sqrt{|h|}\, r_\nu& e^\mu_Ie^\nu_J\,\leftidx{^\gamma}{\omega}{^{IJ}_\mu} \notag\\
&=2\sqrt{|h|} \left[\left(\frac{N^a}{N}n_I\mathcal{E}^b_J\tilde{e}^\mu_a\tilde{e}^\nu_b-\frac{n_I}{N}\mathcal{E}^b_Jt^\mu\tilde{e}^\nu_b\right)\left(r_\nu\leftidx{^\gamma}{\omega}{^{IJ}_\mu}-r_\mu\leftidx{^\gamma}{\omega}{^{IJ}_\nu}\right)
+r_\nu\mathcal{E}^a_I\mathcal{E}^b_J\,\leftidx{^\gamma}{\omega}{^{IJ}_\mu}\tilde{e}^\mu_a\tilde{e}^\nu_b\right]\,.  \label{surfaceDecomp1}
\end{align}

Following \cite{Holst}, we now partially fix the gauge by imposing the \emph{time-gauge}. This consists in choosing an orthonormal basis in internal Minkowski space such that
\begin{equation} \label{internalNormal}
n^I=\eta^{IJ}n_J=\delta^I_0
\end{equation}
or in different notation $n^I=(1,0,0,0)$ and $n_I=(-1,0,0,0)$. 
The time-gauge is equivalent to first choosing an arbitrary orthonormal basis for the internal space and then restricting to those tetrads for which 
\begin{equation}  \label{timeGauge}
e_0^\mu=n^\mu\,.
\end{equation}
(Indeed $n^\mu=e^\mu_In^I=e^\mu_I\delta^I_0=e_0^\mu$). 
It follows from (\ref{timeGauge}) and (\ref{tetradDecomposition}) that $\mathcal{E}^a_0=0$ and $e^\mu_i=\mathcal{E}^a_i\tilde{e}^\mu_a$, for $i=1, 2, 3$. In words, we `align' $e_0^\mu$ with the normal to the foliation so the remaining vectors $e^\mu_i$ are purely spatial and the orthonormal reference co-frame $e_\mu^I$ coincides with that of Eulerean observers. Further demanding gauge transformations to preserve the vector (\ref{internalNormal}) effectively reduces the Lorentz gauge group to $SO(3)$.

Before we expand  each of the bulk and surface terms imposing the time-gauge (\ref{timeGauge}), we additionally restrict to coordinates adapated  to the foliation. In these coordinates $t^\mu=\partial x^\mu/\partial t=\delta^\mu_t$, $\tilde{e}^\mu_a=\delta^\mu_a$, and the projections
\[
\leftidx{^\gamma}{F}{^{IJ}_{tb}}:=t^\mu\,\tilde{e}^\nu_b\,\leftidx{^\gamma}{F}{^{IJ}_{\mu\nu}}   \qquad \text{ and } \qquad
\leftidx{^\gamma}{F}{^{IJ}_{ab}}:=\tilde{e}^\mu_a\,\tilde{e}^\nu_b\,\leftidx{^\gamma}{F}{^{IJ}_{\mu\nu}}
\]
are precisely the $tb$- and $ab$-components of the curvature tensor $\leftidx{^\gamma}{F}{^{IJ}_{\mu\nu}}$. Similarly 
$\leftidx{^\gamma}{\omega}{^{IJ}_{t}}:=t^\mu\,\leftidx{^\gamma}{\omega}{^{IJ}_{\mu}}$  and $
\leftidx{^\gamma}{\omega}{^{IJ}_{a}}:=\tilde{e}^\mu_a\,\leftidx{^\gamma}{\omega}{^{IJ}_{\mu}}$ are the time $t$- and spatial $a$-components of  $\leftidx{^\gamma}{\omega}{^{IJ}_{\mu}}$.  One defines
\begin{align}
E^a_i:&=\sqrt{\det q}\,\mathcal{E}^a_i \label{defE}\\
K_a^i:&=\omega_a^{0i} \label{defK} \\
\Gamma_a^i:&=\frac{1}{2}\epsilon^{0i}\,_{jk}\,\omega_a^{jk}=-\frac{1}{2}\epsilon^i\,_{jk}\,\omega_a^{jk}  \label{defGamma}\\
A_a^i:&=\gamma\,\leftidx{^\gamma}{\omega}{^{0i}_a}=\frac{1}{2}\epsilon^{0i}\,_{jk}\,\omega_a^{jk}+\gamma\,\omega_a^{0i}
=\Gamma_a^i+\gamma K_a^i  \label{defA}\\
\Lambda^i:&=\gamma\,\leftidx{^\gamma}{\omega}{^{0i}_t}=\frac{1}{2}\epsilon^{0i}\,_{jk}\,\omega_t^{jk}+\gamma\,\omega_t^{0i} \label{defLambda}
\end{align}
(Notice we could have written these definitions to include the zeroth or temporal internal index $I=0$, e.g. $E^a_I:=\sqrt{\det q}\,\mathcal{E}^a_I$, $K_a^I:=\omega_a^{0I}$, etc., but these `zeroth-components' all vanish). 
We use the convention $\epsilon_{0123}=1$ consistent with the covariant formulation and define the three-dimensional Levi-Civita symbol as $\epsilon_{ijk}:=\epsilon_{0ijk}$, so that $\epsilon^{0i}\,_{jk}=\eta^{0l}\eta^{im}\epsilon_{lmjk}=-\delta^{im}\epsilon_{mjk}=-\epsilon^i\,_{jk}$ and additionally $\epsilon^{ij}\,_{k0}=-\epsilon^{ij}\,_k$.

As shown in appendix \ref{a::fromHyper2CylExpansion}, since by assumption our adapted coordinates are asymptotically Lorentzian, the 3+1 decomposition and gauge fixing, with  definition (\ref{defE}), along with expansion (\ref{tetradExpansion2}), imply $E^a_i$ admits an asymptotic expansion to order one
\begin{equation} \label{Efalloff1}
E^a_i=\bar{E}^a_i(x^c/r)+\frac{F^a_i(x^c/r)}{r}+\ord{r}{-1}\,, 
\end{equation}
 or up to order two  if we impose (\ref{tetradExpansion3}). 
This defines an asymptotic densitized triad
\begin{equation}
\bar{E}^a_i:=\sqrt{\det\delta}\,\bar{e}^a_i=\bar{e}^a_i\,,
\end{equation}
with asymptotic triad\footnote{We switch to the `bar' notation to facilitate comparison with \cite{Thiemann}.} 
\begin{equation}
\bar{e}^a_i:=\leftidx{^0}{\mathcal{E}}{^a_i}=\leftidx{^0}{e}{^a_i}\,.
\end{equation}
Similarly (\ref{defA}) and (\ref{defLambda}) together with (\ref{connectionExpansion2}) imply an asymptotic expansion to order two for $\Lambda^i$ and $A_a^i$ of the form
\begin{equation} \label{Lambdafalloff}
\Lambda^i=\frac{G^i(x^c/r)}{r^2}+\ord{r}{-2}\,,
\end{equation}
and
\begin{equation} \label{Afalloff1}
A^i_a=\frac{G^i_a(x^c/r)}{r^2}+\ord{r}{-2}\,,
\end{equation} 
or to order three if we use (\ref{connectionExpansion3}).
Analogous asymptotic expansions also follow for extrinsic curvature $K_a^i$ and spin connection $\Gamma_a^i$  from (\ref{defK}) and (\ref{defGamma}).

The vector potential  $\bar{\omega}_\mu^I\,_J$ of the spin connection compatible with the asymptotic tetrad: $\bar{\partial}_\mu \leftidx{^0}e^\nu_i=0$ gives, by a formula analogous to  (\ref{defGamma}):  $\bar{\Gamma}_a^i:=-\frac{1}{2}\epsilon^i\,_{jk}\,\bar{\omega}_a^{jk}$ with its inverse $\bar{\omega}_a^{ij}=:-\epsilon^{ij}\,_k\bar{\Gamma}_a^k$ , the vector potential of the three-dimensional spin connection compatible with the asymptotic triad. Indeed
\[
\bar{\partial}_a\bar{e}^b_i=\bar{\partial}_a\leftidx{^0}{e}{^b_i}
=\partial_a\leftidx{^0}{e}{^b_i}-\bar{\omega}_a^J\,_i\,\leftidx{^0}{e}{^b_J}
=\partial_a\leftidx{^0}{e}{^b_i}-\bar{\omega}_a^j\,_i\,\leftidx{^0}{e}{^b_j}
=\partial_a\bar{e}^b_i-\bar{\omega}_a^j\,_i\,\bar{e}^b_j
=\partial_a\bar{e}^b_i+\epsilon^j\,_{ik}\bar{\Gamma}_a^k\,\bar{e}^b_j
=0\,,
\]
and solving for $\bar{\Gamma}_a^k$ in the last expression, one can further see  $\bar{\Gamma}_a^k$ is $\Ord{r}{-1}{\!}$.

Following the same argument, if $\omega_a^I\,_J$ matches the (vector potential of the) spin connection annihilating  the tetrad: $\leftidx{^\omega}{\mathcal{D}}{_\mu}e^\nu_I=0$ then
\[
0=\leftidx{^\omega}{\mathcal{D}}{_a}e^b_i
=\bar{\partial}_ae^b_i+\Gamma^b_{a\sigma}e^\sigma_i-\omega_a^J\,_i\,e^b_J
=\bar{\partial}_ae^b_i+\Gamma^b_{ac}e^c_i-\omega_a^j\,_i\,e^b_j
\]
so definition (\ref{defGamma}), with its inverse:
\begin{equation} \label{omegaij}
\omega_a^{ij}=-\epsilon^{ij}\,_k\Gamma_a^k\,,
\end{equation}
substituted above, does correspond to the vector potential for the three-dimensional spin connection.

Imposing the time-gauge and substituting definitions (\ref{defE})-(\ref{defLambda}) in the action, the bulk term (\ref{bulkDecomp1}) becomes (in adapted coordinates) after various but solely algebraic manipulations shown in appendix \ref{a::HolstDecomposition}:  
\begin{align}
|e|e^\mu_Ie^\nu_J\,\leftidx{^\gamma}{F}{^{IJ}_{\mu\nu}}
=&\frac{2}{\gamma}\left(E^b_j\bar{\partial}_tA_b^j+\Lambda^j\mathcal{G}_j-(1+\gamma^2)\omega_t^{0j}\mathcal{S}_j-\bar{\partial}_b(\Lambda^jE^b_j)  \right) -N^a\mathcal{C}_a -N\mathcal{C}  \label{bulkDecomp2}
\end{align}
where
\begin{align}
\mathcal{S}_i:&=\epsilon_{ij}\,^k\,K_a^{j}E^a_k\,, \\
\mathcal{G}_i:&=\bar{\partial}_aE^a_i+\epsilon_{ij}\,^kA_a^jE^a_k\,,  \\
\mathcal{C}_a:&=\frac{2}{\gamma}\left(F_{ab}^jE^b_j-(1+\gamma^2)K_a^k\mathcal{S}_k\right)  \label{VectorConstraint}\\
\mathcal{C}:&=\frac{E^a_iE^b_j}{\sqrt{\det q}}\bigg[\epsilon^{ij}\,_k\,F_{ab}^k-2(1+\gamma^2)K_{[a}^i\,K_{b]}^j-2\left(\frac{1+\gamma^2}{\gamma}\right)\epsilon^{ij}\,_k\,\leftidx{^\Gamma}{\mathcal{D}}{_{[a}}K_{b]}^k \bigg]  \label{CConstraint}
\end{align}
and
\begin{equation}
F_{ab}^i:=2\bar{\partial}_{[a}A_{b]}^i+\epsilon^i\,_{jk}\,A_a^jA_b^k\,.
\end{equation}
\begin{equation}
\leftidx{^\Gamma}{\mathcal{D}}{_{a}}K_{b}^k:=\bar{\partial}_{[a}K_{b]}^{k}+\epsilon^k\,_{lm}\,\Gamma_{[a}^lK_{b]}^{m}\,.
\end{equation}

The boundary term (\ref{surfaceDecomp1}) reads:
\begin{align}
2\sqrt{|h|}\,&r_\nu e^\mu_Ie^\nu_J\,\leftidx{^\gamma}{\omega}{^{IJ}_\mu} \notag\\
&=\frac{2\sqrt{|h|}}{\gamma N\sqrt{\det q}}\bigg[2r_aN^{[a}E^{b]}_jA_b^j
+E^b_j\left(r_b\,\Lambda^j-r_t\,A_b^j\right)   
-r_b\frac{NE^b_j}{\sqrt{\det q}}\left(\gamma\epsilon^{ij}\,_k E^a_iA_a^k-(1+\gamma^2)\mathcal{S}^j\right)\bigg]  \label{surfaceDecomp2}
\end{align}
with $r_t:=t^\mu r_\mu$   and $r_a:=\tilde{e}^\mu_ar_\mu$ the temporal and spatial projections of the co-normal.

We now consider a cylindrical cut-off with cylindrical slabs  which is adapted to the 3+1 decomposition.
We take the convention where $r^\mu$ points \emph{inwards} if the boundary is space-like and \emph{outwards} if the boundary is time-like.
On $\Sigma_{1}$ and $\Sigma_{2}$,  $r_\mu=\pm n_\mu$, and since in the asymptotic region $t^\mu=n^\mu$, it follows that $r_a=0$, i.e. $r_t=\mp N=\mp 1+\Ord{r}{-1}{\!}$ is the only nonzero component of the normal.  
On the time-like cylinder $\tau_\infty\approx\mathbb{R}\times S^2_\infty$, $r_\mu$ is spatial and orthogonal to $t^\mu=n^\mu$, so $r_t=0$. Furthermore, on the cylinder one has $\sqrt{|h|}=N\sqrt{\det\boldsymbol{\sigma}}$, where $\boldsymbol{\sigma}$ is the induced metric and $\sqrt{\det\boldsymbol{\sigma}}\,\epsilon_{\alpha\beta}$  the induced volume element on the two-sphere at infinity. The second term proportional to $\Lambda^j$ in (\ref{surfaceDecomp2}) then exactly cancels the additional surface integral arising from the total divergence in  (\ref{bulkDecomp2}).
The 3+1 decomposition of the Holst action is therefore:
\begin{align}
S_\text{Holst}=&\frac{1}{2\kappa}\int_{t_1}^{t_2} \md t\Bigg[\int_{\Sigma_t}\md^3x\,
\frac{2}{\gamma}\left(E^b_j\bar{\partial}_tA_b^j+\Lambda^j\mathcal{G}_j-(1+\gamma^2)\omega_t^{0j}\mathcal{S}_j  \right) -N^a\mathcal{C}_a -N\mathcal{C} \notag \\
&+\frac{2}{\gamma}\int_{\partial\Sigma_t}\md S_a\,2N^{[a}E^{b]}_jA_b^j\,-\int_{\partial\Sigma_t}\md S_b
\frac{2NE^b_j}{\gamma\sqrt{\det q}}\left(\gamma\epsilon^{ij}\,_k E^a_iA_a^k-(1+\gamma^2)\mathcal{S}^j\right) \Bigg]  \notag\\
&+\frac{1}{\kappa\gamma}\int_{\Sigma_{t_1}}\md^3x\,E^a_iA_a^i\,-\frac{1}{\kappa\gamma}\int_{\Sigma_{t_2}}\md^3x\,E^a_iA_a^i \,,  \label{Holst3plus1}
\end{align}
with the surface element on the two-sphere
\begin{equation}
\md S_a:=\frac{1}{2}\epsilon_{abc}\md x^b\wedge\md x^c=\frac{1}{\sqrt{\det q}}\left(r_a\sqrt{|\det\boldsymbol{\sigma}|}\,\md^2y\right)\,,
\end{equation}
from which it follows that $\md S_a$ is of order $\Ord{r}{2}{}$ and of odd parity, i.e. antisymmetric under reflexions $x^a\to -x^a$ on the sphere. A property much used in asymptotic calculations.

Notice that with the fall-off conditions (\ref{falloffCondLapseShift1}) for the shift vector and lapse, the first surface integral in (\ref{Holst3plus1}) actually vanishes at $r\to\infty$.  We have deliberately kept this term because for constant $N^a$ it matches the ADM momentum $\mathbf{P}^a_\text{ADM}$, while the second (non-zero) surface term recovers, as we shall see shortly, the ADM energy $\mathbf{E}_\text{ADM}$.

Notice also that in this form the action is not manifestly finite:  the kinetic term $E^b_j\bar{\partial}_tA_b^j$ and the term $N\mathcal{C}$ are $\Ord{r}{-3}{\!}$, while the integrands $E^a_iA_a^i$ on the Cauchy surfaces $\Sigma_1$ and $\Sigma_2$ are $\Ord{r}{-2}{\!}$ and hence they all give seemingly divergent integrals.

Up to this point equations above are taken as mere redefinitions of the covariant dynamical fields with respect to the 3+1 foliation. Definition (\ref{defE}) does correspond  in the time gauge to the densitized triad in the canonical formulation, but the identification of the remaining variables with the corresponding canonical variables in the connection formulation will not be established until the constraint analysis is performed. In particular (\ref{defGamma}) does not correspond to the (three-dimensional) spin connection off-shell.
Expression (\ref{Holst3plus1}) is not quite a Hamiltonian action  yet. The split and partially gauge-fixed action does exhibit $A_a^i$ and $E^a_i$ as a canonical dynamical pair and shows which constraints arise from variation of the non-dynamical variables  $\Lambda^i$, $\omega_t^{0j}$, $N$ and $N^a$. However, there are more constraints because we are supposed to take independent variations of the connection components $\omega_a^{0i}$ and $\omega_a^{jk}$. If one is to take variations of the fixed combination $\delta A_a^i$, one must also take variations of the independent combination $\delta\Gamma_a^i$ to account for all degrees of freedom or directions in configuration space. It can be shown that  the equation of motion or constraint $\delta S_\text{Holst}/\delta\Gamma_a^i=0$ fixes (\ref{defGamma}) to be the (three-dimensional) spin connection (for details see \cite{BojowaldBook}).

If we take this on-shell condition, the constraint $\mathcal{S}_i\approx 0$, which from the discussion following equation (\ref{Kextension}) in appendix \ref{a::AshtekarVariables},  amounts to tensor $K_a^ie_b^i$ being symmetric, is equivalent to the Gauss constraint. This is because if $\Gamma_a^i$ is the spin connection then $\bar{\partial}_aE^a_i=-\epsilon_{ij}\,^k\,\Gamma_a^jE^a_k$, and hence  $\mathcal{S}_i:=\epsilon_{ij}\,^k\,K_a^jE^a_k=\gamma^{-1}\epsilon_{ij}\,^k(A_a^j-\Gamma_a^j)E^a_k=\gamma^{-1}\mathcal{G}_i$. 
Thus, in this case, variation with respect to $\omega_t^{0j}$ does not contribute independent conditions and we may drop the bulk term proportional to $\omega_t^{0j}\mathcal{S}_j$ from the action. This is inconsequential for considerations of finiteness and differentiability since such term is $\Ord{r}{-4}{}$.
Action becomes (dropping also the space-like boundary terms\footnote{Discarding these terms amounts to a canonical transformation because these integrals on $\Sigma_1$ and $\Sigma_2$ exactly cancel the surface term that appears if we integrate by parts the kinetic term to change it to $-A_b^j\bar{\partial}_tE^b_j$.}):
\begin{align}
S'_\text{Holst}=&\frac{1}{2\kappa}\int_{t_1}^{t_2} \md t\Bigg[\int_{\Sigma_t}\md^3x\,
\frac{2}{\gamma}\left(E^b_j\bar{\partial}_tA_b^j+\Lambda^j\mathcal{G}_j  \right) -N^a\mathcal{C}_a -N\mathcal{C} \notag \\
&+\frac{2}{\gamma}\int_{\partial\Sigma_t}\md S_a\,2N^{[a}E^{b]}_jA_b^j\,+\int_{\partial\Sigma_t}\md S_b
\frac{2NE^b_j}{\sqrt{\det q}}\left(\bar{\partial}_aE^a_j+\frac{1}{\gamma^2}\mathcal{G}^j\right) \Bigg]\,.   \label{HolstHamiltonian}
\end{align}
This is already in Hamiltonian form:
\begin{equation} \label{generalHamiltonianAction}
S'_\text{Holst}=\int_{t_0}^{t_1} \md t\left[\frac{1}{\kappa\gamma}\int_\Sigma \md^3x\, \left(E^a_i\bar{\partial}_tA_a^i\right)\,-\mathbf{H_\text{T}}\right]\,.
\end{equation}
However, to be a well defined Hamiltonian action giving Hamilton's equations plus constraints:
\[
\frac{\delta S'_\text{Holst}}{\delta A_a^i(t,x^a)}=-(\kappa\gamma)^{-1}\bar{\partial}_tE^a_i-\frac{\delta\mathbf{H_\text{T}}}{\delta A_a^i}=0, \qquad  \kappa\gamma\frac{\delta S'_\text{Holst}}{\delta E^a_i(t,x^a)}=\bar{\partial}_tA_a^i-\kappa\gamma\frac{\delta\mathbf{H_\text{T}}}{\delta E^a_i}=0, 
\]
and 
\[ 
  \frac{\delta S'_\text{Holst}}{\delta N(t,x^a)}=-(2\kappa)^{-1}\mathcal{C}\approx 0, \quad   \frac{\delta S'_\text{Holst}}{\delta N^a(t,x^a)}=-(2\kappa)^{-1}\mathcal{C}_a\approx 0, \quad  \frac{\delta S'_\text{Holst}}{\delta \Lambda^i(t,x^a)}=-(\kappa\gamma)^{-1}\mathcal{G}_i\approx 0
\]
implies (\ref{HolstHamiltonian}) must be finite (at least on shell) and differentiable with respect to $\Lambda^i$, $N$, $N^a$, $A_a^i$ and $E^a_i$.
Finiteness follows from finiteness of the complete Holst action (\ref{Holst1}), since (\ref{HolstHamiltonian}) is a partially on-shell and gauge-fixed version of it. Now that one is partially on-shell and has taken a different linear combination of independent variations one should re-check differentiability directly.  It is straight forward to check action is differentiable with respect to the non-dynamical fields for variations consistent with the original fall-off conditions:
\begin{equation} \label{variationLapseShift1}
\delta N=\frac{\delta f_{tt}}{2r}+\ord{r}{-1}\,, \qquad \delta N^a=\frac{\delta f_{ta}}{r}+\ord{r}{-1}\,.
\end{equation}
and
\[
\delta \Lambda^i=\frac{\delta G^i}{r^2}+\ord{r}{-1}\,.
\]
Variation with respect to $A$ and $E$ gives
\[
\delta S'_\text{Holst}=\int\md^3x \left(-(\kappa\gamma)^{-1}\bar{\partial}_tE^a_i-\frac{\delta\mathbf{H_\text{T}}}{\delta A_a^i}\right)\delta A_a^i \,+\left((\kappa\gamma)^{-1}\bar{\partial}_tA_a^i-\frac{\delta\mathbf{H_\text{T}}}{\delta E^a_i}\right)\delta E^a_i\,.
\]
There are no surface terms  on $\Sigma_1$ and $\Sigma_2$ coming from the integration by parts of the kinetic term  since we are assuming variations are zero there. 
By our assumptions, $\bar{\partial}_tA_a^i$ falls-off as $\Ord{r}{-3}{\!}$, so both terms $\bar{\partial}_tE^a_i\delta A_a^i$ and $\bar{\partial}_tA_a^i\delta E^a_i$ are $\Ord{r}{-4}{\!}$ and give finite integrals. As we shall verify in the next section and in more detail in appendix \ref{a::AshtekarVariables}, the surface term in (\ref{HolstHamiltonian}) is precisely what is needed to make the total Hamiltonian $\mathbf{H_\text{T}}$ differentiable (with respect to $A$ and $E$) on all of phase space. 
We therefore have a well defined Hamiltonian action without additional assumptions.
 
Expression (\ref{HolstHamiltonian}) is our main result, it shows the Hamiltonian for asymptotically flat solutions of general relativity in Ashtekar-Barbero variables may be derived directly from the Holst action (\ref{Holst1}). As expected, this Hamiltonian is the sum of standard constraints plus the ADM momentum and energy.

As we have already stated, action (\ref{HolstHamiltonian}) is nevertheless not manifestly finite. Even on-shell, finiteness is not easily seen: the total Hamiltonian $\mathbf{H_\text{T}}$ equals the non-zero but finite second surface integral giving the ADM energy, but the kinetic term $E^a_i\bar{\partial}_tA_a^i\approx\kappa\gamma E^a_i\frac{\delta\mathbf{H_\text{T}}}{\delta E^a_i}$ is only manifestly $\Ord{r}{-3}{\!}$.
As we shall see, additional requirements on dynamical fields for explicit finiteness of this action (on-shell) are closely related to conditions for a well-posed Hamiltonian formulation. These requirements are parity conditions (\ref{AEparity}) on the leading order terms of $A$ and $E$ fields.  In this case, action (\ref{HolstHamiltonian}) is also finite and differentiable for more general conditions for lapse and shift (corresponding to odd space and time supertranslations and up to constant time translations).


\section{Hamiltonian formulation and Poincar\'e charges}  
\label{s::HamiltonianFormulation}

The well-posed variational principle encoded in action (\ref{HolstHamiltonian}) already provides us with the gravitational Hamiltonian in Ashtekar-Barbero variables along with Einstein's field equations written in Hamiltonian form and expressions for the ADM energy and momentum. We now turn to the related but different problem of constructing or specifying conditions necessary to have a well defined Hamiltonian formulation in terms of connection variables and accommodating asymptotically flat solutions. That is, we look to specify the extended phase space for general relativity $\Gamma_{(A,E)}$ admitting a well defined symplectic structure $\Omega_{(A,E)}$ and canonical generators whose Hamiltonian flows on $\Gamma_{(A,E)}$ correspond to gauge and asymptotic Poincar\'e symmetries.
For simplicity and without loss of generality, we will continue to assume the adapted coordinates to the 3+1 foliation coincide with asymptotic coordinates in
(\ref{cartesianAFCondition2}), so the spatial coordinates $\{x^a\}_{a=1,2,3}$ are the Cartesian coordinates associated to the fixed background asymptotic spatial metric which therefore just looks like the delta $\delta_{ab}$ in these coordinates\footnote{Following \cite{Thiemann} one may consider more general coordinates $\{\bar{x}^a\}_{a=1,2,3}$ on the hypersurfaces $\Sigma_t$ of the 3+1 foliation. One may define asymptotic expansions and perform calculations using  coordinates $\{\bar{x}^a\}_{a=1,2,3}$ which are related to the Cartesian ones `at zeroth order'. This implies in particular the transition matrix $\frac{\partial x^a}{\partial\bar{x}^{\bar{b}}}$ is of order one or equivalently a function of the angular coordinates $\bar{x}^a/\bar{r}$, but there are additional necessary conditions not stated in \cite{Thiemann} to ensure asymptotic expansions with respect to these coordinates are equivalent to Cartesian ones.  We work in Cartesian coordinates to avoid such ambiguities and to simplify expressions for asymptotic Killing fields.}.

\subsection{Phase space}

As the action (\ref{HolstHamiltonian}) already shows, variables (\ref{defE}) and (\ref{defA}) are conjugate canonical pairs corresponding to the densitized triad and the Ashtekar-Barbero connection. We remind the reader that to accommodate spinors, the residual $SO(3)$-gauge symmetry of the action is trivially extended to $SU(2)$-gauge using  the isomorphism of Lie algebras $\mathfrak{su}(2)\simeq\mathfrak{so}(3)$ (and the corresponding two to one projection $SU(2)\to SO(3)$). This isomorphism is given by the map $\tau_i\mapsto J_i$, for $\tau_i:=-\frac{i}{2}\sigma^i$, a basis of $\mathfrak{su}(2)$, with $\sigma_i$ the standard  Pauli matrices, and  $(J_j)^i\,_k=\epsilon_{ijk}$ the basis of $3\times 3$ matrix generators of rotations in the standard representation of $SO(3)$.
Accordingly, the flat $\mathbb{R}^3$ internal indices for the densitized triad (and co-triad) are identified with (dual) $\mathfrak{su}(2)$ indices using the vector isomorphism  $\mathbb{R}^3\ni(X^i,X^2,X^3)\mapsto X^i\sigma_i$, so that an $SO(3)$ rotation of $E^a_i$ is equivalent to the adjoint action of $SU(2)$ on $E^a=E^a_i\tau^i$.

In short, the extended phase space for General Relativity is coordinatized by the $\mathfrak{su}(2)$-valued connection $A_a=A_a^i\tau_i$ and the $\mathfrak{su}(2)$-valued densitized triad vector $E^a=E^a_i\tau^i$ satisfying the canonical relations
\begin{equation}
\{A_a^i(\vec{x}),E^b_j(\vec{y})\}=\kappa\gamma\,\delta^i_j\,\delta^b_a\,\delta(\vec{x},\vec{y})\,, \qquad 
\{A_a^i(\vec{x}),A_b^j(\vec{y})\}=\{E^a_i(\vec{x}),E^b_j(\vec{y})\}=0
\end{equation}
and subject to additional gauge freedom
\begin{equation}  \label{su2Gauge}
A_a \to gA_ag^{-1}+g\partial_ag^{-1},   \qquad  E^a \to gE^ag^{-1},   
\end{equation}
for $g(x^a)\in SU(2)$.

To fully characterize the extended phase space so that it includes asymptotically flat solutions, the general fall-off conditions for the canonical fields may be taken to be (\ref{Afalloff1}) and (\ref{Efalloff1}), the same as those derived from (\ref{cartesianAFCondition2}) after the 3+1 splitting and partial gauge fixing.
These conditions are  consistent with the conditions first used by ADM in their Hamiltonian formulation of general relativity \cite{ADM1962} and in the first geometric formulation of spatial infinity in terms of initial Cauchy data \cite{Geroch}.  The latter conditions are essentially the same asymptotic expansion to order one  (\ref{falloffqmetric1}) for the metric $q_{ab}$ and an asymptotic expansion to order two for extrinsic curvature $K_{ab}$, or equivalently, an asymptotic expansion to order two for the canonical conjugate momenta $p^{ab}:=\sqrt{\det q}(K_{ab}-K_{cd}q^{cd}q_{ab})$ of the form\footnote{As already noted in section \ref{s::preliminaries}, conditions (\ref{falloffqmetric1}) and (\ref{falloffp}) are slightly more general, since remainders are $\ord{r}{-1}$ and $\ord{r}{-2}$ instead of $\Ord{r}{-1}{\!}$ and $\Ord{r}{-2}{\!}$.}
\begin{equation}  \label{falloffp}
p^{ab}=\frac{h^{ab}(t,x^c/r)}{r^2}+\ord{r}{-2}\,.
\end{equation}
As already mentioned and shown  in \cite{AshtekarMagnon}, conditions (\ref{falloffqmetric1}) and (\ref{falloffp}) also follow directly from the geometric treatment of \cite{AshtekarHansen}. However, it is well-known that these conditions are not enough to guarantee a Hamiltonian formulation in terms of geometro-dynamical variables, with well defined symplectic structure, canonical generators of symmetries and unique conserved charges. A sufficient set of \emph{parity conditions} on the leading order terms of (\ref{falloffqmetric1}) and (\ref{falloffp}) were first spelled out by Regge and Teitelboim \cite{ReggeTeitelboim} and require $f_{ab}$ to be \emph{even}, i.e. symmetric with respect to reflections on the sphere, and $h^{ab}$ to be \emph{odd} or anti-symmetric with respect to such reflections:
\begin{equation} \label{qpParity}
f_{ab}(-x^c/r)=f_{ab}(x^c/r)\,, \qquad \text{and}  \qquad h^{ab}(-x^c/r)=-h^{ab}(x^c/r).
\end{equation}
These conditions are not only necessary for the well-posedness of the Hamiltonian formulation in terms of ADM variables $(q_{ab},p^{ab})$, but also to correctly define Poincar\'e charges\footnote{For a related treatment without parity conditions see \cite{CompereDehouck}.}.  

The parity conditions for the next-to-leading order terms of the metric, and correspondingly for the connection variables, are not necessary for the well-posedness of the action (\ref{Holst1}) with cylindrical cut-offs of spacetime or its partially on-shell and gauge-fixed version (\ref{HolstHamiltonian}), hence they cannot be inferred from it. However, the parity conditions can be (partially) read-off from  the more restrictive conditions  necessary for  differentiability of (\ref{Holst1}) for hyperbolic cut-offs and for the well-posedness of the covariant Hamiltonian formulation. These conditions are (\ref{covariantTetradCoef}) and (\ref{covariantConnectionCoef}) with  mass function $\sigma$ reflexion symmetric (\ref{BeigAshtekarCond}). 
As shown in appendix \ref{a::fromHyper2CylExpansion}, to derive what the latter conditions imply for the spatial fall-off behavior of the canonical fields we may write the  $r$-expansions corresponding to $\rho$-expansions of covariant fields evaluated at $t=0$.
Additionally, the condition $\eta_{IJ}\leftidx{^0}e{_\mu^I}\leftidx{^0}e{_\nu^J}=\eta_{\mu\nu}$ implies the asymptotic co-tetrad and tetrad $\leftidx{^0}{e}{^\mu_I}$ must have definite (even or odd) parity with respect to reflections on the hyperboloid and consequently with respect to reflexions on the sphere in the limit $t/r\to 0$. The order $r^{-1}$ spatial component (\ref{cotriadParityForm}) of the co-tetrad, which corresponds to (\ref{covariantTetradCoef}) in an  $r$-expansion at $t=0$, has also the same parity as the asymptotic tetrad $\leftidx{^0}{e}{^\mu_I}$, and therefore from the detailed expansion formula (\ref{detailedEfalloff}) for the densitized triad, one sees $\bar{E}^a_i$ and  $F^a_i$ in the general expansion (\ref{Efalloff1}) must have the same definite parity (even or odd). On the other hand, the  $r$-expansion coefficient (\ref{connectionParityForm}) for the Lorentz connection -corresponding to (\ref{covariantConnectionCoef})-  has definite odd parity regardless of the parity of the asymptotic tetrad. Consequently, $G_a^i$ in asymptotic expansion of the Ashtekar-Barbero connection   (\ref{Afalloff1})  has definite odd parity and the same is true for the corresponding coefficients in the expansions for $K_a^i$ and $\Gamma_a^i$.

Similar fall-off conditions were first stated in \cite{AshtekarLectures} for the spinorial variables and the parity conditions in \cite{Thiemann} for the complex self-dual variables, but they were derived by directly asking compatibility with the ADM asymptotic behavior (\ref{falloffqmetric1}), (\ref{falloffp}) and (\ref{qpParity}). Also in this case, the parity conditions of $q_{ab}$ and $p^{ab}$ alone do not directly fix the parity of expansion coefficients of $A_a^i$ and $E^b_j$. Their parity, as we shall see, is actually fixed by the requirement of a well defined Hamiltonian formulation in terms of these variables.
ADM conditions also imply $E^a_i$ must admit at least an asymptotic expansion to order 1 
with both $\bar{E}^a_i$ and  $F^a_i$ of the same definite parity. 
Compatibility with the triad implies the leading $r^{-2}$ term of the spin connection is always odd, whereas from (\ref{qpParity}) it follows that the corresponding term for $K_a^i=K_{ab}e^{bi}$ has the opposite parity of the triad.
Hence, for the components of the vector potential of the Ashtekar-Barbero connection, the ADM conditions imply $A_a^i$ must admit at least an asymptotic expansion to order 2 with $G^i_a$ of definite odd parity only when $\bar{E}^a_i$ is even.

To summarize, from the restricted conditions (\ref{BeigAshtekarCond}) and (\ref{AshtekarHansenCond}) one may infer the densitized triad must admit an asymptotic expansion at least to order 1 and the Ashtekar-Barbero connection must admit an asymptotic expansion at least to order 2 of the form:
\begin{equation} \label{AEfalloff}
E^a_i=\bar{E}^a_i+\frac{F^a_i}{r}+\ord{r}{-1}, \qquad \qquad A^i_a=\frac{G^i_a}{r^2}+\ord{r}{-2}
\end{equation}
with leading terms $\bar{E}^a_i$, $F^a_i$ of even parity and $G^i_a$ odd:
\begin{equation} \label{AEparity}
\bar{E}^a_i(-x^c/r)=\bar{E}^a_i(x^c/r),  \quad   F^a_i(-x^c/r)=F^a_i(x^c/r),  \quad  G^i_a(-x^c/r)=-G^i_a(x^c/r).
\end{equation}
Accordingly, variations, and tangent vectors on phase space, must be compatible with these conditions:
\[
\delta E^a_i=\frac{\delta F^a_i}{r}+\ord{r}{-1}, \qquad \qquad \delta A^i_a=\frac{\delta G^i_a}{r^2}+\ord{r}{-2}\,.
\]
This behavior for tangent vectors is consistent with the interpretation of the flow of a vector field $X=\leftidx{^A}{X}{^i_a}\frac{\delta}{\delta A^i_a}+\leftidx{^E}{X}{^a_i}\frac{\delta}{\delta E^a_i}$ as the solution to the differential equations
\[
\frac{\partial A_a^i}{\partial t}=\leftidx{^A}{X}{^i_a}(A,E)\,, \qquad  \frac{\partial E^a_i}{\partial t}=\leftidx{^E}{X}{^a_i}(A,E)\,.
\]

The other possibility of  $\bar{E}^a_i$, $F^a_i$ being of odd parity and $G^i_a$ of odd or indefinite parity, is ruled out  by requiring the well-posedness of the Hamiltonian formulation. Indeed, if the symplectic structure $\Omega_{(A,E)}$ acting on tangent vectors $(\delta A_a^i,\delta E^a_i)$ and $(\widetilde{\delta A_a^i},\widetilde{\delta E^a_i})$
\begin{equation}  \label{symplecticStructure}
\Omega_{(A,E)}((\delta A_a^i,\delta E^a_i),(\widetilde{\delta A_a^i},\widetilde{\delta E^a_i}))=\frac{1}{\kappa\gamma}\int_\Sigma\md^3x\,\left(\delta E^a_i\widetilde{\delta A_a^i}-\delta A_a^i\widetilde{\delta E^a_i}\right)
\end{equation}
is to be finite, the leading order $r^{-3}$ term inside parentheses must be of definite odd parity to ensure its integral vanishes and there are no logarithmic divergences.

With the parity conditions, a covector field $\omega=\int\, \leftidx{^A}{\omega}{^a_i}\,\dd A_a^i+\leftidx{^E}{\omega}{_a^i}\,\dd E^a_i$ on the extended phase space must be such that the leading term in $\leftidx{^A}{\omega}{^a_i}$ is $\Ord{r}{-1}{even}$ and the leading term in $\leftidx{^E}{\omega}{_a^i}$ is $\Ord{r}{-2}{odd}$ so that its action on a vector field is well defined (finite):
\[
\omega_{(A,E)}(\delta A_a^i,\delta E^a_i)=\int_\Sigma \md^3x\, \leftidx{^A}{\omega}{^a_i}(A,E)(x)\,\delta A_a^i(x)+\leftidx{^E}{\omega}{_a^i}(A,E)(x)\,\delta E^a_i(x) \; <\infty\,.
\]
In particular for a differentiable function $f(A,E)$ on phase space, the functional derivative $\frac{\delta f}{\delta A_a^i(x)}$ must be of leading order $\Ord{r}{-1}{even}$ and $\frac{\delta f}{\delta E^a_i(x)}$ of leading order $\Ord{r}{-2}{odd}$. One can then check $\Omega_{(A,E)}$ maps allowed vectors into allowed covectors so that Hamiltonian vector fields $X_f=\{\,\cdot\,,f\}$ with Poisson brackets
\[
\{f,g\}=\kappa\gamma\int_\Sigma \md^3x \left(\frac{\delta f}{\delta A_a^i}\frac{\delta g}{\delta E^a_i}-\frac{\delta f}{\delta E^a_i}\frac{\delta g}{\delta A_a^i}\right)
\]
are well defined for differentiable functions $f$ and $g$ on the extended phase space.

Notice that if one does not fix the asymptotic triad $\bar{e}^a_i$ and considers a more general phase space allowing variations
\[
\delta E^a_i=\delta\bar{E}^a_i+\frac{\delta F^a_i}{r}+\ord{r}{-1}
\]
of leading order 1 even (odd), the symplectic structure (\ref{symplecticStructure}) is ill-defined, being  generically  divergent unless we restrict to variations $\delta A_a^i$ of leading order $\Ord{r}{-3}{odd}$ (even) which is inconsistent with ADM fall-off conditions.
A fixed asymptotic triad follows or is consistent with the original assumption of a fixed asymptotic co-tetrad to guarantee differentiability of the Holst action (\ref{Holst1}) and gauge invariance of the boundary term. 
There is no loss of generality or incompleteness of the Hamiltonian treatment in fixing the asymptotic densitized triad $\bar{E}^a_i$ and ultimately freezing the  $SU(2)$-gauge freedom at infinity. Any two asymptotic orthonormal frames $\bar{e}^i_a$ and $\tilde{e}^i_a$ are related by an order 1 even gauge rotation $\tilde{e}^i_a=R^i_j\bar{e}^j_a$, with $R^i_j=\tilde{e}^i_a\bar{e}^a_j\in SO(3)$, so the elements of the infinite family of phase spaces parametrized by different $\bar{e}_a^i$ (or different $\bar{E}^a_i$)  are all gauge equivalent.

Despite the fact that asymptotic conditions (\ref{AEfalloff}) and (\ref{AEparity}) are sufficient for a well defined Hamiltonian formulation, an interpretation of the extended phase space as a cotangent bundle $T^*\mathcal{A}$ is not at hand since the action of cotangent vectors on tangent vectors
\[
E(\delta A):=\int_\Sigma \md^3x\, E^a_i\, \delta A_a^i=\int_\Sigma \md^3x\left(\bar{E}^a_i+\frac{F^a_i}{r}+\ord{r}{-1}\right)\left(\frac{\delta G^i_a}{r^2}+\ord{r}{-2}\right)
\]
is ill-defined due to the generically divergent $\ord{r}{0}$ terms proportional to $\bar{E}^a_i$ in the integrand. By the same token, the kinetic term $\int  E^a_i \bar{\partial}A_a^i$ in a Hamiltonian action may also diverge. To have a well defined action of cotangent vectors on tangent vectors $E(\delta A)$ and a manifestly finite symplectic potential, $A_a^i$ must be required to admit and asymptotic expansion to order three 
\begin{equation} \label{restrictedAExpansion}
A_a^i=\frac{\leftidx{^2}{A}{_a^i}}{r^2}+\frac{\leftidx{^3}{A}{_a^i}}{r^3}+\ord{r}{-3}\,,
\end{equation} 
with both $\leftidx{^2}{A}{_a^i}$ and $\leftidx{^3}{A}{_a^i}$ of odd parity.
This results if we had started from (\ref{connectionExpansion3}) from the very beginning, with parity  also consistent with (\ref{strongAFmetric}) and (\ref{BeigAshtekarCond}) as shown in appendix \ref{a::fromHyper2CylExpansion}.
However, from expressions  (\ref{PDgeneratorConnVariation}) and (\ref{HConnVariation}) in appendix \ref{a::AshtekarVariables}, conditions above alone are not sufficient to guarantee that the parity of the order $r^{-3}$ term is preserved under the Hamiltonian flow of constraints and generators.

\subsection{Gauge vs asymptotic symmetries: constraints and Poincar\'e generators}

We now turn to the constraints and the canonical generators of asymptotic symmetries on this extended phase space.

Unlike conditions (\ref{BeigAshtekarCond})  and  (\ref{AshtekarHansenCond}) for the full spacetime metric (\ref{strongAFLineElement}) which are used in most  covariant treatments, the parity conditions above do not reduce the asymptotic symmetry group to Poincar\'e.
The requirement of the dynamical Hamiltonian flow further preserving the fall-off and parity conditions (\ref{qpParity}), eliminates logarithmic supertranslations but only `halves' the number of  supertranslations  to the \emph{odd supertranslations} \cite{Teitelboim, ourReview}.
In other words, the 3+1 decomposition of the general asymptotic Killing vector field $\xi^\mu=Mn^\mu+M^\mu$, corresponding to coordinate transformations (\ref{coordTransformations}), and further preserving the parity conditions,  gives a 10 parameter family of vector fields which, in our adapted asymptotic Lorentzian coordinates, has the form 
\begin{equation}  \label{PoincareLapseFallOff}
M=\beta_cx^c+T+S_{\text{odd}}(x^c/r)+\ord{r}{0}
\end{equation}
\begin{equation} \label{PoincareShiftFallOff}
M^a=\epsilon^a\,_{bc}\,\alpha^cx^b-\beta^at+T^a+S^a_{\text{odd}}(x^c/r)+\ord{r}{0}\,,
\end{equation}
with additional odd functions:
\begin{equation} \label{oddSuperTranslation}
S_\text{odd}(-x^c/r)=-S_\text{odd}(x^c/r)\qquad \text{and} \qquad S^a_\text{odd}(-x^c/r)=-S^a_\text{odd}(x^c/r)\,,
\end{equation}
generating  (infinitesimal) time and spatial odd supertranslations, constant parameters  $M=T$ and $M_a=T_a$ generating time and space translations, $M=0$, $M^a=\epsilon^a\,_{bc}\alpha^cx^b$ generating rotations around the axis determined by vector $\alpha^a$, and $M=\beta_cx^c$, $M^a=-\beta^at$ corresponding to boosts along the axis determined by $\beta^a$.

To ensure a well defined Hamiltonian formulation, the phase space of full general relativity is restricted to spacetime metrics satisfying (\ref{qpParity}). The restricted phase space admits well defined canonical generators corresponding to symmetries (\ref{PoincareLapseFallOff}) and (\ref{PoincareShiftFallOff}). This reduction, not fully eliminating the supertranslation ambiguities, is different from the Ashtekar-Hansen conditions (\ref{AshtekarHansenCond}) to eliminate supertranslations and the Beig-Ashtekar condition (\ref{BeigAshtekarCond}) to eliminate logarithmic translations. Nevertheless, as we have displayed  -and as first shown for the ADM phase space in \cite{BeigMurchadha} using slightly different methods- the extended phase space $\Gamma_{(A,E)}$ with parity restrictions certainly contains  spacetimes (\ref{strongAFLineElement}) satisfying the Ashtekar-Hansen conditions and (\ref{BeigAshtekarCond}) and which additionally admit a 3+1 decomposition.
Beig and \`O Murchadha have first clarified in \cite{BeigMurchadha} how in a 3+1 Hamiltonian formulation, despite the odd supertranslations, the Poincar\'e group may be regarded as symmetry group of asymptotically flat spacetimes. The ambiguity of odd supertranslations (and terms of order $\ord{r}{0}$) for the Killing fields being regarded as gauge.

Variation of action (\ref{HolstHamiltonian}) with respect to $\Lambda^i$, $N^a$ and $N$ shows the smeared contraints can be taken as
\begin{align}
\mathcal{G}[\lambda^i]&=\frac{1}{\kappa\gamma}\int_\Sigma\lambda^i(\bar{\partial}_aE^a_i+\epsilon_{ij}\,^kA_a^jE^a_k) \label{GaussConstraint1}\\
D[M^a]&:=\frac{1}{\kappa\gamma}\int_\Sigma \md^3xM^a\left(F^i_{ab}E^b_i-A^i_a\mathcal{G}_i\right)
=\frac{1}{\kappa\gamma}\int_\Sigma \md^3x\,M^a\left(2E^b_i\bar{\partial}_{[a}A_{b]}^i-A_a^i\bar{\partial}_bE^b_i\right) \label{diffeomorphismConstraint1}\\
H[M]&:=\frac{1}{2\kappa}\int_\Sigma \md^3x\,\frac{M}{\sqrt{\det E}}E^a_iE^b_j\left(\epsilon^{ij}_kF^k_{ab}-2(1+\gamma^2)K^i_{[a}K^j_{b]}\right)\,. \label{HamiltonianConstraint1}
\end{align}
The first is the Gauss constraint and the others are particular extensions of the ADM diffeomorphism and Hamiltonian constraints.
There is of course and ambiguity in the definition of such extensions. The expressions we have taken  differ from (\ref{VectorConstraint}) and  (\ref{CConstraint}) by multiples of the Gauss constraint but since these terms are all $\Ord{r}{-4}{even}$ they play no role in the discussion of finiteness and differentiability (see appendix \ref{a::AshtekarVariables}).

We already know finiteness and differentiability of the whole action (\ref{HolstHamiltonian}), however this is different from the requirement that each $H[M]$, $D[M^a]$ and $\mathcal{G}[\lambda^i]$ is a well defined gauge generator for corresponding smearing field parameters $M$, $M^a$ and $\lambda^i$ which generically have different asymptotic behavior as (\ref{falloffCondLapseShift1}) and (\ref{Lambdafalloff})\footnote{It is common practice to use the same letters $N$, $N^a$ and $\Lambda^I$ to denote the smearing  parameters  for the gauge generators as the Lagrange multipliers appearing in the action (\ref{generalHamiltonianAction}).
We caution the reader then of the multiple role of smearing parameters $N$ and $N^a$: first as metric components or lapse and shift in a 3+1 decomposition, as Lagrange multipliers in the Hamiltonian action (\ref{HolstHamiltonian})  and in this case also as the normal an tangential components of Killing vector fields. We use different letters for the components of Killing fields to avoid this confusion.}.
We also stress finiteness and differentiability on all of phase space of the three-dimensional integrals over $\Sigma$ for the constraints  is different from finiteness and differentiability of the action over 4-dim region $M$.

If the constraints  are  to canonically generate gauge transformations on all of phase space $\Gamma_{(A,E)}$ for corresponding smearing gauge parameter fields $M$, $M^a$ and $\lambda^i$, they should be finite on all of phase space\footnote{One could try to relax these conditions to a neighborhood of the constraint surface if the constraints are only required to generate gauge on-shell. We will not pursue this here however, since it involves topological considerations on the phase space.}

\[
H[M]=\int_\Sigma\md^3x\, M\mathcal{H}\,<\infty, \quad D[M^a]=\int_\Sigma\md^3x\, M^a\mathcal{D}_a\,<\infty, \quad \mathcal{G}[\lambda^i]=\int_\Sigma\md^3x\, \lambda^i\mathcal{G}_i\,<\infty,
\]
Furthermore, in order for their corresponding Hamiltonian flows  
\[
X_{H[M]}=\{\,\cdot\,,H[M]\}, \qquad X_{D[M^a]}=\{\,\cdot\,,D[M^a]\}, \qquad  X_{\mathcal{G}[\lambda^i]}=\{\,\cdot\,,\mathcal{G}[\lambda^i]\}, 
\]
to be well defined and generate gauge, the constraints must be differentiable on all of phase space, that is variation must be expressible as
\begin{equation} \label{differentiabilityDef2}
\delta H[M]=\int_\Sigma\md^3x\,\vd{H[M]}{A_a^i}\delta A_a^i+\vd{H[M]}{E^a_i}\delta E^a_i
\end{equation}
and similarly for $D[M^a]$ and $\mathcal{G}[\lambda^i]$, for all `directions' on phase space, i.e., for arbitrary variations $\delta A_a^i$, $\delta E^a_i$ consistent with the fall-off conditions of the fields.

\subsubsection{Gauss constraint and $SU(2)$-gauge generator}

We start with the smeared version of the Gauss constraint  (\ref{GaussConstraint1}):
\[
\mathcal{G}[\lambda^i]=\frac{1}{\kappa\gamma}\int_\Sigma \md^3x\,\lambda^i(\underbrace{\bar{\partial}_aE^a_i}_{\Ord{r}{-2}{odd}}+\underbrace{\epsilon_{ij}\,^kA_a^jE^a_k}_{\Ord{r}{-2}{odd}}\;\;)\,,
\]
where -for ease of the reader- we have marked the fall-off behavior of each summand and the parity of its leading term. 
To be finite one must require the smearing field $\lambda^i$ to be at most $\Ord{r}{-1}{even}$. With these fall-off conditions $\mathcal{G}[\lambda^i]$ is also differentiable with respect to $A_a^i$, $E^a_i$ and $\lambda^i$. Indeed, taking the variation
\begin{align*}
\delta\mathcal{G}[\lambda^i]=&\,\mathcal{G}[\delta\lambda^i]\,+\,\int_\Sigma\md^3x\Big[\underbrace{(\,\lambda^i\epsilon_{ij}\,^kE^a_k\,)\,\delta A_a^j}_{\Ord{r}{-3}{odd}}+(\,\underbrace{-\bar{\partial}_a\lambda^k}_{\Ord{r}{-2}{odd}}+\underbrace{\lambda^i\epsilon_{ij}\,^kA_a^j}_{\Ord{r}{-3}{odd}}\,)\,\delta E^a_k\Big]  \\
&\,+\,\int_{\partial\Sigma}\md S_a\underbrace{\lambda^i\delta E^a_i}_{\Ord{r}{-2}{even}}
\end{align*}
one verifies the bulk integral is finite since the potentially divergent order ${r}^{-1}$ leading terms in the bulk integrand are odd and hence integrate to zero. The surface term, with leading term ${\Ord{r}{0}{odd}}$ also integrating to zero, vanishes at infinity $r\to\infty$.  
Hence, with this fall-off behavior for $\lambda^i$, the Hamiltonian vector field of $\mathcal{G}[\lambda^i]$ on $\Gamma_{(A,E)}$ is well defined and generates infinitesimal gauge transformations
\[
\delta A_a^i=\{A_a^i,\mathcal{G}[\lambda^i]\}=-\bar{\partial}_a\lambda^i+\epsilon_{lj}\,^i\lambda^lA_a^j\,, \qquad \qquad
\delta E^a_i=\{E^a_i,\mathcal{G}[\lambda^i]\}=\epsilon_{il}\,^k\lambda^lE^a_k\,,
\]
corresponding to finite $SU(2)$-gauge transformations (\ref{su2Gauge}) for $g=\exp(\lambda^i\tau_i)=\sum_{n=0}^\infty\frac{1}{n!}(\lambda^i\tau_i)^n$. So the allowed gauge transformations become the identity at infinity $r\to\infty$.

Consistent with the fact that the extended phase space only contains fields that asymptote to a fixed frame, there is no canonical infinitesimal generator of asymptotic gauge transformations. As we have already stated, any asymptotic gauge transformation mapping allowed orthonormal frames into allowed orthonormal frames must necessarily be order $r^0$ even which translates (using the correspondence $\exp{\lambda^iJ_i}\to \pm \exp{\lambda^i\tau_i}$) to the leading term of $\lambda^i$ being too order 1 even.  A canonical generator $\mathbf{G_\mathcal{G}}[\lambda^i]$ of asymptotic gauge transformations, if it existed, would have functional derivatives with respect to $E^a_i$ and $A_a^i$ 
\begin{equation}   \label{GaussFDerivatives}
\frac{\delta \mathbf{G_\mathcal{G}}[\lambda^i]}{\delta E^a_i}=-\bar{\partial}_a\lambda^i+\epsilon_{lj}\,^i\lambda^lA_a^j\,, \qquad \qquad    \frac{\delta \mathbf{G_\mathcal{G}}[\lambda^i]}{\delta A_a^i}=-\epsilon_{il}\,^k\lambda^lE^a_k
\end{equation}
of order $\Ord{r}{-1}{odd}$ and $\Ord{r}{0}{even}$ respectively. This is inconsistent with fall-off conditions stated above for the functional derivatives of a differentiable function on this phase space.
Since we have been forced to fix the asymptotic frame, there is no room for a canonical generator of asymptotic $SO(3)$-gauge transformations on phase space and consequently there are no $SO(3)$ charges either.

Finally, note that the fall-off behavior for the smearing angle $\lambda^i$ cannot be derived from (\ref{defLambda}), the Lagrange multiplier $\Lambda^i$ in the action (\ref{HolstHamiltonian}) (although the fall-off behavior of $\Lambda^i$ (\ref{Lambdafalloff}), is consistent with the $\Ord{r}{-1}{\!}$ behavior of $\lambda^i$, the parity of (\ref{connectionParityForm}) -and that of the next higher order term- is `wrong' since the order $r^{-2}$ and $r^{-3}$ terms of $\lambda^i$ must be even if the Gauss constraint is to preserve parity of the more restricted asymptotic expansions (\ref{restrictedAExpansion}) and (\ref{cotriadExpansion2})).

\subsubsection{Proper diffeomorphism gauge generators and improper gauge Poincar\'e generators} \label{s::Pgenerators}

It is straight forward to verify $H[N]$ and $D[N^a]$ are finite and differentiable for smearing fields corresponding to \emph{proper} gauge transformations:
\begin{align}
N&=S_\text{odd}(x^c/r)+\ord{r}{0} \notag\\
N^a&=S_\text{odd}^a(x^c/r)+\ord{r}{0}  \label{properGauge}
\end{align}
and as we shall review at the end of this section, they serve as canonical generators of diffeomorphisms which are supertranslations or the identity at infinity
\begin{align}
\delta q_{ab}&=\epsilon\{q_{ab},H[N]+D[N^c]\}\approx\epsilon\mathcal{L}_{N\vec{n}+\vec{N}}\,q_{ab} \notag\\
\delta p^{ab}&=\epsilon\{p^{ab},H[N]+D[N^c]\}\approx\epsilon\mathcal{L}_{N\vec{n}+\vec{N}}\,p^{ab}\,, \label{gaugeTransformations}
\end{align}
where $\epsilon$ here represents an infinitesimal parameter. The corresponding Hamiltonian vector fields of these generators are hence null directions of the symplectic structure and hence `proper' gauge transformations in the canonical sense.

For spacetime diffeomorphisms which do not vanish at infinity, and in particular for diffeomorphisms associated with asymptotic Poincar\'e transformations (\ref{PoincareLapseFallOff})-(\ref{PoincareShiftFallOff}), the smeared constraints  $H[M]$ and $D[M^a]$  do not act  as the corresponding canonical generators on phase space. As shown in \cite{ReggeTeitelboim} and \cite{BeigMurchadha} for ADM variables, in \cite{Thiemann} for self-dual variables and as reviewed in Appendix \ref{a::AshtekarVariables} for Ashtekar-Barbero variables, a more general fall-off behavior for the smearing fields $M$ and $M^a$, corresponding to asymptotic Killing vector fields (\ref{PoincareLapseFallOff})-(\ref{PoincareShiftFallOff}), makes $H[M]$ and $D[M^a]$ not differentiable or ill-defined. More precisely, a leading constant fall-off for $M$ and $M^a$, amounting to translations, certainly results in finite expressions for $H[M]$ and $D[M^a]$ but these expressions are no longer differentiable, so that no Hamiltonian vector fields can be defined from them. For a linear fall-off behavior for $M$ and $M^a$, that is boosts and rotations, expressions $H[M]$ and $D[M^a]$ are actually divergent. Since the smeared constraints  $H[M]$ and $D[M^a]$ cannot be the generators of Poincar\'e transformations, these asymptotic symmetries are called \emph{improper} gauge transformations which means they are not gauge transformations in the canonical sense, i.e. the corresponding Hamiltonian flows are not null directions of the symplectic structure.

An strategy to find  the correct generators satisfying 
\begin{align}
\delta q_{ab}&=\epsilon\{q_{ab},\mathbf{G_H}[M]+\mathbf{G_D}[M^c]\}\approx\epsilon\mathcal{L}_{M\vec{n}+\vec{M}}\,q_{ab} \notag\\
\delta p^{ab}&=\epsilon\{p^{ab},\mathbf{G_H}[M]+\mathbf{G_D}[M^c]\}\approx\epsilon\mathcal{L}_{M\vec{n}+\vec{M}}\,p^{ab}\,. \label{KillingTransformations}
\end{align}  
for the more general fall-off behavior (\ref{PoincareLapseFallOff}) and (\ref{PoincareShiftFallOff}) for $M$ and $M^a$,
 is to construct new well defined (finite) and differentiable expressions starting from $H[M]$ and $D[M^a]$ and adding counter terms that exactly cancel the `problematic' terms that make them not differentiable or that diverge for the more general asymptotic conditions for $M$ and $M^a$. This prescription results in the correct generators and uniquely  determines the conserved charges. The procedure -sometimes referred as the method of counter-terms- was first used by Regge and Teitelboim in \cite{ReggeTeitelboim} and corrected in \cite{BeigMurchadha} to construct the generators in ADM variables. Later in \cite{Thiemann}, Thiemann used it to find generators in self-dual variables starting from the already re-normalized ADM generators. This is the construction we generalize in appendix \ref{a::AshtekarVariables} for general real (or complex) connection variables.

The constraint (\ref{diffeomorphismConstraint1}) could also be taken as the canonical generator of infinitesimal transformations corresponding to the action on the canonical fields of pure spatial diffeomorphisms  which become  supertranslations or the identity at infinity. This is to say
\[
\delta A_a^i=\{A_a^i,D[N^c]\}=\bar{\mathcal{L}}_{\vec{N}}A_a^i \qquad \text{and} \quad
\delta E^a_i=\{E^a_i,D[N^c]\}=\bar{\mathcal{L}}_{\vec{N}}E^a_i\,.
\]
As can easily be verified. Here the `extended' Lie derivative $\bar{\mathcal{L}}$ is as defined in (\ref{barLieE}).
Note however that (\ref{diffeomorphismConstraint1}) is not quite the same as
\[
G_\text{Spatial Diff}[N^a]:=\frac{1}{\kappa\gamma}\int_\Sigma \md^3x\,N^a\left(2E^b_i\partial_{[a}A_{b]}^i-A_a^i\partial_bE^b_i\right)\,.
\]
This functional is also well defined and differentiable for $N^a$ of the form (\ref{properGauge}) so that\footnote{The two generators differ by the term $\epsilon^i\,_{jk}N^a\bar{\Gamma}_a^jA_b^kE^b_i$, with $\bar{\Gamma}_a^j$ the (vector potential of the) spin connection compatible with the asymptotic triad $\bar{E}^a_i$ ($G_\text{Spatial Diff}[N^a]$ and $D[N^a]$ only match when $\bar{E}^a_i=\delta^a_i$ is the `constant' triad, so that  $\bar{\Gamma}_a^j=0$).}
\[
\delta A_a^i=\{A_a^i,G_\text{Spatial Diff}[N^c]\}=\mathcal{L}_{\vec{N}}A_a^i \qquad \text{and} \quad
\delta E^a_i=\{E^a_i,G_\text{Spatial Diff}[N^c]\}=\mathcal{L}_{\vec{N}}E^a_i\,.
\]
Again, neither $D[M^a]$ nor $G_\text{Spatial Diff}[M^a]$ is finite and differentiable for the more general fall-off (\ref{PoincareShiftFallOff}) corresponding to asymptotic Poincar\'e transformations.

In contrast to the ADM phase space, on the extended phase space $\Gamma_{(A,E)}$ there cannot be a canonical generator corresponding to the action of spatial translations and rotations on the canonical variables, specifically on $E^a_i$. Any such generator  $G_\text{Trans/Rot}[M^a]$ would necessarily satisfy
\[
\frac{\delta G_\text{Trans/Rot}}{\delta A_a^i}=-\mathcal{L}_{\vec{M}}E^a_i
\]
but, since $\mathcal{L}_{\vec{M}}E^a_i$ is $\Ord{r}{-1}{odd}$ for `constant' vector field $M^a$ corresponding to pure translations or $\Ord{r}{0}{even}$ for rotations, this
contradicts the consistency condition stated before that any functional derivative with respect to $A_a^i$ must be at most of leading order $\Ord{r}{-1}{even}$.
Roughly speaking, a translation or rotation `takes us out of the extended phase space'.
This of course does not mean that the Poincar\'e generator $\mathbf{G_D}[M^a]$ is not well defined on the extended phase space because $\mathbf{G_D}[M^a]$ only has to satisfy (\ref{KillingTransformations}), that is, it only has to generate infinitesimal translations and rotations on the composite variables $q_{ab}$ and $p^{ab}$.

The canonical Poincar\'e generator corresponding to spatial translations and rotations and extending the renormalized ADM generator (\ref{ADMDgenerator}) is
 \begin{align}
     \mathbf{G_D}[M^a]=\frac{1}{\kappa\gamma}&\int_\Sigma \md^3x\left[M^a\left(F^i_{ab}E^b_i-A^i_a\mathcal{G}_i\right)
     +\frac{\epsilon_{abc}}{2 \det E}E^a_iE^b_j\mathcal{G}_i\bar{\mathcal{L}}_{\bar{M}}E^c_j\right]   \notag  \\
     &+\frac{2}{\kappa\gamma}\int_{\partial\Sigma}dS_a(A^i_b-\Gamma^i_b)M^{[b}E^{a]}_i\,. \label{AshtekarDgenerator}
\end{align}
As reviewed in appendix \ref{a::AshtekarVariables}, the surface term above is an extension of the surface term in (\ref{ADMDgenerator}) so it matches the ADM momentum and angular momentum on the constraint surface determined by $\mathcal{C}\approx 0$, $\mathcal{C}_a\approx 0$ and $\mathcal{G}_i\approx 0$. For proper gauge transformations (\ref{properGauge}) this surface term vanishes and $\mathbf{G_D}[N^a]$ coincides with the vector constraint $D[N^a]$ on the Gauss constraint surface $\mathcal{G}_i\approx 0$ (the last bulk term in (\ref{AshtekarDgenerator}) proportional to $\bar{\mathcal{L}}_{\bar{M}}E^c_j$ is necessary to cancel the divergences of $D[M^a]$: Unlike the ADM case, here these divergences are not expressed as a single surface term).  The Hamiltonian flow of $\mathbf{G_D}[M^a]$ is given in (\ref{PDgeneratorConnVariation}).

The canonical Poincar\'e generator corresponding to time translations and boosts and extending the  ADM generator (\ref{ADMHgenerator}) is
\begin{align}
     \mathbf{G_H}[M]=\frac{1}{2\kappa}\int_\Sigma \md^3x\frac{M}{\sqrt{\det E}}\left[E^a_iE^b_j\left(\epsilon^{ij}\,_kF^k_{ab}-2(1+\gamma^2)K^i_{[a}K^j_{b]}\right)+2\bar{\partial}_a(E^a_i\mathcal{G}_i)\right] \notag \\
     -\frac{1}{\kappa}\int_{\partial\Sigma}dS_a\left[\frac{M}{\sqrt{\det E}}E^a_i\bar{\partial}_b E^b_i-\bar{\partial}_b\left(\frac{M}{\sqrt{\det E}}\right)E^b_i(E^a_i-\bar{E}^a_i)\right]\,.
\end{align}
For proper gauge transformations (\ref{properGauge}) the surface terms vanish and $\mathbf{G_H}[N]$ coincides with the ADM Hamiltonian constraint $H[N]$ on the Gauss constraint surface $\mathcal{G}_i\approx 0$.
For the general fall-off (\ref{PoincareLapseFallOff}) the surface terms above are also extensions of the corresponding ADM surface terms in (\ref{ADMHgenerator}). However, we point out here that the proof given in \cite{Thiemann} only works if we restrict to co-triads admitting an asymptotic expansion to order two
\begin{equation} \label{cotriadExpansion2}
e^i_a=\bar{e}^i_a+\frac{f_a^i}{r}+\frac{\leftidx{^2}{f}{_a^i}}{r^2}+\ord{r}{-2}
\end{equation}
with $\leftidx{^2}{f}{_a^i}$ of even parity. As revised in appendix \ref{a::AshtekarVariables}, following the arguments in \cite{Thiemann}, one can see that for constant $M=1$, the surface term proportional to $M$ 
 matches the ADM energy on the constraint surface determined by $\mathcal{C}\approx 0$, $\mathcal{C}_a\approx 0$ and $\mathcal{G}_i\approx 0$, even for  the more general asymptotic expansion to order one for the co-triad  (or equivalently (\ref{Efalloff1})). However, for linear $M=\alpha_ax^a$, the surface terms equal the ADM `center of mass'  or relativistic angular momentum only if we restrict to (\ref{cotriadExpansion2})\footnote{This would seem consistent with expansion(\ref{tetradExpansion3}) and covariant results \cite{AshtekarES} where well defined relativistic angular momentum requires an asymptotic expansion to order two for the co-tetrad.}.  
 
From expressions like (\ref{PDgeneratorConnVariation}) and (\ref{HConnVariation})  it is not possible to guarantee the parity of the order $r^{-2}$ term in (\ref{cotriadExpansion2}) or in the corresponding densitized triad expansion will be preserved by the flow of the generators. Furthermore, (\ref{cotriadExpansion2}) implies an expansion to order two for the spatial metric and requires tuning gauge transformations (\ref{properGauge}) accordingly (see \cite{ourReview}). We will hence give an alternate argument based on the Poisson algebra satisfied by the generators and showing the surface terms do indeed match the Poincar\'e generators on-shell.

Also, from expressions (\ref{PDgeneratorConnVariation}) and (\ref{HConnVariation}) it is not immediately obvious that the flow of $\mathbf{G_D}[M^a]$ and $\mathbf{G_H}[M]$ actually reproduces (\ref{KillingTransformations}).  Even for the flow of $\mathbf{G_D}[M^a]$, which on the Gauss constraint surface may be written as
\begin{align}
\delta \mathbf{G_D}[M^a]
\approx 2\gamma^{-1}\int&\md^3x\, \left(-\bar{\mathcal{L}}_{\vec{M}}E^a_i + \Lambda_j[\vec{M}]\,\epsilon_{ji}\,^kE^a_k\right)\delta A_a^i \notag\\
+&\bigg(\bar{\mathcal{L}}_{\vec{M}}A_a^i+\left(-\bar{\partial}_a \Lambda_i[\vec{M}]\,+\Lambda_j[\vec{M}]\,\epsilon_{jk}\,^iA_a^k\right) 
\bigg)\delta E^a_i \,, \label{onShellGDvariation}
\end{align}
an infinitesimal diffeomorphism transformation plus an infinitesimal  gauge rotation on the canonical variables, this interpretation is only formal since the `phase space dependent' rotation parameter (\ref{gaugeAngle1}) is order 1, so the infinitesimal gauge transformation alone would rotate the asymptotic frame and take us out of the phase space. Nevertheless, using (\ref{onShellGDvariation}), it is relatively straight forward to check by direct calculation that $\mathbf{G_D}[M^a]$ generates rotations and translations on the composite ADM variables. 
For the other generator, the key is to first observe that on the Gauss constraint surface $\mathbf{G_H}[M]$ is a $SU(2)$-gauge invariant extension  of the corresponding generator in ADM variables and hence it is constant on the gauge orbits of the Gauss constraint. Furthermore, from
\[
\{\mathcal{G}_i,\mathbf{G_H}[M]\}\approx 0,
\]
it also follows that the flow of $\mathbf{G_H}[M]$ is tangent to the Gauss constraint so it has to coincide with the flow of the corresponding ADM generator for $SU(2)$-gauge invariant functions. 

Observations above  also show that on the Gauss constraint surface the Poisson algebra of the extended generators reproduces the algebra of the corresponding ADM generators computed in \cite{BeigMurchadha}:
\begin{align}
\{\mathbf{G_D}[M^a],\mathbf{G_D}[\tilde{M}^b]\}\approx\mathbf{G_D}[[M^a,\tilde{M}^b]], &\quad \{\mathbf{G_H}[M],\mathbf{G_D}[M^a]\}\approx-\mathbf{G_H}[M^a\partial_aM], \notag \\
\{\mathbf{G_H}[M],\mathbf{G_H}[\tilde{M}]\}\approx&\mathbf{G_D}[q^{ab}(M\partial_a\tilde{M}-\tilde{M}\partial_aM)]. \label{GeneratorAlgebra}
\end{align} 
The first two follow because $\mathbf{G_D}[M^a]$ generates spatial diffeomorphisms and the last one because the flow of $\mathbf{G_H}[M]$ on itself has to match the ADM flow.

Algebra (\ref{GeneratorAlgebra}) is the same \emph{Hypersurface deformation algebra}  satisfied by the constraints.  Linearizing and specializing the smearing fields to each of the ten linearly independent Killing vector fields corresponding to translations, rotations and boosts,  it gives a Poisson representation for the Poincar\'e Lie algebra. From the Hamiltonian point of view, (\ref{GeneratorAlgebra}) is the key to showing all the basic properties of the generators, including the fact that their corresponding surface terms are the conserved charges and that they are well defined despite the residual odd supertranslation ambiguity. 

In particular the conserved charges are defined  for each of the ten combinations of smearing fields corresponding to the ten linearly independent Killing vector fields (\ref{PoincareLapseFallOff})-(\ref{PoincareShiftFallOff}):
\begin{align}
\mathbf{E}_{\text{ADM}}&:\approx \mathbf{G_H}[M=1]\,, \notag\\
\mathbf{P}^b_{\text{ADM}}&:\approx \mathbf{G_D}[M^a=\delta^a_b]\,, \notag\\
\mathbf{J}^c_{\text{ADM}}&:\approx \mathbf{G_D}[M^a=\epsilon^a_{bc}x^b]\,, \notag\\
\mathbf{K}^b_{\text{ADM}}&:\approx \mathbf{G_H}[M=\delta^b_cx^c]+\mathbf{G_D}[M^a=-\delta^a_bt]\,, \label{PoincareGenerators}
\end{align}
where $:\approx$ means equality with the fields  satisfying all the constraints. On the constraint surface we have $H[M]\approx 0$ and $D[M^a]\approx 0$ so the values for the generators of Poincar\'e transformations are given precisely by the surface integrals. Just as $H[N]$ and $D[N^a]$ play a dual role as canonical generators of gauge transformations and  as constraints,  $\mathbf{G_H}[M]$ and $\mathbf{G_D}[M^a]$ play a double role as canonical generators of Poincar\'e transformations on phase space and as conserved quantities of the corresponding spacetime.
Indeed, due to the anti-symmetry of the Poisson bracket relations (\ref{GeneratorAlgebra}) one can see that the surface integrals (\ref{PoincareGenerators}) are conserved under canonical transformations of the corresponding Poincar\'e generators.
Explicitly, using the short hand notation for the independent Poincar\'e generators:
\begin{align*}
\mathbf{G_T}:=\mathbf{G_H}[M=1],  \quad  \mathbf{G_{Tx^b}}:=\mathbf{G_D}&[M^a=\delta^a_b], \quad \mathbf{G_{Rx^c}}:=\mathbf{G_D}[M^a=\epsilon^a_{bc}x^b], \\
\mathbf{G_{Bx^b}}:=\mathbf{G_H}[M=\delta^b_cx^c]&+\mathbf{G_D}[M^a=-\delta^a_bt],
\end{align*}
algebra (\ref{GeneratorAlgebra}) gives
\begin{align*}
\{\mathbf{E}_{\text{ADM}}, \mathbf{G_T}\}\approx\{\mathbf{G_T},\mathbf{G_T}\}=0\,, \quad
\{\mathbf{P}^b_{\text{ADM}},\mathbf{G_{Tx^b}}\} \approx \{\mathbf{G_{Tx^b}},\mathbf{G_{Tx^b}}\}=0\,, \notag\\
\{\mathbf{J}^c_{\text{ADM}},\mathbf{G_{Rx^c}}\}\approx\{\mathbf{G_{Rx^c}},\mathbf{G_{Rx^c}}\}=0\,, \quad
\{\mathbf{K}^b_{\text{ADM}},\mathbf{G_{Bx^b}}\}\approx\{\mathbf{G_{Bx^b}},\mathbf{G_{Bx^b}}\}=0\,, 
\end{align*}
In the spacetime picture this means the surface integrals (\ref{PoincareGenerators}) are indeed the conserved quantities associated to the Poincar\'e symmetries of the asymptotically flat spacetime\footnote{As shown in \cite{AshtekarMagnon}, the ADM charges coincide with the Ashtekar Hansen charges \cite{AshtekarHansen}, which are also defined as two dimensional surface integrals but on a cross section of $\mathcal{H}_1$, in terms of the direction dependent limits of the electric and magnetic parts of the Weyl tensor at spatial infinity $\iota^0$. These charges may also be seen as the components of a \emph{four- momentum} vector (and a \emph{relativistic angular momentum} tensor) on the tangent space at $\iota^0$.}.

Lastly, we comment that the action (\ref{HolstHamiltonian}) is well defined and gives as its critical points, solutions to Einstein's equations with lapse and shift with asymptotic fall-off of at most constant for lapse and at most containing odd supertranslations for the shift. However, the most general Hamiltonian flow on phase space, giving Einstein's equations and consistent or preserving the parity conditions is generated by $\mathbf{G_H}[M]$ and $\mathbf{G_D}[M^a]$. A Hamiltonian action using $\mathbf{G_H}[M]$ and $\mathbf{G_D}[M^a]$ nevertheless does not render a well defined variational principle since, as shown in Appendix \ref{a::AshtekarVariables}, they are not differentiable with respect to $M$ and $M^a$ for the more general linear fall-off conditions.


\section{Summary and discussion}
\label{sec:6}

In this work we have considered the issue of having well defined gravitational actions for which a 3+1 
splitting yields a consistent Hamiltonian formulation, for asymptotically flat configurations. In 
particular, we have
deepened the study of the first order action (\ref{Holst1}) for general relativity. This action is well adapted to configuration spaces containing solutions which are asymptotically flat at spatial infinity.

Let us now summarize our findings.
We have considered a generalized variational principle where variations at spatial infinity are not fixed but instead they are consistent with an interpretation as generalized tangent vectors on the phase space admitting such asymptotically flat solutions. 
While this generalization may not be strictly necessary from a purely classical point of view, with an eye towards quantization such condition is certainly required, particularly for the semiclassical approximation of the path integral.

Once again the richness and complexity of structure at infinity pervades and is reflected in the different behaviour of the action depending on how one approaches spatial infinity. 
In section \ref{sec:3}, we have shown that for cylindrical cut-offs of spacetime (with cylindrical slabs) --which is the geometric setting most adapted to a 3+1 decomposition-- the action gives a well defined variational principle as we have required in section \ref{s::preliminaries}, i.e. the limit expression (\ref{Holst1}) is finite (even off-shell) and differentiable for the generic fall-off conditions (\ref{tetradExpansion2}) and (\ref{connectionExpansion2}) of the basic fields. These conditions are those directly derived from or compatible with the more geometric and coordinate independent definition of asymptotic flatness given by Ashtekar and Hansen \cite{AshtekarHansen}. In contrast, a limiting procedure or approach to spatial infinity using hyperbolic cut-offs of spacetime --the geometric setting most adapted to covariant analyses--  requires specialization to a metric of the form (\ref{strongAFmetric}) to guarantee differentiability of the action. This restriction of the configuration space is the so called Ashtekar-Hansen condition which in addition reduces the asymptotic symmetry group by eliminating supertranslations.

In section \ref{sec:4}, we have generalized the well-known result of \cite{Holst} to configuration spaces containing asymptotically flat solutions. We have performed the 3+1 decomposition of the action using a 3+1 foliation of spacetime asymptotically orthogonal to our cylindrical cut-offs and, as a consequence, with Cauchy surfaces giving cylindrical temporal cut-offs.  We have shown that the 3+1 decomposition plus time-gauge fixing ({\ref{timeGauge}) of action (\ref{Holst1}) renders the finite and differentiable Hamiltonian action (\ref{HolstHamiltonian}) in Ashtekar-Barbero variables which are compatible with asymptotically flat solutions. The boundary term in the Hamiltonian in (\ref{HolstHamiltonian}) yielding the ADM energy and momentum expressions.

While Hamiltonian action (\ref{HolstHamiltonian}) --being a re-writing of (\ref{Holst1}) with a partial gauge fixing-- is necessarily finite and differentiable for the asymptotic behavior (\ref{cartesianAFCondition2}) directly derived from the more geometric definition, from a purely `frozen time' or 3+1 perspective, taking it as a starting point, it is not obvious at all how to show its finiteness and differentiability for such general fall-off conditions. This is related --but strictly different-- to the problem of properly defining the corresponding phase space along with geometric structures on it:  a symplectic structure --or more generally a Poisson bracket structure-- and Hamiltonian vector fields generating `evolution', that is gauge and asymptotic symmetries.

There is no known clean way to construct the phase space of general relativity accommodating  the full class of asymptotically flat solutions  consistent with (\ref{cartesianAFCondition2})\footnote{For one possibility see \cite{CompereDehouck}.}. Additional restrictions need to be imposed. For the covariant phase space formulation these are the Ashtekar-Hansen condition (\ref{strongAFmetric}), which eliminates supertranslations, and the Beig-Ashtekar parity condition (\ref{BeigAshtekarCond}), which in turn rules out logarithmic translations. For the 3+1 Hamiltonian formulation, one has to restrict to spatial metrics satisfying the Regge-Teitelboim parity conditions (\ref{qpParity}).  
With the advantage of hindsight, the corresponding parity conditions for Ashtekar-Barbero variables may be directly inferred from some of the conditions (\ref{covariantTetradCoef}) and (\ref{covariantConnectionCoef}) on the spacetime connection and tetrad fields, derived from the requirement for a well defined covariant phase space formulation. Indeed, it is known that the ADM phase space contains such solutions satisfying the Beig-Ashtekar and the Ashtekar-Hansen conditions. As we have seen in section \ref{s::HamiltonianFormulation} and appendix \ref{a::fromHyper2CylExpansion}, starting from covariant or hyperbolic fall-off conditions for the spacetime fields and deriving cylindrical expansions from them, one is able to infer, at least partially, the parity conditions for the canonical Hamiltonian framework. The only remaining ambiguity is resolved from the requirement of a well defined symplectic structure.

Additionally, in section \ref{s::HamiltonianFormulation} we have displayed the Poincar\'e generators in Ashtekar-Barbero variables directly generalizing those constructed in \cite{Thiemann} in self-dual variables. We have included a complete re-derivation in appendix \ref{a::AshtekarVariables}. We remark here the key role of the hypersurface deformation algebra (\ref{GeneratorAlgebra}) that the generators satisfy in order to ensure the surface terms actually match the ADM charges. A point not sufficiently stressed or overlooked in \cite{Thiemann}. In particular, a direct proof showing  equality for the angular momentum requires a refined expansion  (\ref{cotriadExpansion2}) for the triad  (and correspondingly and expansion (\ref{restrictedAExpansion}) for the Ashtekar connection). This would seem consistent with expansions (\ref{tetradExpansion3}) and (\ref{connectionExpansion3}) used in covariant treatments and would also allow a cotangent bundle interpretation of the canonical phase space. However, this appears to overly restrict not only the asymptotic form of the spatial metric but also the asymptotic symmetries. The parity of the next leading order terms in the asymptotic expansion of the fields does not seem to be preserved under the Hamiltonian flow of the standard generators of gauge and symmetries.

Finally, we want to comment on two possible avenues for future research. The first one has to do with the lack of a generator of `infinitesimal spatial diffeomorphisms' of the canonical fields on the extended phase space. This is linked to the standard but naive definition of Lie derivatives on the associated vector bundles. It may be instructive to try to generalize this notion to better accommodate the principal and associated vector bundle structures associated with the Ashtekar-Barbero fields and to construct generators of such gauge symmetries.
The other point concerns the full decomposition and Dirac analysis prior to gauge fixing of the action (\ref{Holst1}). Such analysis was carried out in \cite{BarroseSa} for compact Cauchy slices and it would be interesting to see how it generalizes in the case of asymptotically flat spacetimes. We shall leave those investigations for future publications.




\begin{acknowledgments}
We would like to thank I. Rubalcava-Garc\'ia, T. Vukasinac and E. Wilson-Ewing for valuable comments.
This work was in part supported by DGAPA-UNAM IN103610 grant, by CONACyT 0177840 
and 0232902 grants, by the PASPA-DGAPA program, by NSF
PHY-1403943 and PHY-1205388 grants, and by the Eberly Research Funds of Penn State.
JDR thanks the Institute for Gravitation and the Cosmos for their hospitality during partial completion of this work.

\end{acknowledgments}


\appendix


\section{Comparison of hyperbolic and cylindrical asymptotic expansions} \label{a::fromHyper2CylExpansion}  
One can show that an expansion of the form (\ref{rhoExpansion}) implies the existance and coincides asymptotically with an expansion of the form (\ref{rExpansion}) for cylindrical temporal cut-offs in the limit where ADM time $t_0\leq t\leq t_1$ is kept 
fixed.  From (\ref{hyperCylRelations}) the limit $\rho\to\infty$ implies $r\to\infty$, and $t/r\to 0$ or equivalently $\chi\to 0$ and $\rho\sim r$.

Assuming a Tylor series expansion for $f(\Phi)=f(\chi,\vartheta,\varphi)$ around $\chi=0$:
\[
f(\chi,\vartheta,\varphi)\approx f(0,\vartheta,\varphi)+\left(\partial_\chi f|_{\chi=0}\right)\chi+\frac{1}{2}\left(\partial^2_\chi f|_{\chi=0}\right)\chi^2+\cdots
\]
and expanding $\chi(t/r)=\tanh^{-1} t/r$ so that $\chi=t/r+\ord{r}{-2}$:
\begin{equation}
f(\chi(t/r),\vartheta,\varphi)= f(0,\vartheta,\varphi)+\left(\partial_\chi f|_{\chi=0}\right)\frac{t}{r}+\frac{1}{2}\left(\partial^2_\chi f|_{\chi=0}\right)\left(\frac{t}{r}\right)^2+\ord{r}{-2}
\end{equation}
Similarly
\[
\frac{1}{\rho}=\frac{1}{r}\frac{1}{\;\left(1-(t/r)^2\right)^{1/2}}=\frac{1}{r}\left(1+\frac{1}{2}\left(\frac{t}{r}\right)^2+\ord{r}{-2}\right)
\]
and
\[
\frac{1}{\rho^2}=\frac{1}{r^2}\left(1+\left(\frac{t}{r}\right)^2+\ord{r}{-2}\right)
\]
so
\begin{align}
\leftidx{^0_r}{f}(t,\vartheta,\varphi)&=\leftidx{^0_\rho}{f}(\chi,\vartheta,\varphi)|_{\chi=0}\,, \label{rho2r0}\\
\leftidx{^1_r}{f}(t,\vartheta,\varphi))&=\leftidx{^1_\rho}{f}(\chi,\vartheta,\varphi)|_{\chi=0}+\left(\partial_\chi\,\leftidx{^0_\rho}{f}|_{\chi=0}\right)t\,, \label{rho2r1}\\
\leftidx{^2_r}{f}(t,\vartheta,\varphi))&=\leftidx{^2_\rho}{f}(\chi,\vartheta,\varphi)|_{\chi=0}+\left(\partial_\chi\,\leftidx{^1_\rho}{f}|_{\chi=0}\right)t+\frac{1}{2}\left(\partial^2_\chi\,\leftidx{^0_\rho}{f}|_{\chi=0}\right)t^2\,,
\end{align}
where we have added a left subindex $r$ or $\rho$  to expansion functions to stress that functions on the left hand side correspond to a cylindrical  expansion  (\ref{rExpansion}) while functions on the right correspond to a hyperbolic expansion (\ref{rhoExpansion}).  We may compute similar expansions for $\rho^{-m}$ and derive the corresponding formulas for $\leftidx{^m_r}{f}(t,\vartheta,\varphi)$ in an asymptotic expansion to order $n$. 

Using formulas above one can infer from $\rho$-expansions the form of asymptotic $r$-expansions for the co-tetrad and Lorentz connection and subsequently for the canonical Ashtekar-Barbero connection and densitized triad. Indeed, $\rho$-expansion (\ref{tetradExpansion1}) for the tetrad implies
\[
e_\mu^I=\leftidx{^0_r}{e}{_\mu^I}(x^a/r)+\frac{\leftidx{^1_r}{e}{_\mu^I}(t,x^a/r)}{r}+\ord{r}{-1}\,.
\]
with $\leftidx{^0_r}{e}{_\mu^I}(x^a/r)=\leftidx{^0_\rho}{e}{_\mu^I}(\Phi)|_{\chi=0}$ and for the particular case (\ref{covariantTetradCoef}): 
\[
\leftidx{^1_\rho}{e}{_\mu^I}(\Phi)=\,\leftidx{^0_\rho}{e}{^\nu_J}(\Phi)\,\sigma(\Phi)(2\rho_\mu\rho_\nu-\eta_{\mu\nu})\eta^{IJ}
\]
using
\begin{align*}
\frac{t}{\rho}=\sinh\chi, \qquad
\frac{x^1}{\rho}=\cosh\chi\sin\vartheta\cos\varphi, \qquad
\frac{x^2}{\rho}=\cosh\chi\sin\vartheta\sin\varphi, \qquad
\frac{x^3}{\rho}=\cosh\chi\cos\vartheta, 
\end{align*}
and hence
\[
\rho^0|_{\chi=0}=\left.\frac{t}{\rho}\right|_{\chi=0}=0, \qquad \rho^a|_{\chi=0}=\left.\frac{x^a}{\rho}\right|_{\chi=0}=\frac{x^a}{r}
\]
one has
\[
\leftidx{^1_r}{e}{_\mu^I}(t,x^a/r)=\left.\leftidx{^0_\rho}{e}{^\nu_J}(\Phi)\,\sigma(\Phi)\right|_{\chi=0}\left(2\eta_{\mu c}\,\eta_{\nu d}\,\frac{x^c}{r}\frac{x^d}{r}-\eta_{\mu\nu}\right)\eta^{IJ}+\left.\partial_\chi\, \leftidx{^0_\rho}{e}{^\nu_J}(\Phi)\right|_{\chi=0}\,t\,.
\]
At $t=0$ or in the limit $t/r\to\infty$ the last term drops or is negligible and one has in particular
\begin{equation}  \label{cotriadParityForm}
\leftidx{^1_r}{e}{_a^i}(x^c/r)=\left.\leftidx{^0_\rho}{e}{^b_j}(\Phi)\,\sigma(\Phi)\right|_{\chi=0}\left(2\delta_{a c}\,\delta_{b d}\,\frac{x^c}{r}\frac{x^d}{r}-\delta_{ab}\right)\delta^{ij}\,.
\end{equation}

From the condition $e_\mu^Ie^\nu_I=\delta_\mu^\nu$ one infers an asymptotic expansion to order one (or two) for the tetrad:
\[
\left(\leftidx{^0_r}{e}{_\mu^I}+\frac{\leftidx{^1_r}{e}{_\mu^I}}{r}+\frac{\leftidx{^2_r}{e}{_\mu^I}}{r^2}+\ord{r}{-2}\right)
\left(\leftidx{^0_r}{e}{^\nu_I}+\frac{\leftidx{^1_r}{e}{^\nu_I}}{r}+\frac{\leftidx{^2_r}{e}{^\nu_I}}{r^2}+\ord{r}{-2}\right)=\delta_\mu^\nu
\]
with 
\[
\leftidx{^0_r}{e}{_\mu^I}\,\leftidx{^0_r}{e}{^\nu_I}=\delta_\mu^\nu\,, \qquad  
\leftidx{^1_r}{e}{^\nu_I}=-\leftidx{^0_r}{e}{^\mu_I}\,\leftidx{^0_r}{e}{^\nu_J}\,\leftidx{^1_r}{e}{_\mu^J}\,, \qquad
\leftidx{^2_r}{e}{^\nu_I}=-\,\leftidx{^0_r}{e}{^\nu_J}\left(\leftidx{^2_r}{e}{_\mu^J}\,\leftidx{^0_r}{e}{^\mu_I}\,+\leftidx{^1_r}{e}{_\mu^J}\,\leftidx{^1_r}{e}{^\mu_I}  \right)\,.
\]
In the time gauge (with adapted coordinates) $e^t_i=e_a^0=0$ and $e^a_ie^i_b=\delta^a_b$, so formulas above become conditions for the triad and co-triad:
\[
\bar{e}{_a^i}\,\bar{e}{^b_i}=\delta_a^b\,, \qquad  
\leftidx{^1}{e}{^b_i}=-\bar{e}{^a_i}\,\bar{e}{^b_j}\,\leftidx{^1}{e}{_a^j}\,, \qquad
\leftidx{^2}{e}{^b_i}=-\,\bar{e}{^b_j}\left(\leftidx{^2}{e}{_a^j}\,\bar{e}{^a_i}\,+\leftidx{^1}{e}{_a^j}\,\leftidx{^1}{e}{^a_i}  \right),
\]
where to avoid cluttering notation, we have dropped the right subindex $r$ and renamed the asymptotic co-triad $\bar{e}_a^i:=\leftidx{^0_r}{e}{_a^i}$ and triad $\bar{e}^b_i:=\leftidx{^0_r}{e}{^b_i}$.
It follows $\bar{e}{_a^i}$ and $\bar{e}{^b_i}$ have the same parity on the sphere and from (\ref{cotriadParityForm}) if $\sigma$ is symmetric then also $\leftidx{^1}{e}{_a^i}$ and $\leftidx{^1}{e}{^b_i}$ have the same parity as the asymptotic co-triad. 
Additionally, from the formula $\det B=\epsilon^{a_1a_2\cdots a_n}B^1\,_{a_1}B^2\,_{a_2}\cdots B^n\,_{a_n}$ for the determinant of a $n\times n$ matrix $B$ one infers an expansion for the determinant of the co-triad $e_a^i$:
\begin{align*}
\sqrt{\det q}=|\det e|=1+\frac{\leftidx{^1}{F}{\!}}{r}+\frac{\leftidx{^2}{F}{\!}}{r^2}+\ord{r}{-2}\,,
\end{align*}
with $\leftidx{^1}{F}{\!}$ and $\leftidx{^2}{F}{}$ having the same parity as $\leftidx{^1}{e}{_a^i}$ and $\leftidx{^2}{e}{_a^i}$ respectively.  Hence the densitized triad (\ref{defE}) is expanded as
\begin{equation}  \label{detailedEfalloff}
E^a_i=\bar{e}^a_i+\frac{\leftidx{^1}{e}{^a_i}+\leftidx{^1}{F}{\,}\bar{e}^a_i}{r}+\frac{\leftidx{^2}{e}{^a_i}+\leftidx{^1}{F}{\,}\leftidx{^1}{e}{^a_i}+\leftidx{^2}{F}{\,}\bar{e}^a_i}{r^2}+\ord{r}{-2}\,,
\end{equation}
and if (\ref{cotriadParityForm}) holds, the order $r^{-1}$ term has to have the same fixed parity as the leading order asymptotic triad.

On the other hand, formula (\ref{covariantConnectionCoef}) for the connection gives 
\begin{align} 
\leftidx{^2_r}{\omega}{_{\mu IJ}}&=2\left.\left((2\rho_\mu\rho_{[I}-\leftidx{^0_\rho}{e}{_{\mu[I}})\leftidx{^0_\rho}{e}{^\nu_{J]}}\rho\,\partial_\nu\sigma-\sigma\,\leftidx{^0_\rho}{e}{_{\mu[I}}\rho_{J]}\right)\right|_{\chi=0} \notag\\
&=2\left[\left(2\eta_{\mu b}\,\delta_{cd}\frac{x^b}{r}\frac{x^c}{r}\,\leftidx{^0_r}{e}{^d_{[I}}-\leftidx{^0_r}{e}{_{\mu[I}}\right)\leftidx{^0_r}{e}{^\nu_{J]}}\rho\,\partial_\nu\sigma|_{\chi=0}-\sigma|_{\chi=0}\,\leftidx{^0_r}{e}{_{\mu[I}}\,\leftidx{^0_r}{e}{^b_{J]}}\delta_{bc}\frac{x^c}{r}\right]\,. \label{connectionParityForm}
\end{align}
By Taylor expanding $\sigma$ one can see $\rho\,\partial_\nu\sigma$ is an anti-symmetric function on the hyperboloid if $\sigma$ is symmetric, so $\rho\,\partial_\nu\sigma|_{\chi=0}$ is an anti-symmetric or odd function on the sphere. It then follows from (\ref{connectionParityForm}) that the leading order term of the spacetime connection $\omega_{\mu IJ}$ and hence of the Ashtekar-Barbero connection (\ref{defA}) has odd parity.
From (\ref{spinConnectionFormula1}) it also follows that if $\leftidx{^3_r}{\omega}{_{\mu IJ}}$ and hence $\leftidx{^3}{A}{_a^i}$ are to have definite parity, this parity must be odd and $\leftidx{^2}{e}{_\mu^I}$ must be even.

\section{3+1 decomposition with gauge fixing}  \label{a::HolstDecomposition}
In this appendix, we present the detailed calculations for the 3+1 decomposition plus time gauge fixing, as first done in \cite{Holst}, for the full Holst action with boundary term (\ref{Holst1}). We include the derivation of the kinetic and constraint terms from the bulk integral in  (\ref{Holst1}). This is for completeness and because there is an additional surface term arising from it. For the bulk terms we follow \cite{BojowaldBook}, using similar notation (but slightly different conventions). 

We expand the first term in (\ref{bulkDecomp1}):
\begin{align}
-2t^\mu n_I&\sqrt{\det q}\,\mathcal{E}^b_J\,\tilde{e}^\nu_b\,\leftidx{^\gamma}{F}{^{IJ}_{\mu\nu}}  \notag\\
&=2 \sqrt{\det q}\,\mathcal{E}^b_j\,t^\mu\tilde{e}^\nu_b\,\leftidx{^\gamma}{F}{^{0j}_{\mu\nu}}  \notag\\
&=2E^b_j\,\leftidx{^\gamma}{F}{^{0j}_{tb}}  \notag\\
&=2E^b_j\left(F_{tb}^{0j}+\frac{1}{2\gamma}\epsilon^{0j}\,_{kl}F_{tb}^{kl}\right) \notag\\
&=2E^b_j\left(2\bar{\partial}_{[t}\omega_{b]}^{0j}+2\omega^0_{[t\,|k|}\omega_{b]}^{kj}+\frac{1}{2\gamma}\epsilon^{0j}\,_{kl}\left(2\bar{\partial}_{[t}\omega_{b]}^{kl}+2\omega^k_{[t\,|K|}\omega_{b]}^{Kl}\right)\right)  \notag\\
&=2E^b_j\left(2\left(\frac{1}{2\gamma}\epsilon^{0j}\,_{kl}\,\bar{\partial}_{[t}\omega_{b]}^{kl}+\bar{\partial}_{[t}\omega_{b]}^{0j}\right)+2\omega^0_{[t\,|k|}\omega_{b]}^{kj}+\frac{1}{\gamma}\epsilon^{0j}\,_{kl}\,\omega^k_{[t\,|K|}\omega_{b]}^{Kl}\right) \notag\\
&=2E^b_j\left(\frac{2}{\gamma}\bar{\partial}_{[t}\left(\frac{1}{2}\epsilon^{0j}\,_{kl}\,\omega_{b]}^{kl}+\gamma\omega_{b]}^{0j}\right)+2\omega^0_{[t\,|k|}\omega_{b]}^{kj}+\frac{1}{\gamma}\epsilon^{0j}\,_{kl}\,\omega^k_{t\,K}\omega_{b}^{Kl}\right) \notag\\
&=2E^b_j\left(\frac{1}{\gamma}\left(\bar{\partial}_tA_b^j-\bar{\partial}_b\Lambda^j\right) 
+2\omega^0_{[t\,|k|}\omega_{b]}^{kj}+\frac{1}{\gamma}\epsilon^{0j}\,_{kl}\,\omega^k_{t\,K}\omega_{b}^{Kl}\right) \notag\\
&=\frac{2}{\gamma}\left(E^b_j\bar{\partial}_tA_b^j+\Lambda^j\bar{\partial}_bE^b_j-\bar{\partial}_b(\Lambda^jE^b_j) \right)
+2E^b_j\left(2\omega^0_{[t\,|k|}\omega_{b]}^{kj}+\frac{1}{\gamma}\epsilon^{0j}\,_{kl}\,\omega^k_{t\,K}\omega_{b}^{Kl}\right)  \label{bulk1Decomp1}
\end{align}
where in the last line one `integrates by parts' the term proportional to $\bar{\partial}_b\Lambda^j$. The first term in the last expression is already the kinetic term while the total divergence will give rise to a surface integral. To see that the remaining terms combine to give the Gauss constraint, it is easier to work backwards:
\begin{align}
\Lambda^j\mathcal{G}_j:&=\Lambda^j\left(\bar{\partial}_bE^b_j+\epsilon_{jm}\,^nA_b^mE^b_n\right) \notag\\
&=\Lambda^j\bar{\partial}_bE^b_j+\epsilon_{jm}\,^n\Lambda^jA_b^mE^b_n \notag\\
&=\Lambda^j\bar{\partial}_bE^b_j+\epsilon_{jm}\,^n\left(\frac{1}{2}\epsilon^{0j}\,_{kl}\,\omega_t^{kl}+\gamma\omega_t^{0j}\right) \left(\frac{1}{2}\epsilon^{0m}\,_{pq}\,\omega_b^{pq}+\gamma\omega_b^{0m}\right)E^b_n \notag\\
&=\Lambda^j\bar{\partial}_bE^b_j+\frac{1}{4}\epsilon^{0j}\,_{kl}\,\epsilon^{0m}\,_{pq}\,\epsilon_{jm}\,^n\,\omega_t^{kl}\omega_b^{pq}E^b_n+\frac{1}{2}\gamma\epsilon^{0j}\,_{kl}\,\epsilon_{jm}\,^n\,\omega_t^{kl}\omega_b^{0m}E^b_n \notag\\
&\,\quad+\frac{1}{2}\gamma\epsilon^{0m}\,_{pq}\,\epsilon_{jm}\,^n\,\omega_t^{0j}\omega_b^{pq}E^b_n+\gamma^2\epsilon_{jm}\,^n\omega_t^{0j}\omega_b^{0m}E^b_n
\end{align}
and use identites for contracted Levi-Civita symbols along with anti-symmetry of $\omega^{ij}$ to simplify each term above
\begin{align*}
\frac{1}{4}\epsilon^{0j}\,_{kl}\,\epsilon^{0m}\,_{pq}\,\epsilon_{jm}\,^n\,\omega_t^{kl}\omega_b^{pq}E^b_n 
&=\frac{1}{4}\epsilon^{jkl}\,\epsilon_{mpq}\,\epsilon_{jm}\,^n\,\omega_t^{kl}\omega_b^{pq}E^b_n   \\
&=\frac{1}{4}(3!\,\delta^{[j}_m\,\delta^{k}_p\,\delta^{l]}_q)\epsilon_{jm}\,^n\,\omega_t^{kl}\omega_b^{pq}E^b_n \\
&=-\epsilon^j\,_{kl}\,\omega_t^{km}\omega_b^{ml}E^b_j \,, \\
&=\epsilon^{0j}\,_{kl}\,\omega_t^{km}\omega_b^{ml}E^b_j 
\end{align*}
\begin{align*}
\frac{1}{2}\gamma\epsilon^{0j}\,_{kl}\,\epsilon_{jm}\,^n\,\omega_t^{kl}\omega_b^{0m}E^b_n
&=-\frac{1}{2}\gamma\epsilon^{jkl}\,\epsilon_{jmn}\,\omega_t^{kl}\omega_b^{0m}E^b_n \\
&=-\frac{1}{2}\gamma(2\delta^{[k}_m\,\delta^{l]}_n\,)\omega_t^{kl}\omega_b^{0m}E^b_n \\
&=-\gamma\omega_t^{kj}\omega_b^{0k}E^b_j
\end{align*}
and
\begin{align*}
\frac{1}{2}\gamma\epsilon^{0m}\,_{pq}\,\epsilon_{jm}\,^n\,\omega_t^{0j}\omega_b^{pq}E^b_n
&=\frac{1}{2}\gamma\epsilon^{mpq}\,\epsilon_{mjn}\,\omega_t^{0j}\omega_b^{pq}E^b_n \\
&=\frac{1}{2}\gamma(2\delta^{[p}_j\,\delta^{q]}_n)\omega_t^{0j}\omega_b^{pq}E^b_n \\
&=\gamma\omega_t^{0k}\omega_b^{kj}E^b_j
\end{align*}
so
\begin{align}
\Lambda^j\mathcal{G}_j&=\Lambda^j\bar{\partial}_bE^b_j
+\left(\epsilon^{0j}\,_{kl}\,\omega_t^{km}\omega_b^{ml}
-\gamma\omega_t^{kj}\omega_b^{0k}
+\gamma\omega_t^{0k}\omega_b^{kj}\right)E^b_j
+\gamma^2\epsilon_{jm}\,^n\omega_t^{0j}\omega_b^{0m}E^b_n \notag\\
&=\Lambda^j\bar{\partial}_bE^b_j
+\gamma\left(2\omega_{[t}^{0k}\omega_{b]}^{kj}+\frac{1}{\gamma}\epsilon^{0j}\,_{kl}\,\omega_t^{km}\omega_b^{ml} \right)E^b_j +\gamma^2\epsilon_{jm}\,^n\omega_t^{0j}\omega_b^{0m}E^b_n  \notag
\end{align}
or equivalently
\begin{align*}
\frac{2}{\gamma}\Lambda^j\bar{\partial}_bE^b_j+2E^b_j\left(2\omega_{[t}^{0k}\omega_{b]}^{kj}+\frac{1}{\gamma}\epsilon^{0j}\,_{kl}\,\omega_t^{km}\omega_b^{ml} \right)
=\frac{2}{\gamma}\Lambda^j\mathcal{G}_j-2\gamma\epsilon_{jm}\,^n\,\omega_t^{0j}K_b^{m}E^b_n \,.
\end{align*}
Going back to (\ref{bulk1Decomp1}) and substituting the latter identity
\begin{align}
-2t^\mu &n_I\sqrt{\det q}\,\mathcal{E}^b_J\,\tilde{e}^\nu_b\,\leftidx{^\gamma}{F}{^{IJ}_{\mu\nu}} \notag\\
&=\frac{2}{\gamma}\left(E^b_j\bar{\partial}_tA_b^j+\Lambda^j\bar{\partial}_bE^b_j-\bar{\partial}_b(\Lambda^jE^b_j) \right) \notag\\
&\,\quad+2E^b_j\left(2\omega^0_{[t\,|k|}\omega_{b]}^{kj}+\frac{1}{\gamma}\epsilon^{0j}\,_{kl}\,\omega^k_{t\,0}\omega_{b}^{0l}
+\frac{1}{\gamma}\epsilon^{0j}\,_{kl}\,\omega^k_{t\,m}\omega_{b}^{ml}\right)  \notag\\
&=\frac{2}{\gamma}\left(E^b_j\bar{\partial}_tA_b^j-\bar{\partial}_b(\Lambda^jE^b_j) \right)
+\frac{2}{\gamma}\Lambda^j\bar{\partial}_bE^b_j+2E^b_j\left(2\omega_{[t}^{0k}\omega_{b]}^{kj}+\frac{1}{\gamma}\epsilon^{0j}\,_{kl}\,\omega_t^{km}\omega_b^{ml} \right) +\frac{2}{\gamma}\epsilon^{j}\,_{kl}\,\omega^{0k}_{t}\omega_{b}^{0l}E^b_j  \notag\\
&=\frac{2}{\gamma}\left(E^b_j\bar{\partial}_tA_b^j+\Lambda^j\mathcal{G}_j-(1+\gamma^2)\epsilon_{jm}\,^n\omega_t^{0j}K_b^{m}E^b_n-\bar{\partial}_b(\Lambda^jE^b_j) \right) \notag\\
&=\frac{2}{\gamma}\left(E^b_j\bar{\partial}_tA_b^j+\Lambda^j\mathcal{G}_j-(1+\gamma^2)\omega_t^{0j}\mathcal{S}_j-\bar{\partial}_b(\Lambda^jE^b_j) \right)  \label{bulk1Decompf}
\end{align}
where one defines
\begin{equation}
\mathcal{S}_j:=\epsilon_{jm}\,^n\,K_b^{m}E^b_n\,.
\end{equation}

Now we work on the second term in (\ref{bulkDecomp1}):
\begin{align}
2N^an_I&\tilde{e}^\mu_a\sqrt{\det q}\,\mathcal{E}^b_J\,\tilde{e}^\nu_b\,\leftidx{^\gamma}{F}{^{IJ}_{\mu\nu}}  \notag\\
&=-2N^aE^b_j\,\leftidx{^\gamma}{F}{^{0j}_{ab}}  \notag\\
&=-2N^aE^b_j\left(F_{ab}^{0j}+\frac{1}{2\gamma}\epsilon^{0j}\,_{kl}\,F_{ab}^{kl}\right) \notag\\
&=-2N^aE^b_j\left(2\bar{\partial}_{[a}\omega_{b]}^{0j}+2\omega_{[a\,|k|}^0\omega_{b]}^{kj}+\frac{1}{2\gamma}\epsilon^{0j}\,_{kl}\,\left(2\bar{\partial}_{[a}\omega_{b]}^{kl}+2\omega_{[a\,|K|}^k\omega_{b]}^{Kl}\right)\right) \notag\\
&=-2N^aE^b_j\left[\frac{2}{\gamma}\bar{\partial}_{[a}\left(\frac{1}{2}\epsilon^{0j}\,_{kl}\,\omega_{b]}^{kl}+\gamma\omega_{b]}^{0j}\right)+2\omega_{[a\,|k|}^0\omega_{b]}^{kj}+\frac{1}{\gamma}\epsilon^{0j}\,_{kl}\,\left(\omega_{a\,0}^k\omega_{b}^{0l}+\omega_{a\,m}^k\omega_{b}^{ml}\right)\right] \notag\\
&=-2N^aE^b_j\left[\frac{2}{\gamma}\bar{\partial}_{[a}A_{b]}^j+2K_{[a}^k(-\epsilon^{kj}\,_l\,\Gamma_{b]}^l)+\frac{1}{\gamma}\epsilon^{0j}\,_{kl}\,\left(K_{a}^kK_{b}^{l}+(-\epsilon^k\,_{mn}\,\Gamma_{a}^n)(-\epsilon^{ml}\,_p\,\Gamma_{b}^p)\right)\right] \notag\\
&=-2N^aE^b_j\left[\frac{2}{\gamma}\bar{\partial}_{[a}A_{b]}^j-2\epsilon^j\,_{lk}\,K_{[a}^k\Gamma_{b]}^l+\frac{1}{\gamma}\epsilon^{0j}\,_{kl}\,\left(K_{a}^kK_{b}^{l}+(2\delta^{[l}_n\,\delta^{p]}_k)\Gamma_{a}^n\Gamma_{b}^p\right)\right] \notag\\
&=-2N^aE^b_j\left[\frac{2}{\gamma}\bar{\partial}_{[a}A_{b]}^j+2\epsilon^j\,_{kl}\,\Gamma_{[a}^kK_{b]}^l+\frac{1}{\gamma}\epsilon^{j}\,_{kl}\,\left(\Gamma_{a}^k\Gamma_{b}^l-K_{a}^kK_{b}^{l}\right)\right]  \notag\\
&=-\frac{2}{\gamma}N^aE^b_j\left[2\bar{\partial}_{[a}A_{b]}^j+2\gamma\epsilon^j\,_{kl}\,\Gamma_{[a}^kK_{b]}^l+\epsilon^{j}\,_{kl}\,\left(\Gamma_{a}^k\Gamma_{b}^l-K_{a}^kK_{b}^{l}\right)\right] 
\end{align}
which upon substituting the second and third terms using
\begin{align*}
\epsilon^j\,_{kl}\,A_a^kA_b^l&=\epsilon^j\,_{kl}\left(\Gamma_a^k+\gamma K_a^k\right)\left(\Gamma_b^l+\gamma K_b^l\right) \\
&=2\gamma\epsilon^j\,_{kl}\,\Gamma_{[a}^kK_{b]}^l+\epsilon^{j}\,_{kl}\,\Gamma_{a}^k\Gamma_{b}^l+\gamma^2\epsilon^{j}\,_{kl}\,K_{a}^kK_{b}^l
\end{align*}
results in
\begin{align}
2N^an_I\tilde{e}^\mu_a\sqrt{\det q}\,\mathcal{E}^b_J\,\tilde{e}^\nu_b\,\leftidx{^\gamma}{F}{^{IJ}_{\mu\nu}}
&=-\frac{2}{\gamma}N^aE^b_j\left(2\bar{\partial}_{[a}A_{b]}^j+\epsilon^j\,_{kl}\,A_a^kA_b^l-(1+\gamma^2)\epsilon^{j}\,_{kl}\,K_{a}^kK_{b}^{l}\right)   \notag\\
&=-\frac{2}{\gamma}N^a\left(F_{ab}^jE^b_j-(1+\gamma^2)K_a^k\mathcal{S}_k\right)\,, \label{bulk2Decompf}
\end{align}
with
\begin{equation}
F_{ab}^i:=2\bar{\partial}_{[a}A_{b]}^i+\epsilon^i\,_{jk}\,A_a^jA_b^k\,.
\end{equation}

For the third and last bulk term we have:
\begin{align}
N\mathcal{E}^a_I\tilde{e}^\mu_a&\sqrt{\det q}\,\mathcal{E}^b_J\,\tilde{e}^\nu_b\,\leftidx{^\gamma}{F}{^{IJ}_{\mu\nu}}  \notag\\
&=\frac{N}{\sqrt{\det q}}E^a_iE^b_j\,\leftidx{^\gamma}{F}{_{ab}^{ij}} \notag\\
&=\frac{N}{\sqrt{\det q}}E^a_iE^b_j\left(F_{ab}^{ij}+\frac{1}{\gamma}\epsilon^{ij}\,_{k0}F_{ab}^{k0}\right) \notag\\
&=\frac{N}{\sqrt{\det q}}E^a_iE^b_j\left(2\bar{\partial}_{[a}\omega_{b]}^{ij}+2\omega_{[a\,0}^i\,\omega_{b]}^{0j}+2\omega_{[a\,|k|}^i\omega_{b]}^{kj}-\frac{1}{\gamma}\epsilon^{ij}\,_k\left(2\bar{\partial}_{[a}\omega_{b]}^{k0}+2\omega_{[a\,|l|}^k\omega_{b]}^{l0}\right)\right) \notag\\
&=\frac{N}{\sqrt{\det q}}E^a_iE^b_j\bigg(-2\epsilon^{ij}\,_k\,\bar{\partial}_{[a}\Gamma_{b]}^k+2K_{[a}^i\,K_{b]}^j+2\epsilon^i\,_{km}\,\epsilon^{kj}\,_n\Gamma_{[a}^m\,\Gamma_{b]}^n \notag\\
&\qquad\qquad\qquad\qquad-\frac{1}{\gamma}\epsilon^{ij}\,_k\left(-2\bar{\partial}_{[a}K_{b]}^{k}+2\epsilon^k\,_{lm}\,\Gamma_{[a}^mK_{b]}^{l}\right)\bigg) \notag\\
&=\frac{N}{\sqrt{\det q}}E^a_iE^b_j\bigg(-2\epsilon^{ij}\,_k\,\bar{\partial}_{[a}\Gamma_{b]}^k+2K_{[a}^i\,K_{b]}^j-2\Gamma_{[a}^i\,\Gamma_{b]}^j 
+\frac{1}{\gamma}\epsilon^{ij}\,_k\left(2\bar{\partial}_{[a}K_{b]}^{k}+2\epsilon^k\,_{lm}\,\Gamma_{[a}^lK_{b]}^{m}\right)\bigg) \notag\\
&=\frac{N}{\sqrt{\det q}}E^a_iE^b_j\bigg(-\epsilon^{ij}\,_k\,R_{ab}^k+2K_{[a}^i\,K_{b]}^j+\frac{2}{\gamma}\epsilon^{ij}\,_k\,\leftidx{^\Gamma}{\mathcal{D}}{_{[a}}K_{b]}^k \bigg)   \label{bulk3Decomp1}
\end{align}
where we have taken definitions
\[
\epsilon^{ij}\,_k\,R_{ab}^k:=\epsilon^{ij}\,_k\left(2\bar{\partial}_{[a}\Gamma_{b]}^k+\epsilon^{kl}\,_m\,\Gamma_{[a}^l\,\Gamma_{b]}^m\right)=2\epsilon^{ij}\,_k\,\bar{\partial}_{[a}\Gamma_{b]}^k+2\Gamma_{[a}^i\,\Gamma_{b]}^j
\]
and
\[
\leftidx{^\Gamma}{\mathcal{D}}{_{a}}K_{b}^k:=\bar{\partial}_{[a}K_{b]}^{k}+\epsilon^k\,_{lm}\,\Gamma_{[a}^lK_{b]}^{m}\,.
\]
Then we can use (\ref{FRIdentity}) here, which is merely an algebraic identity:
\[
F_{ab}^i=R_{ab}^i+2\gamma\,\leftidx{^\Gamma}{\mathcal{D}}{_{[a}}K_{b]}^i+\gamma^2\epsilon_{ijk}K_a^jK_b^k\,.  
\]
to rewrite (\ref{bulk3Decomp1}) as
\begin{align}
N\mathcal{E}^a_I\tilde{e}^\mu_a&\sqrt{\det q}\,\mathcal{E}^b_J\,\tilde{e}^\nu_b\,\leftidx{^\gamma}{F}{^{IJ}_{\mu\nu}}  \notag\\
&=-\frac{N}{\sqrt{\det q}}E^a_iE^b_j\bigg[\epsilon^{ij}\,_k\,F_{ab}^k-2(1+\gamma^2)K_{[a}^i\,K_{b]}^j-2\left(\frac{1+\gamma^2}{\gamma}\right)\epsilon^{ij}\,_k\,\leftidx{^\Gamma}{\mathcal{D}}{_{[a}}K_{b]}^k \bigg]  \label{bulk3Decompf}
\end{align}

Putting (\ref{bulk1Decompf}), (\ref{bulk2Decompf}) and (\ref{bulk3Decompf}) together
\begin{align}
|e|e^\mu_Ie^\nu_J\,\leftidx{^\gamma}{F}{^{IJ}_{\mu\nu}}
=&\left(-2t^\mu n_I+2N^an_I\tilde{e}^\mu_a+N\mathcal{E}^a_I\tilde{e}^\mu_a\right)\sqrt{\det q}\,\mathcal{E}^b_J\,\tilde{e}^\nu_b\,\leftidx{^\gamma}{F}{^{IJ}_{\mu\nu}}  \notag\\
=&\frac{2}{\gamma}\left(E^b_j\bar{\partial}_tA_b^j+\Lambda^j\mathcal{G}_j-(1+\gamma^2)\omega_t^{0j}\mathcal{S}_j-\bar{\partial}_b(\Lambda^jE^b_j)  \right) \notag\\ 
&-\frac{2}{\gamma}N^a\left(F_{ab}^jE^b_j-(1+\gamma^2)K_a^k\mathcal{S}_k\right)  \notag\\
&-\frac{N}{\sqrt{\det q}}E^a_iE^b_j\bigg[\epsilon^{ij}\,_k\,F_{ab}^k-2(1+\gamma^2)K_{[a}^i\,K_{b]}^j-2\left(\frac{1+\gamma^2}{\gamma}\right)\epsilon^{ij}\,_k\,\leftidx{^\Gamma}{\mathcal{D}}{_{[a}}K_{b]}^k \bigg] 
\end{align}

Finally, we turn to the boundary term in (\ref{HolstActionIndex}).
We start with the general decomposition
\begin{align}
2\sqrt{|h|}\, r_\nu& e^\mu_Ie^\nu_J\,\leftidx{^\gamma}{\omega}{^{IJ}_\mu} \notag\\
&=2\sqrt{|h|}\, r_\nu \left(\frac{\left(N^\mu-t^\mu\right)}{N}n_I+\mathcal{E}^\mu_I\right)\left(\frac{\left(N^\nu-t^\nu\right)}{N}n_J+\mathcal{E}^\nu_J\right)\leftidx{^\gamma}{\omega}{^{IJ}_\mu} \notag\\
&=2\sqrt{|h|}\,  \left(\frac{\left(N^\mu-t^\mu\right)}{N}n_I\mathcal{E}^\nu_J(r_\nu\leftidx{^\gamma}{\omega}{^{IJ}_\mu}-r_\mu\leftidx{^\gamma}{\omega}{^{IJ}_\nu})+r_\nu\mathcal{E}^\mu_I\mathcal{E}^\nu_J\,\leftidx{^\gamma}{\omega}{^{IJ}_\mu}\right) \notag\\
&=2\sqrt{|h|}\,  \left[\left(\frac{N^a}{N}n_I\mathcal{E}^b_J\tilde{e}^\mu_a\tilde{e}^\nu_b-\frac{n_I}{N}\mathcal{E}^b_Jt^\mu\tilde{e}^\nu_b\right)\left(r_\nu\leftidx{^\gamma}{\omega}{^{IJ}_\mu}-r_\mu\leftidx{^\gamma}{\omega}{^{IJ}_\nu}\right)
+r_\nu\mathcal{E}^a_I\mathcal{E}^b_J\,\leftidx{^\gamma}{\omega}{^{IJ}_\mu}\tilde{e}^\mu_a\tilde{e}^\nu_b\right]\,.
\end{align}
Imposing the time-gauge (\ref{timeGauge}), and using adapted coordinates along with definitions (\ref{defE})-(\ref{defLambda}) and 
\[
\leftidx{^\gamma}{\omega}{_a^{ij}}=\omega_a^{ij}+\frac{1}{\gamma}\epsilon^{ij}\,_{k0}\omega_a^{k0}
=\epsilon^{ij}\,_k\left(\frac{1}{\gamma}K_a^k-\Gamma_a^k\right)\,,
\]
the boundary term reads:
\begin{align}
2\sqrt{|h|}\,&r_\nu e^\mu_Ie^\nu_J\,\leftidx{^\gamma}{\omega}{^{IJ}_\mu} \notag\\
&=2\sqrt{|h|}\,\bigg[-\frac{N^a}{N}\mathcal{E}^b_j\left(r_b\,\leftidx{^\gamma}{\omega}{_a^{0j}}-r_a\,\leftidx{^\gamma}{\omega}{_b^{0j}}\right)
+\frac{1}{N}\mathcal{E}^b_j\left(r_b\,\leftidx{^\gamma}{\omega}{_t^{0j}}-r_t\,\leftidx{^\gamma}{\omega}{_b^{0j}}\right)
+r_b\mathcal{E}^a_i\mathcal{E}^b_j\,\leftidx{^\gamma}{\omega}{_a^{ij}}\bigg] \notag\\
&=2\sqrt{|h|}\,\bigg[-\frac{N^a E^b_j}{\gamma N\sqrt{\det q}}\left(r_b\,A_a^j-r_a\,A_b^j\right)
+\frac{E^b_j}{\gamma N\sqrt{\det q}}\left(r_b\,\Lambda^j-r_t\,A_b^j\right)  \notag\\
&\qquad\qquad\quad+r_b\frac{E^a_iE^b_j}{(\det q)}\epsilon^{ij}\,_k\left(\frac{1}{\gamma}K_a^k-\Gamma_a^k\right)\bigg]  \notag\\
&=\frac{2\sqrt{|h|}}{\gamma N\sqrt{\det q}}\bigg[-N^a E^b_j\left(r_b\,A_a^j-r_a\,A_b^j\right)
+E^b_j\left(r_b\,\Lambda^j-r_t\,A_b^j\right)  \notag\\
&\qquad\qquad\quad+r_b\frac{NE^a_iE^b_j}{\sqrt{\det q}}\epsilon^{ij}\,_k\left(K_a^k-\gamma\Gamma_a^k\right)\bigg] \notag\\
&=\frac{2\sqrt{|h|}}{\gamma \sqrt{|\det g|}}\bigg[2r_aN^{[a}E^{b]}_jA_b^j
+E^b_j\left(r_b\,\Lambda^j-r_t\,A_b^j\right)   
-r_b\frac{NE^a_iE^b_j}{\sqrt{\det q}}\epsilon^{ij}\,_k\left(\gamma A_a^k-(1+\gamma^2)K_a^k\right)\bigg]
\end{align}
with $r_t:=t^\mu r_\mu$   and $r_a:=\tilde{e}^\mu_ar_\mu$.

\section{Generators in Ashtekar-Barbero variables}  \label{a::AshtekarVariables}

In this appendix we construct Poincar\'e generators in  Ashtekar-Barbero variables and verify their well-posedness (finiteness) and differentiability on all of phase space. As first done in \cite{Thiemann} for self-dual variables, we think of the phase space $\Gamma_{(A,E)}$ as an extension of the ADM phase space $\Gamma_\text{ADM}$ coordinatized by geometrodynamical variables $(q_{ab},p^{cd})$. We construct the generators of spatial translations and rotations  by extending to $\Gamma_{(A,E)}$  the corresponding generators on $\Gamma_\text{ADM}$. For the generators of time translations and boosts we use a particular extension of the Hamiltonian constraint and apply the method of counter terms. The ADM expressions for the generators are:
\begin{equation}  \label{ADMDgenerator}
   \mathbf{G_D}[M^a]=D[M^a]+\frac{1}{\kappa}\int_{r=\infty}dS_b\,M^ap^b\,_a=\frac{1}{2\kappa}\int\md^3x\,p^{ab}\mathcal{L}_{\vec{M}}q_{ab}  
\end{equation}
\begin{align}
    \mathbf{G_H}[M]=H[M]+\frac{1}{2\kappa}\int_{r=\infty}&{\md S}_d\,\sqrt{\det q}\Big[M q^{ab}q^{cd}(\partial_bq_{ac}-\partial_cq_{ab}) \notag \\
           &-\partial_bMq^{ab}q^{cd}(q_{ac}-\delta_{ac})+\partial_cMq^{ab}q^{cd}(q_{ab}-\delta_{ab})\Big]  \label{ADMHgenerator}
\end{align}
with
\begin{align}
D[N^a]&:=-\frac{1}{2\kappa}\int_\Sigma\md^3x\,2N^aD_bp^b\,_a \notag\\
H[N]&:=\frac{1}{2\kappa}\int_\Sigma\md^3x\,N\left[\frac{1}{\sqrt{\det q}}\left(q_{ac}q_{bd}-\frac{1}{2}q_{ab}q_{cd}\right)p^{ab}p^{cd}-\sqrt{\det q}\,R\right] \label{ADMHconstraint}
\end{align}
the diffeomorphism and Hamiltonian constraint expressions.  Here $D_a$ denotes the Levi-Civita connection: $D_cq_{ab}=0$. 

Given a function $F(q,p)$ on the ADM phase space (which may also depend on some additional smearing function(s) on space $\Sigma$), there is of course no unique extension $\widetilde{F}(A,E)$ of such function to the larger phase space $\Gamma_{(A,E)}$ of connection and triad variables. All that is required is that $\widetilde{F}$ coincides with $F$ on the Gauss constraint surface $\mathcal{G}_i\approx 0$ and hence any two such extensions $\widetilde{F}$ and $\widetilde{F}'$  necessarily differ by a multiple of the Gauss constraint, e.g. $\widetilde{F}'(A,E)=\widetilde{F}(A,E)+\int \Lambda^i(A,E)\mathcal{G}_i$.
For an expression to serve solely as an extension of an ADM constraint, this matching on the $\mathcal{G}_i\approx 0$ surface  is all that is needed. However, if a given expression is also to serve as a canonical generator on the extended phase space, the Hamiltonian flow of $X_{\widetilde{F}}=\{\,\cdot\,,\widetilde{F}\}$ must also reproduce, on the constraint surface $\mathcal{G}_i\approx 0$, the Hamiltonian flow of $F$ on the ADM phase space.
In particular $\widetilde{F}$ must also be differentiable (and therefore finite).

Finiteness and differentiability of a function on the ADM phase space do not guarantee finiteness and differentiability of its extension. Depending on how one performs the extension of a well defined and differentiable ADM function, one may or may not arrive at a finite and differentiable expression.
Of course on the constraint surface $\mathcal{G}_i\approx 0$ any extension of a well defined generator is finite (it matches the finite ADM expression) but this does not guarantee a priori even that its flow on the Gauss constraint surface is  well defined and furthermore that it corresponds to Poincar\'e transformations or any other given flow. On the extended phase space there are extra `directions to move' compared to the ADM phase space.

On the other hand, given any two well defined and differentiable extensions $\widetilde{F}(A,E)$ and $\widetilde{F}'(A,E)=\widetilde{F}(A,E)+\int \Lambda^i(A,E)\mathcal{G}_i$, one may ask (at least formally) how do their Hamiltonian flows deviate. By linearity of the bracket, the infinitesimal transformations generated by $\widetilde{F}$ and $\widetilde{F}'$ on the canonical variables $(A,E)$ differ by those transformations generated by the functional $\int \Lambda^i(A,E)\mathcal{G}_i$:
\begin{align*}
\left\{A_a^i,\widetilde{F}'\right\}&=\left\{A_a^i,\widetilde{F}\right\}+\left\{A_a^i,\int \Lambda^j\mathcal{G}_j\right\} =\left\{A_a^i,\widetilde{F}\right\}+\int \frac{\delta \Lambda^j}{\delta E^a_i}\mathcal{G}_j+\int \Lambda^j\frac{\delta \mathcal{G}_j}{\delta E^a_i}\\
&=\left\{A_a^i,\widetilde{F}\right\}+\int \frac{\delta \Lambda^j}{\delta E^a_i}\mathcal{G}_j+ \left(-\bar{\partial}_a\Lambda^i+\epsilon^i\,_{jk}\Lambda^jA_a^k\right) \\ 
&\approx \left\{A_a^i,\widetilde{F}\right\}+\left(-\bar{\partial}_a\Lambda^i+\epsilon^i\,_{jk}\Lambda^jA_a^k\right)
\end{align*} 
and similarly
\begin{align*}
\left\{E^a_i,\widetilde{F}'\right\}=\left\{E^a_i,\widetilde{F}\right\}-\int\frac{\delta \Lambda^j}{\delta A_a^i}\mathcal{G}_j+\left(\epsilon_{ji}\,^k\Lambda^jE^a_k\right)\approx\left\{E^a_i,\widetilde{F}\right\}+\left(\epsilon_{ji}\,^k\Lambda^jE^a_k\right)\,.
\end{align*}
So `on-shell', that is on the Gauss constraint surface $\mathcal{G}_i\approx 0$, flows differ only by `phase-space-dependent infinitesimal $SU(2)$-gauge transformations'. These gauge transformations vanish at infinity since by assumption $\widetilde{F}$ and $\widetilde{F}'$, and hence $\int \Lambda^i(A,E)\mathcal{G}_i$ are finite, so $\Lambda^i(A,E)$ is necessarily of leading order $\Ord{r}{-1}{even}$.

The ambiguity in the extension of an ADM function comes from the relation of phase space variable $K_a^i$ with extrinsic curvature $K_{ab}$ on the Gauss constraint surface: $K_a^i=K_{ab}E^b_j\delta^{ij}/\sqrt{\det E}$, 
which solving for $K_{ab}$ gives 
\[
K_{ab}=\sqrt{\det E}\,K_a^iE_b^i\,.
\]
If we use the right hand side of the above relation as an extension of the function representing the extrinsic curvature tensor and define
\begin{equation}  \label{Kextension}
\widetilde{K}_{ab}:=\sqrt{\det E}\,K_a^iE_b^i\,,
\end{equation}
we get a generally non-symmetric tensor. To match extrinsic curvature, at the very least this tensor has to be symmetric. The requirement that its anti-symmetric part
\begin{align}
\widetilde{K}_{[ab]}&=\frac{1}{2}\epsilon_{abc}\,\epsilon^{cde}\widetilde{K}_{de}\notag\\
&=\frac{1}{2}\epsilon_{abc}\,\epsilon^{cde}\sqrt{\det E}\,K_d^iE_e^i \notag\\
&=-\frac{1}{2\sqrt{\det E}}\epsilon_{abc}E^c_i\epsilon^{ijk}K_d^jE^d_k    \label{antiSymmK}
\end{align}
vanishes, gives precisely the Gauss constraint $\gamma^{-1}\mathcal{G}^i=\epsilon^{ijk}K_d^jE^d_k\approx 0$ since $E^c_i$ is generally invertible. (In (\ref{antiSymmK}) we have used the identity $\epsilon^{abc}(\det E)=\epsilon^{ijk}E^a_iE^b_jE^c_k$). So
\begin{equation}
\widetilde{K}_{[ab]}=-\frac{1}{2\gamma\sqrt{\det E}}\epsilon_{abc}E^c_i\mathcal{G}^i\,.
\end{equation}

The ambiguity in extending the ADM momentum function $p_{ab}=\sqrt{\det q}(K_{ab}-Kq_{ab})$ may then be translated into the ambiguity of extending $K_{ab}$. The simplest possibility is to take $\widetilde{K}_{ab}$ and define $\widetilde{p}_{ab}:=\sqrt{\det q}(\widetilde{K}_{ab}-\widetilde{K}q_{ab})$. Unfortunately, for the Poincar\'e generator $\widetilde{\mathbf{G_D}}[M^a]$ this prescription will result in a divergent expression on the extended phase space. As done in \cite{Thiemann}, to find the correct generator one then has to isolate the divergent term in the latter expression and subtract the appropriate counter-term. However, as also noted in \cite{Thiemann},  one may directly arrive at the correct generator by using a different prescription that  more accurately reflects the symmetry property of $p_{ab}$. We therefore define the extension of $p_{ab}$ using the symmetric part of $\widetilde{K}_{ab}$:
\begin{equation} \label{pExtension}
\widetilde{p}_{ab}:=\sqrt{\det q}(\widetilde{K}_{(ab)}-\widetilde{K}q_{ab})=\sqrt{\det E}(\widetilde{K}_{ab}-\widetilde{K}_{[ab]}-\widetilde{K}q_{ab})
\end{equation}
with
\[
\widetilde{K}=\widetilde{K}_{(cd)}\,q^{cd}=\widetilde{K}_{cd}\,q^{cd}=K_c^ie_d^iq^{cd}=K_c^iE^c_i/\sqrt{\det E}.
\]

In the following, as it is customary, abusing notation we will drop the tilde on extended functions.
Also for simplicity, we will set $2\kappa=1$ in all expressions for the generators.
Using these definitions and $q^{ab}=E^a_iE^b_i/\det E$,  we have
\begin{align*}
p^b\,_a=p_{ca}q^{cb}&=K_a^iE^b_i-K_c^iE^c_i\,\delta^b_a\,-\sqrt{\det E}\,K_{[ac]}q^{cb}
\end{align*}
with
\[
\sqrt{\det E}\,K_{[ac]}q^{cb}=-\frac{1}{2\gamma(\det E)}E^b_j\epsilon_{acd}E^c_jE^d_i\mathcal{G}^i\,.
\]
If we ignore this contribution from the subtraction of the anti-symmetric part of $K_{ab}$,  
the full ADM Poincar\'e generator with surface integral counter term would be
\begin{align*}
\mathbf{G'_D}[M^a]&=-2\int\md^3x\,M^aD_b\,p^b\,_a+2\int\md S_b\,M^ap^b\,_a \\
&=4\int\md^3x\,M^aD_{[a}(K_{b]}^iE^b_i)+4\int\md S_b\,M^{[a}\,K_a^{|i|}E^{b]}_i \\
&=2\gamma^{-1}\int\md^3x\,M^a\left[F_{ab}^iE^b_i-(A_a^i-\Gamma_a^i)\mathcal{G}_i\right]
+4\gamma^{-1}\int\md S_b\,M^{[a}\,(A_a^{|i}-\Gamma_a^{i|})E^{b]}_i \,,  
\end{align*}
where for the standard form in the last line one uses
$2D_{[a}(K_{b]}^iE^b_i)=2\,\leftidx{^\Gamma}{\mathcal{D}}{_{[a}}(K_{b]}^iE^b_i)=\gamma^{-1}F_{ab}^iE^b_i-K_a^i\mathcal{G}_i\,$\footnote{This is derived by contracting  with $E^b_i$ the well-known identity (see e.g. \cite{ThiemannBook})
\begin{align}
F_{ab}^i&=2\bar{\partial}_{[a}A_{b]}^i+\epsilon_{ijk}A_a^jA_b^k \notag\\
&=2\bar{\partial}_{[a}(\Gamma_{b]}^i+\gamma K_{b]}^i)+\epsilon_{ijk}(\Gamma_a^j+\gamma K_a^j)(\Gamma_b^k+\gamma K_b^k) \notag \\
&=2\bar{\partial}_{[a}\Gamma_{b]}^i+\epsilon_{ijk}\Gamma_a^j\Gamma_b^k+2\gamma(\bar{\partial}_{[a}K_{b]}^i+\epsilon_{ijk}\Gamma_{[a}^jK_{b]}^k)+\gamma^2\epsilon_{ijk}K_a^jK_b^k \notag \\
&=R_{ab}^i+2\gamma\,\leftidx{^\Gamma}{\mathcal{D}}{_{[a}}K_{b]}^i+\gamma^2\epsilon_{ijk}K_a^jK_b^k\,.  \label{FRIdentity}
\end{align}
 and then applying the Bianchi identity $R_{ab}^iE^b_i=0$, the compatibility of the spin connection with the triad $\leftidx{^\Gamma}{\mathcal{D}}{_a}E^b_i=0$, and substituting the Gauss constraint $\mathcal{G}_i=\gamma\,\epsilon_{ij}\,^kK_b^jE^b_k$.}. 
As already mentioned however, this expression for the generator  in terms of connection variables turns out to be divergent on the extended phase space.  The contribution from the subtraction of the anti-symmetric part of $K_{ab}$ (we get two additional terms, one from the bulk and one from the surface integral that combine):
\begin{align*}
-&2\int\md^3x\,\left[M^aD_b\left(-\sqrt{\det E}\,K_{[ac]}q^{cb}\right)-D_b\left(-M^a\sqrt{\det E}\,K_{[ac]}q^{cb}\right)\right]=\\
&=-2\int\md^3x\,D_bM^a\sqrt{\det E}\,K_{[ac]}q^{cb}
=2\int\md^3x\,\frac{\epsilon_{acd}}{2\gamma(\det E)}E^b_jE^c_jE^d_i\mathcal{G}^iD_bM^a
\end{align*}
is precisely what is needed to make the extended generator finite and differentiable. If we re-insert this term then the extended (renormalized) generator is
\begin{align}
\mathbf{G_D}[M^a]=&-2\int\md^3x\,M^aD_b\,p^b\,_a+2\int\md S_b\,M^ap^b\,_a \notag\\
=&\,2\gamma^{-1}\int\md^3x\,\left[M^a\left[F_{ab}^iE^b_i-(A_a^i-\Gamma_a^i)\mathcal{G}_i\right]
+\frac{\epsilon_{acd}}{2(\det E)}E^b_jE^c_jE^d_i\mathcal{G}^iD_bM^a\right]  \notag\\
&+4\gamma^{-1}\int\md S_b\,M^{[a}\,(A_a^{|i}-\Gamma_a^{i|})E^{b]}_i \,.   \label{RenPDgenerator1}
\end{align}
In this form, the surface term exactly matches the ADM surface term and hence corresponds to ADM momentum and angular momentum on-shell. 

An expression better suited to check finiteness of the extended generator is obtained by substituting (\ref{pExtension}) in the ADM expression (\ref{ADMDgenerator}) for $\mathbf{G_D}[M^a]$ in terms of a single bulk integral.
For that we define the Lie derivative\footnote{The formula above is only valid for the Levi-Civita connection $D_a$. For a general connection, e.g. $\partial_a$ there is an additional term: $\mathcal{L}_{\vec{M}}E^a_i=-\frac{1}{2}M^bE_c^jE^a_i\partial_bE^c_j+M^b\partial_bE^a_i-E^b_i\partial_bM^a+E^a_i\partial_bM^b$,  so also $\bar{\mathcal{L}}_{\vec{M}}E^a_i=-\frac{1}{2}M^bE_c^jE^a_i\bar{\partial}_bE^c_j+M^b\bar{\partial}_bE^a_i-E^b_i\bar{\partial}_bM^a+E^a_i\bar{\partial}_bM^b$. }
\begin{equation}   \label{barLieE}
\bar{\mathcal{L}}_{\vec{M}}E^a_i:=M^b\bar{D}_bE^a_i-E^b_i\bar{D}_bM^a+E^a_i\bar{D}_bM^b
\end{equation}
where $\bar{D}_a$ is an extension of the Levi-Civita connection defined by $\bar{D}_aq_{bc}=0$ and $\bar{D}_av^i=\bar{\partial}_av^i$, then $\mathcal{L}_{\vec{M}}(E^a_iE^b_i)=\bar{\mathcal{L}}_{\vec{M}}(E^a_iE^b_i)$.  After these substitutions and some algebraic manipulations the generator becomes
\begin{align}
\mathbf{G_D}[M^a]&=\int\md^3x\,p^{ab}\mathcal{L}_{\vec{M}}q_{ab} \notag\\
&=-\int\md^3x\,(K_a^i\bar{\mathcal{L}}_{\vec{M}}E^a_i+E_b^iE^a_jK_a^i\bar{\mathcal{L}}_{\vec{M}}E^b_j) \notag\\
&=-\int\md^3x\,E_b^iK_a^i\bar{\mathcal{L}}_{\vec{M}}(E^a_jE^b_j) \notag\\
&=-\gamma^{-1}\int\md^3x\,\underbrace{E_b^i(A_a^i-\Gamma_a^i)}_{\Ord{r}{-2}{odd}}\,\underbrace{\mathcal{L}_{\vec{M}}((\det q)q^{ab})}_{\Ord{r}{-1}{even}}  \label{explicitlyFiniteGD}
\end{align}
and from this expression one can explicitly see  that the extended generator is finite.

One can also explicitly check that from (\ref{explicitlyFiniteGD}) one  recovers (\ref{RenPDgenerator1}). 
One additionally has equivalent expressions:
\begin{align}
\mathbf{G_D}[M^a]=&-\gamma^{-1}\int\md^3x\,E_b^i(A_a^i-\Gamma_a^i)\,\mathcal{L}_{\vec{M}}((\det q)q^{ab}) \notag \\
=&-2\int\md^3x\,\left(K_a^i\,\bar{\mathcal{L}}_{\vec{M}}E^a_i-\frac{\epsilon_{abc}}{2\gamma(\det E)}E^a_iE^b_j\mathcal{G}_i\bar{\mathcal{L}}_{\vec{M}}E^c_j\,\right) \notag \\
=&2\gamma^{-1}\int\md^3x\,\left[M^a\left[F_{ab}^iE^b_i-A_a^i\mathcal{G}_i\right]
+\frac{\epsilon_{abc}}{2(\det E)}E^a_iE^b_j\mathcal{G}_i\bar{\mathcal{L}}_{\vec{M}}E^c_j\,\right] \notag\\
& -4\gamma^{-1}\int\md S_{[b}[M^b(A_{a]}^i-\Gamma_{a]}^i)E^a_i]\,.   \label{RenPDgenerator2}
\end{align}

We now turn to checking differentiability of the generator. For that one takes the variation of (\ref{RenPDgenerator2}). Variation of the `standard' bulk term gives
\begin{align}
2\gamma^{-1}&\delta\int\md^3x\,M^a\left(F_{ab}^iE^b_i-A_a^i\mathcal{G}_i\right) \notag\\
&=2\gamma^{-1}\delta\int\md^3x\,M^a\left(2E^b_i\bar{\partial}_{[a}A_{b]}^i-A_a^i\bar{\partial}_bE^b_i\right)  \notag\\
&=2\gamma^{-1}\int\md^3x\,\left(-\bar{\mathcal{L}}_{\vec{M}}E^a_i\right)\delta A_a^i+\left(\bar{\mathcal{L}}_{\vec{M}}A_a^i\right)\delta E^a_i \notag\\
&\,\quad+2\gamma^{-1}\int 2\,\md S_{[a}\left(M^aE^b_i\delta A_{b]}^i\right) -\md S_b\left(M^aA_a^i\delta E^b_i\right)\,. \label{DConnVariation1}
\end{align}
Variation of the counter-term proportional to the Gauss constraint results in terms proportional to the Gauss constraint and terms that formally resemble a `phase-space-dependent' infinitesimal gauge transformation with gauge angle
\begin{equation}  \label{gaugeAngle1}
\Lambda_i[\vec{M}]:=\frac{\epsilon_{abc}}{2(\det E)}E^a_iE^b_j\bar{\mathcal{L}}_{\vec{M}}E^c_j
=\frac{1}{2}\epsilon_{ijk}E^k_c\bar{\mathcal{L}}_{\vec{M}}E^c_j \,.
\end{equation}
The variation of the complete generator is:
\begin{align}
\delta \mathbf{G_D}[M^a]
=2\gamma^{-1}\int&\md^3x\, \bigg(-\underbrace{\bar{\mathcal{L}}_{\vec{M}}E^a_i}_{\Ord{r}{0}{even}} +\underbrace{\Lambda_j[\vec{M}]\,\epsilon_{ji}\,^kE^a_k}_{\Ord{r}{0}{even}}\bigg)\delta A_a^i \notag\\
+&\bigg(\underbrace{\bar{\mathcal{L}}_{\vec{M}}A_a^i}_{\Ord{r}{-2}{odd}}+\bigg(-\underbrace{\bar{\partial}_a \Lambda_i[\vec{M}]}_{\Ord{r}{-2}{odd}}\,+\underbrace{\Lambda_j[\vec{M}]\,\epsilon_{jk}\,^iA_a^k}_{\Ord{r}{-2}{odd}}\bigg) 
-\frac{1}{2}\epsilon_{jkl}\underbrace{E_a^kE_c^i\,\mathcal{G}_l\,\bar{\mathcal{L}}_{\vec{M}}E^c_j}_{\Ord{r}{-2}{odd}}  \notag\\
&\quad+\frac{1}{2}\epsilon_{ijk}\bigg(\frac{1}{2}\underbrace{M^bE_a^k\,\mathcal{G}_j(\bar{\partial}_bE^c_l)E_c^l}_{\Ord{r}{-3}{odd}}+\underbrace{M^b\bar{\partial}_b(E_a^k\,\mathcal{G}_j)}_{\Ord{r}{-2}{odd}}+\underbrace{E_b^k\,\mathcal{G}_j\bar{\partial}_aM^b}_{\Ord{r}{-2}{odd}}\bigg)\bigg)\delta E^a_i \notag\\
+2\gamma^{-1}&\int\md S_b\, \frac{1}{2}\epsilon_{ijk}\underbrace{M^bE_a^k\,\mathcal{G}_i\delta E^a_j}_{\Ord{r}{-2}{even}}  -\md S_a\,\underbrace{M^aA_b^i\delta E^b_i}_{\Ord{r}{-2}{even}} \notag\\
&\quad+\md S_a\,\underbrace{\Lambda_i[\vec{M}]\,\delta E^a_i}_{\Ord{r}{-1}{even}} + \md S_{[a}\,\underbrace{M^a \delta(\Gamma_{b]}^iE^b_i)}_{\Ord{r}{-1}{even}} \label{PDgeneratorConnVariation}
\end{align}
Differentiability (with respect to $A_a^i$ and $E^a_i$) will follow if  all the bulk terms are finite and all the surface terms are shown to vanish in the limit $r\to\infty$.  To verify this we needed the fall-off behavior of $\bar{\mathcal{L}}_{\vec{M}}E^a_i$ which may be easily inferred from (\ref{barLieE}). Its leading term is $\Ord{r}{-1}{even}$ for odd supertranslations and $M^a$ containing up to constant terms (translations) and it is $\Ord{r}{0}{even}$ for the general fall-off (\ref{PoincareShiftFallOff}) containing rotations. (Similarly $\bar{\mathcal{L}}_{\vec{M}}A_a^i=M^b\bar{\partial}_bA_a^i+A_b^i\bar{\partial}_aM^b$ has leading term $\Ord{r}{-3}{odd}$ for odd supertranslations and up to translations, and $\Ord{r}{-2}{odd}$ for rotations). It is now easy to see that for the most general fall-off for the Killing fields (\ref{PoincareShiftFallOff}), each of the factors multiplying $\delta E^a_i$ in the bulk integral has leading term at most of order $\Ord{r}{-2}{odd}$, as required for $\int \frac{\delta \mathbf{G_D}[M^c]}{\delta E^a_i}\delta E^a_i$ to converge.
The term proportional to $\delta A_a^i$ corresponding to $\frac{\delta\mathbf{G_D}[M^c]}{\delta A_a^i}$ is not manifestly finite but it may be easily re-expressed in such form
\begin{align*}
-\bar{\mathcal{L}}_{\vec{M}}E^a_i + \Lambda_j[\vec{M}]\,\epsilon_{ji}\,^kE^a_k&=
-\bar{\mathcal{L}}_{\vec{M}}E^a_i +\frac{1}{2}\left(\bar{\mathcal{L}}_{\vec{M}}E^a_i - E_b^iE^a_j\bar{\mathcal{L}}_{\vec{M}}E^b_j\right) \\
&=-\frac{1}{2}E_b^i\,\bar{\mathcal{L}}_{\vec{M}}(E^a_jE^b_j)= -\frac{1}{2}\underbrace{E_b^i\,\mathcal{L}_{\vec{M}}((\det q)q^{ab})}_{\Ord{r}{-1}{even}}\,.
\end{align*}
This form can also be obtained by taking the variation $\delta A_a^i$ directly from  the manifestly finite form of the generator (\ref{explicitlyFiniteGD}).

It remains to check the surface integrals in (\ref{PDgeneratorConnVariation}) cancel. The first two surface terms have integrands of leading order $\Ord{r}{0}{odd}$, so they vanish as the boundary $\partial\Sigma$ is taken to infinity. Showing that the last two terms are also zero at infinity is less trivial. For the details we refer the reader to \cite{ourReview}.
This concludes the proof of differentiability (with respect to $A_a^i$ and $E^a_i$) of the generator of asymptotic spatial translations and rotations. For completeness, taking the variation of $\mathbf{G_D}[M^a]$ with respect to $M^a$, using the manifestly finite form (\ref{explicitlyFiniteGD}), one gets
\begin{align*}
\delta_{\vec{M}}\mathbf{G_D}[M^a]&=-\delta_{\vec{M}}\int\md^3x\,E_b^iK_a^i\mathcal{L}_{\vec{M}}\left((\det q)q^{ab}\right) \\
&=-\delta_{\vec{M}}\int\md^3x\,E_b^iK_a^i(\det q)\left(2q^{ab}D_cM^c-q^{bc}D_cM^a-q^{ac}D_cM^b\right) \\
&=\int\md^3x\,(\det q)q^{bc}\left(2E_b^i\,\leftidx{^\Gamma}{\mathcal{D}}{_a}K_c^i-E_b^i\,\leftidx{^\Gamma}{\mathcal{D}}{_c}K_a^i - E_a^i\,\leftidx{^\Gamma}{\mathcal{D}}{_c}K_b^i\right)\delta M^a \\
&\quad+\int\md S_c\,E_b^iK_a^i(\det q)\left(2q^{ab}\delta M^c-q^{bc}\delta M^a - q^{ac}\delta M^b\right)\,,
\end{align*}
which shows $\mathbf{G_D}[M^a]$ is not differentiable with respect to $M^a$ for variations of $M^a$ consistent with Killing field components $\delta M^a=\epsilon^a\,_{bc}\,\delta\alpha^cx^b-\delta\beta_at+\delta T^a+\delta S^a_{\text{odd}}(x^c/r)+\ord{r}{0}$.

Lastly we come to the Hamiltonian constraint and the generator of time translations and boosts. The easiest way to arrive at the generator is to first perform a simple extension of the ADM Hamiltonian constraint (\ref{ADMHconstraint}) and then use the method of counter-terms to construct the generator from it. From the ADM expression
\begin{align*}
H[N]&=\int\md^3x\,N\left(\sqrt{\det E}(K_{ab}K^{ab}-K^2)-\sqrt{\det E}\,R\right)
\end{align*}
we may extend the Hamiltonian as
\begin{align}
H[N]&=\int\md^3x\,N\left(\sqrt{\det E}(\widetilde{K}_{ba}\widetilde{K}_{cd}\,q^{ac}q^{bd}-\widetilde{K}^2)-\sqrt{\det E}\,R\right) \notag\\
&=\int\md^3x\,N\left(\frac{1}{\sqrt{\det E}}\left(K_a^iK_b^jE^a_jE^b_i-(K_a^iE^a_i)^2\right)-\sqrt{\det E}\,R\right) \notag\\
&=\int\md^3x\,N\left(-\frac{2}{\sqrt{\det E}}E^a_iE^b_jK_{[a}^iK_{b]}^j-\sqrt{\det E}\,R\right) \label{extendedH1}
\end{align}
with $\widetilde{K}_{ab}$ as defined before in (\ref{Kextension}). Notice this does not correspond to the symmetric extension of extrinsic curvature. However, since $\widetilde{K}_{ba}=\widetilde{K}_{(ab)}-\widetilde{K}_{[ab]}$, we have
\[
\widetilde{K}_{ba}\widetilde{K}^{ab}=\widetilde{K}_{(ab)}\widetilde{K}^{(ab)}-\widetilde{K}_{[ab]}\widetilde{K}^{[ab]} 
\]
so the symmetric extension differs from this one by the term $N\sqrt{\det E}\,\widetilde{K}_{[ab]}\widetilde{K}^{[ab]}$, quadratic in the Gauss constraint. Unlike the case of the diffeomorphism constraint and the generator of rotations, this term is $\Ord{r}{-3}{odd}$ for boosts and plays no role in the discussion of finiteness and differentiability of $\mathbf{G_H}[M]$.

To write the Hamiltonian in standard form, one uses  (\ref{FRIdentity}) again to derive an identity for the curvature scalar\footnote{This is obtained by contracting (\ref{FRIdentity}) with $\epsilon_{ijk}E^a_jE^b_k$ (and using anti-symmetry of $ij$ in the second term on the right hand side):
\[
\epsilon_{ijk}E^a_jE^b_kF_{ab}^i=\epsilon_{ijk}E^a_jE^b_kR_{ab}^i+2\gamma\,\epsilon_{ijk}E^a_jE^b_k\,\leftidx{^\Gamma}{\mathcal{D}}{_{a}}K_{b}^i+\gamma^2\epsilon_{ijk}\epsilon_{ilm}E^a_jE^b_kK_a^lK_b^m\,.
\]
Substituting the identities
\[
\epsilon_{jik}R_{ab}^ie^a_je^b_k=R_{abjk}e^a_je^b_k=R_{abcd}e^c_je^d_ke^a_je^b_k=R
\]
and
\[
\leftidx{^\Gamma}{\mathcal{D}}{_a}\mathcal{G}_j=\leftidx{^\Gamma}{\mathcal{D}}{_a}(\gamma\epsilon_{jik}K_b^iE^b_k)=\gamma\epsilon_{jik}(\leftidx{^\Gamma}{\mathcal{D}}{_a}K_b^i)E^b_k
\]
gives
\[
\epsilon_{ijk}E^a_jE^b_kF_{ab}^i=-(\det E)R-2(\leftidx{^\Gamma}{\mathcal{D}}{_a}\mathcal{G}_j)E^a_j+2\gamma^2E^a_iE^b_jK_{[a}^iK_{b]}^j
\].}:
\[
-\sqrt{\det E}\,R=\frac{1}{\sqrt{\det E}}\left(\epsilon_{ijk}E^a_jE^b_kF_{ab}^i-2\gamma^2E^a_iE^b_jK_{[a}^iK_{b]}^j+2\,\leftidx{^\Gamma}{\mathcal{D}}{_a}(\mathcal{G}_jE^a_j)\right)\,.
\]
Substituting for curvature in (\ref{extendedH1}) gives the Hamiltonian in standard form
\[
H[N]=\int\md^3x\frac{N}{\sqrt{\det E}}\left[E^a_iE^b_j\left(\epsilon_{ijk}F_{ab}^k-2(1+\gamma^2)K_{[a}^iK_{b]}^j\right)+2\,\leftidx{^\Gamma}{\mathcal{D}}{_a}(\mathcal{G}_iE^a_i)\right]
\]
It is more convenient to use a slightly different extension for the Hamiltonian where the covariant derivative in the last term above is replaced by $\bar{\partial}_a$ ($\mathcal{G}_iE^a_i$ is a vector of density weight two and, contrary to what is stated in \cite{Thiemann}, its divergence does depend on the connection used:
\[
\leftidx{^\Gamma}{\mathcal{D}}{_a}(\mathcal{G}_iE^a_i)=\bar{\partial}_a(\mathcal{G}_iE^a_i)-\Gamma^b_{ba}(\mathcal{G}_iE^a_i)
\]
so by replacing $\leftidx{^\Gamma}{\mathcal{D}}{_a}$ with $\bar{\partial}_a$ in the expression for the Hamiltonian above we are dropping the term $-2N\Gamma^b_{ba}(\mathcal{G}_iE^a_i)/\sqrt{\det E}$. This term is at most $\Ord{r}{-3}{odd}$ for linear smearing field and plays no significant role in the discussions of finiteness and differentiability).
The expression for the Hamiltonian is manifestly finite for the fall-off behavior of the smearing field $N$ corresponding to gauge transformations (\ref{properGauge}). As expected however, this expression is not manifestly finite for smearing field $M$ corresponding to translations and it actually diverges for general $M$ containing boosts (\ref{PoincareLapseFallOff}). The trick is as in the ADM case to integrate by parts the seemingly divergent term $dA$ coming from curvature $F$, and the last term containing the derivative of the Gauss constraint:
\begin{align*}
H[M]=&\int\md^3x\frac{M}{\sqrt{\det E}}\bigg[E^a_iE^b_j\big(\epsilon_{ijk}\big(\underbrace{2\bar{\partial}_{[a}A_{b]}^k}_{\Ord{r}{-3}{even}}+\underbrace{\epsilon_{klm}A_a^lA_b^m}_{\Ord{r}{-4}{even}}\,\big) \\
&\qquad\qquad\qquad\quad-\underbrace{2(1+\gamma^2)K_{[a}^iK_{b]}^j}_{\Ord{r}{-4}{even}}\big)+2\,\underbrace{\bar{\partial}_a(\mathcal{G}_iE^a_i)}_{\Ord{r}{-3}{even}}\bigg]  \\
=&\int\md^3x\bigg[-2\bar{\partial}_a\left(\frac{M}{\sqrt{\det E}}\epsilon_{ijk}E^a_iE^b_j\right)A_b^k
-2\,\bar{\partial}_a\left(\frac{M}{\sqrt{\det E}}\right)E^a_i\mathcal{G}_i \\
&\qquad\qquad+\frac{M}{\sqrt{\det E}}E^a_iE^b_j\left(\epsilon_{ijk}\epsilon_{klm}A_a^lA_b^m-2(1+\gamma^2)K_{[a}^iK_{b]}^j\right)\bigg] \\
&+2\int\md S_a \frac{M}{\sqrt{\det E}}\left(\epsilon_{ijk}E^a_iE^b_jA_b^k+E^a_i\mathcal{G}_i\right) \\
=&\int\md^3x\bigg[-2\bar{\partial}_a\left(\frac{M}{\sqrt{\det E}}\right)\epsilon_{ijk}E^a_iE^b_jA_b^k 
-\frac{2M}{\sqrt{\det E}}\epsilon_{ijk}\bar{\partial}_a(E^a_iE^b_j)A_b^k \\
&\qquad\qquad-2\bar{\partial}_a\left(\frac{M}{\sqrt{\det E}}\right)E^a_i(\bar{\partial}_bE^b_i+\epsilon_{ijk}A_b^jE^b_k) \\
&\qquad\qquad+\frac{M}{\sqrt{\det E}}E^a_iE^b_j\left(2A_{[a}^iA_{b]}^j-2(1+\gamma^2)K_{[a}^iK_{b]}^j\right) \,\bigg] \\
&+2\int\md S_a \frac{M}{\sqrt{\det E}}E^a_i\bar{\partial}_bE^b_i \\
=&\int\md^3x\bigg[
\frac{M}{\sqrt{\det E}}\bigg(-\underbrace{2\epsilon_{ijk}\bar{\partial}_a(E^a_iE^b_j)A_b^k}_{\Ord{r}{-4}{even}} 
+\underbrace{E^a_iE^b_j\left(2A_{[a}^iA_{b]}^j-2(1+\gamma^2)K_{[a}^iK_{b]}^j\right)}_{\Ord{r}{-4}{even}} \bigg) \\
&\qquad\qquad-2\bar{\partial}_a\left(\frac{M}{\sqrt{\det E}}\right)\underbrace{E^a_i\bar{\partial}_bE^b_i}_{\Ord{r}{-2}{odd}} \,\bigg] \,+\,2\int\md S_a \frac{M}{\sqrt{\det E}}\underbrace{E^a_i\bar{\partial}_bE^b_i}_{\Ord{r}{-2}{odd}}  
\end{align*} 
For the most general fall-off behavior (\ref{PoincareLapseFallOff}) for $M$ containing boosts, the bulk integrand terms  proportional to $M/\sqrt{\det E}$ are $\Ord{r}{-3}{odd}$ hence giving convergent integrals. $\bar{\partial}_a(M/\sqrt{\det E})$ is $\Ord{r}{-1}{even}$ for $M$ containig up to time translations and $\Ord{r}{0}{even}$ for boosts. So for time translations the bulk integrand proportional to $\bar{\partial}_a(M/\sqrt{\det E})$ gives a manifestly finite integral but it diverges for boosts. Similarly, the surface integral is finite for translations and will be proportional to the ADM energy, but it is divergent for boosts. So just like in the ADM case, the extended $H[M]$ is also finite for $M$ corresponding to translations but as we will also see shortly it is not differentiable either.

Just like in the ADM case, directly performing a second integration by parts  on the divergent term will result in more divergent terms. One needs to `subtract' the asymptotic frame first to isolate the divergence in a surface integral:
\begin{align*}
&-2\int\md^3x\,\bar{\partial}_a\left(\frac{M}{\sqrt{\det E}}\right)E^a_i\bar{\partial}_bE^b_i=
-2\int\md^3x\,\bar{\partial}_a\left(\frac{M}{\sqrt{\det E}}\right)E^a_i\bar{\partial}_b(E^b_i-\bar{E}^b_i) \\
&=2\int\md^3x\,\bar{\partial}_b\left[\bar{\partial}_a\left(\frac{M}{\sqrt{\det E}}\right)E^a_i\right](E^b_i-\bar{E}^b_i)
\,-\,2\int\md S_b\,\bar{\partial}_a\left(\frac{M}{\sqrt{\det E}}\right)E^a_i(E^b_i-\bar{E}^b_i) \\
&=2\int\md^3x\,\Bigg[\underbrace{\bar{\partial}_b\,\bar{\partial}_a\left(\frac{M}{\sqrt{\det E}}\right)E^a_i(E^b_i-\bar{E}^b_i)}_{\Ord{r}{-3}{odd}}
+\underbrace{\bar{\partial}_a\left(\frac{M}{\sqrt{\det E}}\right)(\bar{\partial}_bE^a_i)(E^b_i-\bar{E}^b_i)}_{\Ord{r}{-3}{odd}} \Bigg]\\
&\quad-\,2\int\md S_b\,\underbrace{\bar{\partial}_a\left(\frac{M}{\sqrt{\det E}}\right)E^a_i(E^b_i-\bar{E}^b_i)}_{\Ord{r}{-1}{even}}
\end{align*}
In this form the bulk integral is manifestly finite and the divergence is contained in the surface term. Hence
\begin{align*}
H[M]&=\int\md^3x\frac{M}{\sqrt{\det E}}\left[E^a_iE^b_j\left(\epsilon_{ijk}F_{ab}^k-2(1+\gamma^2)K_{[a}^iK_{b]}^j\right)+2\bar{\partial}_a(\mathcal{G}_iE^a_i)\right] \\
&=\;\text{ convergent bulk terms } 
\;+\,2\int\md S_a \frac{M}{\sqrt{\det E}}E^a_i\bar{\partial}_bE^b_i \\
&\quad\quad-\,2\int\md S_b\,\bar{\partial}_a\left(\frac{M}{\sqrt{\det E}}\right)E^a_i(E^b_i-\bar{E}^b_i)
\end{align*}

One may therefore define
\begin{align}
\mathbf{G_H}[M]:&=H[M] -2\int\md S_a \left[ \frac{M}{\sqrt{\det E}}E^a_i\bar{\partial}_bE^b_i 
-\,\bar{\partial}_b\left(\frac{M}{\sqrt{\det E}}\right)E^b_i(E^a_i-\bar{E}^a_i)\,\right]  \notag\\
=&\int\md^3x\Bigg[
\frac{M}{\sqrt{\det E}}\bigg(-2\epsilon_{ijk}\bar{\partial}_a(E^a_iE^b_j)A_b^k 
+E^a_iE^b_j\left(2A_{[a}^iA_{b]}^j-2(1+\gamma^2)K_{[a}^iK_{b]}^j\right) \bigg) \notag\\
&\qquad\qquad+\bar{\partial}_b\,\bar{\partial}_a\left(\frac{M}{\sqrt{\det E}}\right)E^a_i(E^b_i-\bar{E}^b_i)
+\bar{\partial}_a\left(\frac{M}{\sqrt{\det E}}\right)(\bar{\partial}_bE^a_i)(E^b_i-\bar{E}^b_i) \Bigg]  \label{GHConn}
\end{align}
By construction $\mathbf{G_H}[M]$ is manifestly finite and it is relatively straight forward to show it is also differentiable.
One may take the variation directly in the last expression and verify all surface terms vanish at infinity or take the variation of $H[M]$ and the surface terms. 
Using $\delta(\det E)=(\det E)E_a^i\delta E^a_i$, and defining
\[
\Lambda_i[M]:=-2\bar{\partial}_a\left(\frac{M}{\sqrt{\det E}}\right)E^a_i
\]
we have
\begin{align}
\delta H[M]=&H[\frac{1}{2}ME_a^i\delta E^a_i] \notag\\
&+\int\md^3x\bigg[\bigg(\frac{2M}{\sqrt{\det E}}
E^b_j\left(\epsilon_{ijk}F_{ab}^k-2(1+\gamma^2)K_{[a}^iK_{b]}^j\right) -2\bar{\partial}_a\left(\frac{M}{\sqrt{\det E}}\right)\mathcal{G}_i \notag\\
&\qquad\qquad\quad+\left(-\bar{\partial}_a\Lambda_i[M]+\Lambda_j[M]\epsilon_{jk}\,^iA_a^k\right)\bigg)\delta E^a_i \notag\\
&\qquad\qquad\quad+ \frac{M}{\sqrt{\det E}}\bigg(2\Big(\bar{\partial}_b(\epsilon_{ijk}E^a_jE^b_k)+2E^{[a}_iE^{b]}_j\Gamma_b^j\Big)
-\frac{4}{\gamma}E^{[a}_iE^{b]}_jK_b^j
\bigg)\delta A_a^i \notag\\
&\qquad\qquad\quad+4\left(\frac{1+\gamma^2}{\gamma}\right)\frac{M}{\sqrt{\det E}}E^{[a}_iE^{b]}_jK_b^j\delta\Gamma_a^i\bigg] \notag\\
&+2\int\md S_a\ \left[\frac{M}{\sqrt{\det E}}\Big(\delta(E^a_i\bar{\partial}_bE^b_i)+\epsilon_{ijk}A_b^j\delta(E^a_iE^b_k)\Big)\,-\,\bar{\partial}_b\left(\frac{M}{\sqrt{\det E}}\right)E^b_i\delta E^a_i\,\right] \label{HConnVariation}
\end{align}
Already from this expression, it is simple to check that for proper gauge transformations (\ref{properGauge}) $H[N]$ is differentiable, that is, all bulk terms are finite and all the surface terms vanish in the limit $r\to\infty$. For the bulk and surface integrals resulting from the integration by parts of the term proportional to $\delta\Gamma_a^i$, one can see this from the formula for the spin connection in terms of the co-triad 
or the equivalent formula
\[
\Gamma^i_a=-\frac{1}{2}\epsilon^{ijk}E^b_j\left[2\bar{{\partial}}_{[a}E_{b]}^k+E_a^lE^c_k\bar{\partial}_cE_b^l-E^k_aE_c^l\bar{\partial}_bE^c_l\right]\,,
\]
so all resulting bulk integrand terms have either a form proportional to 
\[
\frac{M}{\sqrt{\det E}}\underbrace{(E\cdots E)(\bar{\partial} K)\delta E}_{\Ord{r}{-4}{even}}   \quad \text{or } \quad M\underbrace{(E\cdots E)(\bar{\partial}E)K\delta E}_{\Ord{r}{-5}{even}}
\]
and hence are finite even for the general fall-off  (\ref{PoincareLapseFallOff}) for $M$ corresponding to time translations and boosts. Similarly, the corresponding surface terms have a form proportional to
\[
M\underbrace{(E\cdots E)K\delta E}_{\Ord{r}{-3}{odd}}
\]
and vanish at infinity too, even for $M$ corresponding to time translations and boosts.

As already mentioned above, from (\ref{HConnVariation}) one can see $H[M]$, although finite, is not differentiable for $M$ containing up to constant terms corresponding to time translations since the first surface term
\[
2\int\md S_a\,\underbrace{\frac{M}{\sqrt{\det E}}\delta(E^a_i\bar{\partial}_bE^b_i)}_{\Ord{r}{-2}{odd}}
\]
does not vanish at infinity. Furthermore, as expected, for general $M$ containing terms corresponding to boosts, this and the last term
\[
-2\int\md S_a\,\underbrace{\bar{\partial}_b\left(\frac{M}{\sqrt{\det E}}\right)E^b_i\delta E^a_i}_{\Ord{r}{-1}{even}}
\]
generally diverge. Variation of the surface terms subtracted to $H[M]$ in (\ref{GHConn}) cancels precisely the offending terms and gives additional ones which vanish at infinity:
\begin{align*}
-2\,\delta\int\md S_a \Big[ \frac{M}{\sqrt{\det E}}E^a_i&\bar{\partial}_bE^b_i 
-\,\bar{\partial}_b\left(\frac{M}{\sqrt{\det E}}\right)E^b_i(E^a_i-\bar{E}^a_i)\,\Big]  \\
=-2\,\int\md S_a \bigg[ &\frac{M}{\sqrt{\det E}}\delta(E^a_i\bar{\partial}_bE^b_i) 
-\,\bar{\partial}_b\left(\frac{M}{\sqrt{\det E}}\right)E^b_i\delta E^a_i \\
&-\frac{M}{2\sqrt{\det E}}(E^a_i\bar{\partial}_bE^b_i)E_c^j\delta E^c_j 
-\,\delta\left(\bar{\partial}_b\left(\frac{M}{\sqrt{\det E}}\right)E^b_i\right)(E^a_i-\bar{E}^a_i)\,\bigg] 
\end{align*}
Finally, it is straight forward to check all the bulk terms in  (\ref{HConnVariation}) are also finite for the most general fall-off of the smearing field $M$, so $\mathbf{G_H}[M]$ is indeed differentiable (with respect to $A_a^i$ and $E^a_i$). 
Calculations above also show the extended Hamiltonian constraint plus the ADM energy surface term
\[
H[M]-2\int\md S_a \,\frac{M}{\sqrt{\det E}}E^a_i\bar{\partial}_bE^b_i 
\]
is differentiable with respect to canonical variables too for $M$ containing up to constant terms. 

As expected from the ADM expressions, $\mathbf{G_H}[M]$ is not differentiable with respect to $M$ for general variations consistent with Killing fields of the metric. Variation
\[
\delta_M\mathbf{G_H}[M]=H[\delta M] -2\int\md S_a \left[ \frac{\delta M}{\sqrt{\det E}}E^a_i\bar{\partial}_bE^b_i 
-\,\bar{\partial}_b\left(\frac{\delta M}{\sqrt{\det E}}\right)E^b_i(E^a_i-\bar{E}^a_i)\,\right]
\]
giving generally divergent bulk and nonzero or divergent surface terms.

The very last thing to check is that the surface counter-terms just obtained, indeed match the ADM energy surface term and the ADM relativistic angular momentum or `center of mass' surface terms. For this we note that the proof given in \cite{Thiemann} is incomplete (for more details see \cite{ourReview}).  Indeed, one can show
\begin{align} 
-2\int\md S_d\frac{M}{\sqrt{\det E}}E^d_i\bar{\partial}_cE^c_i 
=&\int\md S_d\,\sqrt{\det q}\,Mq^{ab}q^{cd}\left(\partial_bq_{ac}\,-\,\partial_cq_{ab}\right) \notag\\
&+\int\md S_d\,\sqrt{\det q}\,Mq^{ab}q^{cd}\left(e_a^i\bar{\partial}_be_c^i\,-\,e_c^i\bar{\partial}_be_a^i\right)\,.  \label{energySurfTermEquality}
\end{align}
To match the corresponding ADM surface term in (\ref{ADMHgenerator}) at infinity, the last integral
must vanish in the limit $r\to\infty$. One can easily check that the leading $\Ord{r}{-2}{}$-term in $2e_{[a}^i\bar{\partial}_{| b|}e_{c]}^i$ is identically zero, so for $M$ containing up to constant terms, this integral indeed vanishes and the left hand side of (\ref{energySurfTermEquality}) equals the ADM energy surface term. However, as opposed to what is implied in \cite{Thiemann}, for the general up to linear fall-off behavior (\ref{PoincareLapseFallOff}) for $M$ there could be a generically nonzero contribution from this last surface integral if one considers the full class of co-triads  admitting  an asymptotic expansion only to order one.  For the naive expansion of (\ref{energySurfTermEquality}) to guarantee the surface-counter terms derived here also match the relativistic angular momentum, one must restrict to co-triads admitting an asymptotic expansion to order two
\begin{equation} \label{cotriadExpansion22}
e^i_a=\bar{e}^i_a+\frac{f_a^i}{r}+\frac{\leftidx{^2}{f}{_a^i}}{r^2}+\ord{r}{-2}
\end{equation}
and so that $\leftidx{^2}{f}{_a^i}$ is even as a function on the sphere:
\[
\leftidx{^2}{f}{_a^i}(-x^c/r)=\leftidx{^2}{f}{_a^i}(x^c/r)\,.
\]
The same must be true to prove -from a direct expansion-  that the second surface counter-term matches  at infinity  the corresponding ADM surface term which contributes to define relativistic angular momentum:
 \begin{align*}
2\int\md S_d\,\bar{\partial}_c&\left(\frac{M}{\sqrt{\det E}}\right)E^c_i\left(E^d_i-\bar{E}^d_i\right)  
=\int\md S_d\,\sqrt{\det q}\,\left({\partial}_cM\right)\left(q^{cd}q^{ab}-q^{ca}q^{db}\right)\left(q_{ab}-\delta_{ab}\right)\,. 
\end{align*}
Nevertheless, an argument to show the surface terms match ADM terms is given in section \ref{s::Pgenerators}.



\end{document}